\documentclass[aps,preprint,amsmath,amssymb,amsfonts]{revtex4}
\usepackage{epsfig}
\usepackage{graphicx}
\usepackage{subfigure}
\usepackage{dcolumn}
\usepackage{bm}
\usepackage{amsthm}
\usepackage{enumitem}
\usepackage{slashed}
\usepackage{braket}

\newcommand{\del}{\partial}

\newcommand{\sign}{\operatorname{sign}}
\newcommand{\tr}{\operatorname{tr}}

\newcommand{\bbR}{\mathbb{R}}

\newcommand{\bbZ}{\mathbb{Z}}

\newcommand{\calD}{\mathcal{D}}

\newcommand{\calH}{\mathcal{H}}
\newcommand{\calN}{\mathcal{N}}
\newcommand{\calP}{\mathcal{P}}

\newcommand{\Langle}{\langle\!\langle}
\newcommand{\Rangle}{\rangle\!\rangle}
\newcommand{\Mid}{|\!|}

\begin{document}

\title{Holographic quantization of linearized higher-spin gravity in the de Sitter causal patch}

\author{Yasha Neiman}
\email{yashula@gmail.com}

\affiliation{Okinawa Institute of Science and Technology, 1919-1 Tancha, Onna-son, Okinawa 904-0495, Japan}

\date{\today}

\begin{abstract}
We study the dS/CFT duality between minimal type-A higher-spin gravity and the free $Sp(2N)$ vector model. We consider the bulk spacetime as ``elliptic'' de Sitter space $dS_4/\bbZ_2$, in which antipodal points have been identified. We apply a technique from arXiv:1509.05890, which extracts the quantum-mechanical commutators (or Poisson brackets) of the linearized bulk theory in an \emph{observable patch} of $dS_4/\bbZ_2$ directly from the boundary 2-point function. Thus, we construct the Lorentzian commutators of the linearized bulk theory from the Euclidean CFT. In the present paper, we execute this technique for the entire higher-spin multiplet, using a higher-spin-covariant language, which provides a promising framework for the future inclusion of bulk interactions. Aside from its importance for dS/CFT, our construction of a Hamiltonian structure for a bulk causal region should be of interest within higher-spin theory itself. The price we pay is a partial symmetry breaking, from the full dS group (and its higher-spin extension) to the symmetry group of an observable patch. While the boundary field theory plays a role in our arguments, the results can be fully expressed within a boundary \emph{particle mechanics}. Bulk fields arise from this boundary mechanics via a version of second quantization. 
\end{abstract}

\maketitle
\tableofcontents
\newpage

\section{Introduction and summary} \label{sec:intro}

\subsection{Motivation from dS/CFT} \label{sec:intro:dS}

Higher-spin (HS) gravity \cite{Vasiliev:1995dn,Vasiliev:1999ba} is a theory of interacting massless fields, which in its minimal version includes one field of every even spin. Like string theory, it admits a holographic description \cite{Klebanov:2002ja} within the framework of AdS/CFT \cite{Maldacena:1997re,Witten:1998qj,Aharony:1999ti}. In particular, it admits what is perhaps the simplest holographic dual -- a free vector model of massless scalar fields on the 3d boundary of 4d spacetime. Crucially, this is the only known model of AdS/CFT which appears to extend to the case of a positive cosmological constant \cite{Anninos:2011ui}, i.e. to dS/CFT. My long-term goal is to address, within this working model, the conceptual problems of quantum gravity in de Sitter space. In particular, I'm interested in the holographic emergence of causal structure and quantum-mechanical commutators in an observable bulk patch, i.e. in the region enclosed by a pair of cosmological horizons (also known as a ``causal patch'', or a ``static patch''). Indeed, a crucial difference between AdS and dS is that in the latter, the boundary is spacelike, and thus time and causality exist only in the bulk. Moreover, since commutators in QFT only arise at causal separation, the commutator algebra of quantum operators in de Sitter is also purely a bulk structure. In fact, in our view, the holographic emergence of causality is roughly the same as the emergence of quantum commutators. In practice, it is the bulk commutators that will be our focus in this paper. 

\begin{figure}%
	\centering%
	\includegraphics[scale=0.8]{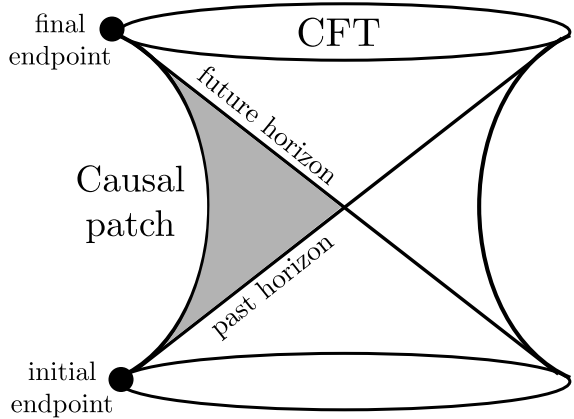} \\
	\caption{The basic problem of dS/CFT. De Sitter space is represented as a hyperboloid in 4+1d embedding space. The CFT is defined on the Euclidean conformal boundary. Lorentzian physics, including commutators of quantum fields, only takes place in the bulk. Moreover, the conformal boundary is unobservable. An observable bulk patch (shaded) intersects the boundary at two points -- the initial and final endpoints of the observer's worldline.}
	\label{fig:Problem} 
\end{figure}

Note that our focus on an \emph{observable} bulk patch is somewhat unusual in recent dS/CFT literature. The latter tends to focus on the Lorentzian bulk physics of either global de Sitter space or a Poincare patch. There, the CFT partition function defines a Hartle-Hawking state \cite{Maldacena:2002vr}. In particular, such is the viewpoint taken in \cite{Anninos:2011ui}. For a more current state of the art, see \cite{Anninos:2017eib}. The approach of \cite{Maldacena:2002vr,Anninos:2011ui,Anninos:2017eib} may be suitable for describing \emph{temporary} de Sitter phases, such as inflation, where the would-be future infinity of de Sitter eventually becomes observable. In contrast, we're considering a \emph{truly} asymptotically de Sitter spacetime. In such a universe, the future boundary is unobservable, and one must focus on a causal patch between a pair of cosmological horizons. This change of focus has considerable implications for the entire dS/CFT project. In particular, one cannot be content with any output that refers only to the unobservable conformal boundary. Instead, one must construct some extra dictionary between this boundary and the causal patch. Specifically, we will be interested in extracting causal-patch commutators from the boundary partition function. Since the causal patch and boundary only intersect at two points, any such dictionary must be non-local; see figure \ref{fig:Problem}. As we will see, in this context, the non-locality of higher-spin theory will be just what the doctor prescribed. 

The problem of extracting causal-patch commutators from the boundary partition function was first formulated and solved in \cite{Halpern:2015zia}. There, we were working with \emph{individual} free massless fields, i.e. with the spectrum of higher-spin gravity, not yet arranged into an HS-covariant multiplet. Note that the commutator algebra for free fields is equivalent to the symplectic structure on their classical phase space. In \cite{Halpern:2015zia}, we extracted this structure (along with the bulk fields' Hamiltonian) out of the boundary CFT's 2-point function, with the aid of some kinematical structures associated with the choice of observer (which themselves can be formulated in boundary terms). A crucial step was to consider, following \cite{Parikh:2002py}, a causal patch embedded not quite in $dS_4$, but in so-called ``elliptic'' de Sitter space $dS_4/\bbZ_2$ \cite{Folacci:1986gr} -- a topologically modified version, in which antipodal points (in both space and time) are identified. In this setup, the phase space in the bulk causal patch can be identified with the space of boundary sources. It is this identification that enables us to express bulk structures in boundary terms.

A central goal of my research is to extend the free-field construction of \cite{Halpern:2015zia} to the full interacting bulk theory. For this, one must leave behind the local language of fields, and adopt one that is more compatible with HS symmetry, and hence with the peculiar interactions of higher-spin gravity. This is the step that will be accomplished in the present paper. We will recast the free-field result of \cite{Halpern:2015zia} into a closed-form expression for the entire HS multiplet, using variables in which the HS symmetry is manifest, and the partition function (not just its 2-point piece) is easy to write down in full. This will hopefully open the door for a Lorentzian bulk interpretation also of the CFT's higher $n$-point functions, which correspond to bulk interactions.

Our variables for encoding the bulk fields and boundary sources will be functions of two $O(3)$ spinors, living at a boundary endpoint of the causal patch. This language is essentially the on-shell (and thus gauge-invariant) version \cite{Koch:2010cy,Das:2012dt,Koch:2014aqa} of the boundary vector model's bilocal description \cite{Das:2003vw}. In particular, our spinors are the square roots of the boundary fields' on-shell momenta. A closely related formalism has appeared e.g. in \cite{Vasiliev:2012vf,Gelfond:2008ur}. While these spinor variables are not unknown, it appears that their utility for dS/CFT has not been fully realized. In our treatment, they provide a first proof-of-concept for the non-local dictionary between the boundary and the causal patch, by encoding the information in both at their intersection point. Furthermore, the momentum spinors' sign ambiguity turns out to be beautifully related to the fact that de Sitter space is ``twice too big''. Ordinarily, the CFT is blind to the fact that $dS_4$ has \emph{two} conformal boundaries, which can be identified via the $\bbZ_2$ antipodal map. As we will see, this antipodal map ends up encoded as a sign flip of the momentum spinors, which does not affect their squares, i.e. the momentum vectors.

As an intermediary between the boundary momentum-spinor variables and the bulk fields, we will use spacetime-independent twistor functions, which were introduced to higher-spin holography in \cite{Neiman:2017mel}. As the most covariant description of higher-spin holography, one might expect that this twistor language should be our exclusive tool. However, it turns out to be ``too complex'' for our purposes: it does not allow for reality conditions compatible with the Lorentzian signature and symmetries of the causal patch. This is what led us to the momentum-spinor formalism.

It is worth clarifying the relationship between the present work and higher-spin AdS/CFT. We treat the HS AdS/CFT duality of \cite{Klebanov:2002ja} as a given. Rather than attempt to prove or test it, our focus is on how to \emph{use} it in the de Sitter context. In the Hartle-Hawking-state approach of \cite{Maldacena:2002vr}, one uses the Euclidean AdS/CFT quite directly: the CFT partition function defines a Euclidean AdS path integral, which, through straightforward analytical continuation, is reinterpreted as a Hartle-Hawking wavefunction for de Sitter space. In our approach, this step is not disputed, but rather put to further use, as we read off from this global Hartle-Hawking wavefunction some useful information about physics in the causal patch.

Finally, we should acknowledge the ``worldline holography'' of \cite{Anninos:2011af} as another tentative approach to quantum gravity in the de Sitter causal patch. Perhaps the central difference between our work and that of \cite{Anninos:2011af} is that the latter uses bulk conformal symmetry (specifically, the $SL(2,\bbR)$ group extending the observer's time translations) as an organizing principle. We avoid the use of bulk conformal symmetry, since it is not a property of HS interactions, or even of the HS formulation of free bulk fields. For us, the organizing principle is HS symmetry. Just like the conformal group, it extends bulk isometries, but it does so in a different direction, and to a greater extent.

\subsection{Motivation from higher-spin gravity} \label{sec:intro:HS}

Due to its extreme non-locality, the relationship between higher-spin theory and spacetime can be recast in a number of forms. As has long been recognized, this makes the theory a promising candidate for a fully tractable realization of holography. On the other hand, this same non-locality poses serious challenges for the understanding of causality in the bulk. One way to phrase this difficulty is to consider the standard formalism of higher-spin gravity, i.e. bulk master fields satisfying unfolded field equations. Since the master field at each point contains the entire Taylor series of the fields' derivatives, at first sight causality becomes meaningless: it is always possible to propagate the entire solution from one point to another, regardless of their causal separation. How can a language that automatically encompasses all of spacetime be used to study the causal structure due to e.g. a bulk horizon?

The causal patch of a de Sitter observer may be the perfect setup for navigating this contradiction. This is particularly true in the context of ``elliptic'' de Sitter space $dS_4/\bbZ_2$. On one hand, an observer's causal patch in $dS_4/\bbZ_2$ covers all of space, so that field solutions inside are in one-to-one correspondence with solutions on the entire spacetime. On the other hand, $dS_4/\bbZ_2$ has no global time orientation, and thus no observer-independent notion of causality. Instead, causal structure is induced by a choice of observer, and is restricted to the interior of their causal patch. Thus, we get to explore a causal structure associated with the observer's horizons, while still dealing with global field solutions on the entire spacetime!

In section \ref{sec:intro:dS}, we argued that the question of causality is intimately related to the question of quantum commutators, or, at the classical level, to the symplectic structure of phase space. For an explorer of higher-spin theory, the situation appears similar. Our lack of understanding of bulk causality appears closely related to the lack of a known Hamiltonian structure. Even at the level of free fields, which individually obey the standard Hamiltonian structure of local field theory, an HS-symmetric formulation of this structure is hard to come by. One notable example is \cite{Koch:2014aqa}, which uses the on-shell bilocal formulation of the CFT to construct the bulk Hamiltonian structure in lightcone coordinates (which, as usual, partially break spacetime symmetry, and thus HS symmetry). In the present paper, we will perform a similar task, with two important differences. First, since we're working in dS, we will not be using an existing causal structure on the conformal boundary. Second, while our global higher-spin symmetry will also be partially broken, this will \emph{not} be an artifact of the formalism, but a true reflection of the reduced symmetry in a given causal patch. 

As an aside, there seems to be \emph{no} symplectic structure for higher-spin fields in (A)dS\textsubscript{4} with the full higher-spin symmetry $\mathfrak{hs}[O(1,4)]$ or $\mathfrak{hs}[O(2,3)]$. Indeed, there are only two bilinears that can be constructed out of a pair of twistor functions $f(Y),g(Y)$ in the adjoint representation of HS symmetry:
\begin{align}
 \begin{split}
   \left.f(Y)\star g(Y)\right|_{Y=0} &= \int d^4U d^4V f(U)g(V) e^{-iUV} \ ; \\ 
   \left.f(Y)\star g(Y)\star\delta(Y)\right|_{Y=0} &= \int d^4U f(U)g(U) \ ,
 \end{split} \label{eq:bilinears}
\end{align}
both of which are symmetric under $f\leftrightarrow g$. Thus, \emph{there is no antisymmetric form compatible with full HS symmetry}. Let us briefly reflect on this, perhaps surprising, conclusion. In $dS_4/\bbZ_2$, it is no surprise at all: there, even ordinary field theories lack a global symplectic structure \cite{Hackl:2014txa}, due to the lack of global time orientation. For $AdS_4$, a similar comment applies: if we assume its $O(2,3)$ symmetry \emph{globally}, it requires a compact time direction, i.e. again no globally consistent causal structure. For ordinary bulk theories, this isn't usually a problem: one can either restricting to a patch or decompactify the time direction, leading to a version of $AdS_4$ that is both causally consistent and locally indinstinguishable from the truly $O(2,3)$-symmetric one. However, with the non-local symmetries of HS theory, we should expect the spacetime's global structure to play a greater role, and for ``merely topological'' violations of $O(2,3)$ to become important. This leaves us with one final candidate for a fully HS-symmetric Hamiltonian structure -- global $dS_4$. This spacetime \emph{does} admit a global time orientation, and thus an $O(1,4)$-symmetric symplectic structure for each field in the HS multiplet. Here, the absence of an HS-symmetric symplectic form is perhaps the most surprising. It may be attributed to the complex nature of twistor space, which prevents the HS-covariant twistor language from seeing the causal structure of real Lorentzian spacetime. 

In contrast, when we focus on a de Sitter causal patch, the $O(1,4)$ de Sitter symmetry is broken down to the observer's $SO(1,1)\times O(3)$ symmetry of time translations and rotations. Higher-spin symmetry is broken down accordingly to $\mathfrak{hs}[SO(1,1)\times O(3)]$ (which, as we'll see, is simpler than it looks, once written in spinor language). Finding a symplectic form that respects this \emph{reduced} HS symmetry is simple: essentially, one can just weight the integrands in \eqref{eq:bilinears} with some odd function $K(\omega)$ of the observer's time translation generator $\omega$. Thus, the invariant symplectic forms are parameterized by a function of one variable. One of these, as we'll identify in this paper, is the correct one for the causal-patch phase space. The choice of causal patch effectively bypasses the ``blindness'' of twistor space to the Lorentzian causal structure, by associating ``past'' and ``future'' not just to an orientation of lightcones, but to a concrete \emph{pair of points} -- the past and future endpoints of the observer's worldline.

In practice, we will not be writing the symplectic form in twistor language: as mentioned previously, twistors are ``too complex''. In particular, the integrals in \eqref{eq:bilinears} are along unspecified complex contours. This is in contrast to the AdS case, where twistor space has a reality structure -- see e.g. \cite{Gelfond:2013xt}. Instead of twistors, we will use momentum spinors, for which a natural real contour will be available. 

\subsection{Summary and structure of the paper} \label{sec:intro:summary}

This paper's main result is a holographic derivation of the Hamiltonian structure -- symplectic form, commutators and Hamiltonian -- of linearized HS fields in a causal patch of $dS_4/\bbZ_2$. The paper can be framed in terms of two nested stories about quantum mechanics in phase space. Thus, we begin in section \ref{sec:WW} by introducing/reviewing some phase space techniques. The section focuses on the Wigner-Weyl transform between quantum operators and phase space functions. An important ingredient will be the treatment of operators on a Hilbert space $\calH$ as \emph{states} in the squared Hilbert space $\calH\otimes\calH^*$.

In section \ref{sec:fields}, we review the main result of \cite{Halpern:2015zia}, while rephrasing it in a phase-space-covariant manner. Specifically, let the boundary sources in dS/CFT be parameterized by a (infinite-dimensional) vector $\xi^I$. The CFT 2-point function is some quadratic form $G_{IJ}$ on this space. This is all the information that the CFT can give us with regard to linearized bulk fields. On the boundary, the $O(1,4)$ de Sitter symmetries are realized kinematically as conformal transformations. When acting on the boundary sources, these take the form of some linear transformations $M^I{}_J$. A \emph{choice of bulk observer} singles out two of these symmetries: the generator $\omega^I{}_J$ of $SO(1,1)$ time translations, and the $O(3)$ parity operation $\calP^I{}_J$. Now, recall that in $dS/\bbZ_2$, the phase space of bulk fields in the causal patch is the same as the space of boundary sources. Thus, the symplectic form on the causal-patch phase space is a matrix $\Omega_{IJ}$ on this same space! The main result of section \ref{sec:fields} is an expression for $\Omega_{IJ}$ in terms of the above boundary inputs -- the 2-point function $G_{IJ}$ and the kinematical symmetries $\omega^I{}_J$ and $\calP^I{}_J$:
\begin{align}
 \Omega_{IJ} = iG_{IK}\left( \coth(\pi\omega) \pm \frac{\calP}{\sinh(\pi\omega)} \right)^K{}_J \ , \label{eq:Omega_summary}
\end{align}
where the sign depends on whether we're considering a single bulk field or the entire higher-spin multiplet (in the latter case, a subtle rearrangement of the boundary sources takes place). From the symplectic form \eqref{eq:Omega_summary}, we can immediately obtain the commutators and Hamiltonian for the causal-patch fields (see eq. \eqref{eq:comm_H_symmary} below).

In section \ref{sec:multiplet}, we proceed to transform the abstract result \eqref{eq:Omega_summary} into an explicit compact expression for the entire HS multiplet. For this purpose, it will be useful to consider HS holography itself as an example of quantum mechanics in phase space -- specifically, the particle mechanics that underlies the CFT's fields. In this view, twistor space is just (a double cover of) the boundary particle's phase space. The twistor metric $I_{ab}$ is the symplectic form on this phase space. A twistor function $F(Y)$ is now a function on the phase space, which corresponds via the Wigner-Weyl transform to a quantum-mechanical operator. Higher-spin algebra is just the quantum-mechanical algebra of these operators. On the other hand, via the Penrose transform \cite{Penrose:1986ca,Ward:1990vs}, a twistor function $F(Y)$ describes a bulk field solution. Combining these two maps, we conclude that the \emph{space of operators} of the boundary particle mechanics is the \emph{classical phase space} of the bulk fields. Viewed in this way, HS holography becomes a ``doubled'' version of second quantization! The CFT partition function in this language becomes a functional on quantum operators $\hat F$ in the boundary particle mechanics:
\begin{align}
 Z_{\text{CFT}}[\hat F] = \exp\left(N\tr\ln[1 + \hat F] \right) \equiv \det[1 + \hat F]^N \ .
\end{align}
Here, the trace is over the particle's Hilbert space, and the log function is the one induced by the operator product. 

The momentum spinors we mentioned above form a ``configuration basis'' for states and operators in the boundary particle mechanics. This basis is singled out by the same symmetry breaking that's associated with a choice of observer in $dS_4$. Using this basis, one can express an operator $\hat F$ in terms of matrix elements $\braket{u|\hat F|u'}$. Here, both the spinors $u$ and $u'$ are complex, but we can choose a ``real contour'' by setting $u' = i\bar u$. In these variables, the CFT 2-point function $G_{IJ}$ takes the form:
\begin{align}
 G[\hat F,\hat F] = \frac{N}{2}\int d^4u\,\Braket{u|\hat F|i\bar u}\Braket{iu|\hat F|\bar u} \ ,
\end{align}
while time translations $\omega^I{}_J$ and parity reflections $\calP^I{}_J$ simply rescale $(u,\bar u)$. Plugging into \eqref{eq:Omega_summary}, we will obtain the causal patch symplectic form as:
\begin{align}
 \Omega_{\text{bulk}}[\hat F_1,\hat F_2] = -\frac{N}{4\pi}\int d^4u \Braket{u|\hat F_1|i\bar u}
    \int_{-\infty}^\infty d\alpha\sign(\alpha)\,\frac{\alpha - 1}{\alpha + 1}\Braket{\sqrt{\alpha}\, u|\hat F_2|\sqrt{\alpha}\,i\bar u} \ . \label{eq:Omega_explicit_summary}
\end{align}
Knowledge of $\Omega_{\text{bulk}}$ immediately gives us the commutators and Hamiltonian for HS fields in the causal patch, via:
\begin{align}
 \left[\hat{\hat\xi}^I,\hat{\hat\xi}^J \right] = i(\Omega^{-1})^{IJ} \ ; \quad \hat{\hat H} = -\frac{i}{2}\,\hat{\hat\xi}^I \Omega_{IJ}\omega^J{}_K \hat{\hat\xi}^K \ , \label{eq:comm_H_symmary}
\end{align}
where the double hats remind us that we're dealing with a second quantization: the \emph{classical phase space} $\xi^I$ in the bulk consists of \emph{quantum operators} $\hat F$ in the boundary particle mechanics. Using the symplectic form \eqref{eq:Omega_explicit_summary}, we will obtain explicit expressions for the bulk commutators:
\begin{align}
 \begin{split}
   &\left[ \widehat{\braket{u|\hat F|\bar u}}, \widehat{\braket{v|\hat F|\bar v}} \right] = \frac{1}{4\pi iN}\int_{-\infty}^\infty d\alpha\,\frac{\alpha - 1}{\alpha + 1} \\
   &\qquad \times \left(\delta^4\!\left(v - \sqrt{\alpha}\,u\right) + \delta^4\!\left(v + \sqrt{\alpha}\,u\right) + \delta^4\!\left(v - \sqrt{\alpha}\,\bar u\right) + \delta^4\!\left(v + \sqrt{\alpha}\,\bar u\right) \right) \ ,
 \end{split}
\end{align}
as well as for the Hamiltonian:
\begin{align}
 \hat{\hat H} = \frac{N}{4\pi}\int d^4u\,\widehat{\Braket{u|\hat F|i\bar u}} \int_{-\infty}^\infty \frac{|\alpha|d\alpha}{(\alpha + 1)^2}\,\widehat{\Braket{\sqrt{\alpha}\,u|\hat F|\sqrt{\alpha}\,i\bar u}} \ . \label{eq:H_summary}
\end{align}
One can also work in a bulk frequency basis, arrange the bulk fields into creation and annihilation operators, and rewrite the Hamiltonian \eqref{eq:H_summary} in normal ordering. This will be accomplished in section \ref{sec:multiplet:symplectic:normal}. The HS symmetry of our construction is analyzed in section \ref{sec:multiplet:symmetry}. Section \ref{sec:discuss} is devoted to discussion and outlook.

\section{The Wigner-Weyl transform and doubled phase space} \label{sec:WW}

The Wigner-Weyl transform \cite{Wigner} between functions on phase space and operators on Hilbert space plays a crucial role in this paper. In this section, we review three useful definitions of this transform. Our focus is on quantum systems that can be obtained by quantizing a linear phase space. One such system is linearized HS gravity in the 4d bulk. Another example, as we'll see in section \ref{sec:multiplet:mechanics}, is a free massless particle on the 3d boundary.

\subsection{Symmetric ordering and the Moyal star product} \label{sec:WW:ordering}

Consider a linear phase space, parameterized by $2\calN$-dimensional vectors $\xi^I$, where $\calN$ is the (possibly infinite) number of degrees of freedom. The phase space is equipped with a symplectic form:
\begin{align}
 \Omega = \frac{1}{2}\Omega_{IJ} d\xi^I d\xi^J \ ,
\end{align} 
with inverse $(\Omega^{-1})^{IJ}$. The linear canonical transformations that preserve this form make up the symplectic group $Sp(2\calN)$. The Heisenberg commutators are given by:
\begin{align}
 [\hat\xi^I,\hat\xi^J] = i(\Omega^{-1})^{IJ} \ . \label{eq:commutators}
\end{align}
Any operator on the system's Hilbert space can be constructed as a sum of products of the fundamental operators $\hat\xi^I$. Given some ordering convention for such products, we can treat their coefficients as the Taylor expansion of a function on the \emph{classical} phase space. Thus, any ordering convention yields a 1-to-1 map between operators $\hat\Psi$ on the Hilbert space and functions $\Psi(\xi^I)$. The only ordering convention that preserves the full $Sp(2\calN)$ group is Weyl ordering, in which the classical product $\xi^{I_1}\dots\xi^{I_n}$ is mapped to the fully symmetrized operator product:
\begin{align}
 \xi^{I_1}\dots\xi^{I_n} \quad \longleftrightarrow \quad \hat\xi^{(I_1}\dots\hat\xi^{I_n)} \ . \label{eq:Weyl_ordering}
\end{align}
To see that \eqref{eq:Weyl_ordering} is indeed unique, note that the LHS is completely symmetric in its indices, while any $Sp(2\calN)$-symmetric corrections to the RHS would have to be constructed using the antisymmetric $\Omega_{IJ}$. Operator orderings other than \eqref{eq:Weyl_ordering}, such as normal ordering, always involve some $Sp(2\calN)$-breaking extra structure on the phase space, such as a decomposition into positive and negative frequencies. The map \eqref{eq:Weyl_ordering} between operators $\hat\Psi$ and phase space functions $\Psi(\xi^I)$ is our first definition of the Wigner-Weyl transform. The transform defines a non-commutative product on phase space functions, which reproduces the product of the corresponding operators. This is the Moyal star product $\ast$, defined by the following equivalent formulas:
\begin{align}
 \begin{split}
   \xi^I\ast\xi^J &= \xi^I\xi^J + \frac{i}{2}(\Omega^{-1})^{IJ} \ ; \\
   \Phi(\xi)\ast\Psi(\xi) &= \Phi(\xi) \exp\left(\frac{i}{2}(\Omega^{-1})^{IJ}\overleftarrow{\frac{\del}{\del\xi^I}}\overrightarrow{\frac{\del}{\del\xi^J}} \right) \Psi(\xi) \\
    &= 2^{2\calN}\int d^{2\calN}\!\xi' d^{2\calN}\!\xi''\, \Phi(\xi + \xi')\,\Psi(\xi + \xi'')\,e^{-2i\Omega_{IJ}\xi'^I\xi''^J} \ .
 \end{split} \label{eq:Moyal}
\end{align} 
In the last formula, we use the phase space measure derived from $\Omega_{IJ}$:
\begin{align}
 d^{2\calN}\!\xi = \pm\frac{\Omega_{I_1 I_2}\dots\Omega_{I_{2\calN-1} I_{2\calN}} d\xi^{I_1}d\xi^{I_2}\dots d\xi^{I_{2\calN-1}}d\xi^{I_{2\calN}}}{2^\calN \calN!(2\pi)^\calN} \ . \label{eq:phase_space_measure}
\end{align}
where we inserted a slightly nonstandard $1/(2\pi)^\calN$ factor, effectively switching from units of $\hbar$ to units of $h$. This will prevent any $2\pi$ factors in our Fourier and Gaussian integrals. In particular, with this definition, we have:
\begin{align}
 \int d^{2\calN}\!\xi' e^{i\Omega_{IJ}\xi'^I\xi^J} \equiv \delta(\xi) \ ; \quad \int d^{2\calN}\!\xi\,\delta(\xi) = 1 \ . \label{eq:phase_space_integrals}
\end{align}
Note also that we left the sign in \eqref{eq:phase_space_measure} undetermined. Normally, this sign is arbitrary: one always has the complementary freedom to choose the order of integration limits, so that e.g. the integral of 1 comes out positive. However, the freedom to choose the sign of $d^{2\calN}\xi$ will be important for us in section \ref{sec:multiplet}, where we'll consider twistors as a \emph{complex} phase space.

Finally, Hermitian conjugation of an operator $\hat \Psi\rightarrow \hat\Psi^\dagger$ is represented by ordinary complex conjugation of the phase space function $\Psi(\xi) \rightarrow \bar\Psi(\xi)$. Operator products are reversed by Hermitian conjugation, and the star product \eqref{eq:Moyal} transforms accordingly as $\Phi(\xi)\ast\Psi(\xi) \rightarrow \bar\Psi(\xi)\ast \bar\Phi(\xi)$.

\subsection{Doubled phase space and the Hilbert space of operators} \label{sec:WW:doubled}

For our second definition of the Wigner-Weyl transform, we consider a doubling of the phase space $\xi^I$ into a $4\calN$-dimensional space with coordinates $(\xi_+^I,\xi_-^I)$, equipped with the symplectic form:
\begin{align}
\Omega_{\text{double}} = \frac{1}{2}\Omega_{IJ}(d\xi_+^I d\xi_+^J - d\xi_-^I d\xi_-^J) \ . \label{eq:Omega_future_past}
\end{align}
In other words, under $\Omega_{\text{double}}$, the $\xi_+^I$ form an isomorphic copy of the original phase space, while $\xi_-^I$ form a time-reversed copy. We will later see this picture realized geometrically in de Sitter space. 

At the quantum level, the doubled phase space \eqref{eq:Omega_future_past} corresponds to a doubled Hilbert space of the form $\calH\otimes\calH^*$. The anti-isomorphism $(\xi_+^I,0)\leftrightarrow(0,\xi_-^I)$ between the ``upright'' and time-reversed copies becomes the complex conjugation $\ket{\psi} \leftrightarrow \bra{\psi}$ between bras and kets. To see why this map should be anti-linear, note that flipping $\Omega_{IJ}\rightarrow-\Omega_{IJ}$ in the commutator \eqref{eq:commutators} can be canceled by flipping $i\rightarrow -i$. Thus, we are led to two equivalent interpretations of the space $\calH\otimes\calH^*$. On one hand, it is the Hilbert space of states $\Mid\Psi\Rangle$ of the doubled system \eqref{eq:Omega_future_past}. On the other hand, it is of course the space of operators $\hat\Psi$ on the original Hilbert space $\calH$. The inner product on the doubled Hilbert space can be expressed via the operator algebra on $\calH$ as:
\begin{align}
\Langle\Psi\Mid\Phi\Rangle = \tr(\Psi^\dagger\Phi) \ .
\end{align}

Explicitly, one often likes to define states as wavefunctions on some ``configuration space'', i.e. on a polarization of phase space. In the original phase space $\xi^I$, we do not have a preferred choice of polarization. However, in the doubled phase space \eqref{eq:Omega_future_past}, we do! Indeed, let us define the even and odd combinations:
\begin{align}
 \xi_{\text{even}}^I = \frac{\xi_+^I + \xi_-^I}{2} \ ; \quad \xi_{\text{odd}}^I = \frac{\xi_+^I - \xi_-^I}{2} \ . \label{eq:xi_even_odd}
\end{align}
The doubled symplectic form \eqref{eq:Omega_future_past} now reads:
\begin{align}
 \Omega_{\text{double}} = 2\Omega_{IJ} d\xi_{\text{even}}^I d\xi_{\text{odd}}^J \ . \label{eq:Omega_even_odd}
\end{align}
Thus, either $\xi_{\text{even}}^I$ or $\xi_{\text{odd}}^I$ can serve as a preferred polarization of the doubled phase space, with the other set then serving as the canonically conjugate ``momenta''.

One can now express states $\Mid\Psi\Rangle$ on the doubled Hilbert space $\calH\otimes\calH^*$ as wavefunctions in either the $\xi_{\text{even}}^I$ basis or the $\xi_{\text{odd}}^I$ basis:
\begin{align}
 \Psi(\xi^I) &= \Langle\xi_{\text{even}}^I=\xi^I\Mid\Psi\Rangle \ ; \label{eq:Psi_even} \\ 
 \tilde\Psi(\xi^I) &= \Langle\xi_{\text{odd}}^I=\xi^I\Mid\Psi\Rangle \ . \label{eq:Psi_odd}
\end{align} 
Our notation here emphasizes that, since all the copies of our original phase space are isomorphic, the same coordinates $\xi^I$ can denote a point either in the $\xi^I_{\text{even}}$ subspace, or in the $\xi^I_{\text{odd}}$ subspace. These two options can be visualized by putting the classical history described by $\xi^I$ in the ``upright'' phase space copy $\xi_+^I$, and then putting either $+\xi^I$ or $-\xi^I$ in the ``time-reversed'' copy $\xi_-^I$.

Combining the configuration basis \eqref{eq:Psi_even} for states in $\calH\otimes\calH^*$ with the interpretation of these states as \emph{operators} on $\calH$, we arrive at a map between phase space functions $\Psi(\xi^I)$ and operators $\hat\Psi$:
\begin{align}
 \Psi(\xi^I) \ \longleftrightarrow \ \Mid\Psi\Rangle \ \longleftrightarrow \ \hat\Psi \ . \label{eq:WW_doubled}
\end{align}
As we'll see in the next subsection, this map is once again the Wigner-Weyl transform defined by \eqref{eq:Weyl_ordering}. We will return to the role of $\tilde\Psi(\xi^I)$ in section \ref{sec:WW:delta_Gaussians}.

\subsection{Explicit expression in canonical coordinates} \label{sec:WW:explicit}

In this subsection, we present an explicit expression for the Wigner-Weyl transform in terms of canonical coordinates on the (original, non-doubled) phase space.  We will start from the construction of section \ref{sec:WW:doubled}, and show that it agrees with that of section \ref{sec:WW:ordering}. 

To begin, we decompose the system's phase space into ``coordinates'' and ``momenta'' $\xi^I = (q^i,p_i)$, such that the symplectic form reads $\Omega = dp_i dq^i$. This decomposition is not covariant under the symplectic group $Sp(2\calN)$, but it will help us connect the two covariant definitions \eqref{eq:Weyl_ordering},\eqref{eq:WW_doubled} of the Wigner-Weyl transform. To construct wavefunctions using $q^i$ and $p_i$, we must also decompose the phase space measure $d^{2\calN}\!\xi$ as defined by $\Omega$ into a product $d^{2\calN}\!\xi = d^\calN\!q\,d^\calN\!p$, such that:
\begin{align}
 \begin{split}
   &\hat 1 = \int d^\calN\!q \ket{q}\!\bra{q} = \int d^\calN\!p \ket{p}\!\bra{p} = \int d^{2\calN}\!\xi\,e^{ip_i q^i}\ket{q}\!\bra{p} \ ; \\
   &\braket{q'|q} = \delta(q-q') \ ; \quad \braket{p'|p} = \delta(p-p') \ ; \quad \braket{q|p} = e^{ip_i q^i} \ ,
 \end{split} \label{eq:q_p_basis}
\end{align}
where the $2\pi$ factors hidden in \eqref{eq:phase_space_measure} are again working in our favor. In the doubled phase space of section \ref{sec:WW:doubled}, we similarly decompose $\xi_\pm^I = (q_\pm^i,p^\pm_i)$, and likewise with $\xi_{\text{even}}^I$ and $\xi_{\text{odd}}^I$. The doubled symplectic form \eqref{eq:Omega_future_past} becomes:
\begin{align}
 \Omega_{\text{double}} = dp_i^+ dq^i_+ - dp_i^- dq^i_- = 2(dp_i^{\text{odd}} dq^i_{\text{even}} + dp^{\text{even}}_i dq^i_{\text{odd}})  \ . \label{eq:Omega_double_pq}
\end{align}
Thus, we now have two possible polarizations on the doubled phase space. On one hand, as in the previous subsection, we can choose the ``configuration space'' $\xi^I_{\text{even}} = (q^i_{\text{even}},p_i^{\text{even}})$, with conjugate momenta $-2\Omega_{IJ} \xi_{\text{odd}}^J = (2p_i^{\text{odd}},-2q^i_{\text{odd}})$. On the other hand, we can choose the ``naive'' configuration space $(q_+^i,q_-^i)$, with conjugate momenta $(p^+_i,-p^-_i)$. Note that the latter choice respects the factorization $\calH\otimes\calH^*$ of the doubled Hilbert space.

The Wigner-Weyl transform can now be constructed as follows. Start from an operator $\hat\Psi$ on $\calH$, which can be represented by matrix elements $\langle q_+|\hat\Psi|q_-\rangle$ in the configuration basis. In the doubled picture, this corresponds to a state $\Mid\Psi\Rangle$, with wavefunction $\Langle q_+^i,q_-^i\Mid\Psi\Rangle$ in the $(q_+^i,q_-^i)$ basis. The transformation of this wavefunction into the $(q_{\text{even}}^i,p^{\text{even}}_i)$ basis will yield, by definition, the Wigner-Weyl transform \eqref{eq:WW_doubled}. To accomplish this transformation, we simply leave $q_{\text{even}}^i = (q_+^i + q_-^i)/2$ untouched, and Fourier transform between $2q^i_{\text{odd}} = q_+^i - q_-^i$ and its canonical conjugate $p_i^{\text{even}}$:
\begin{align}
 \Psi(q^i,p_i) = 2^\calN \int d^\calN\!q_{\text{odd}} \Braket{q + q_{\text{odd}} | \hat\Psi | q - q_{\text{odd}}} e^{-2i p_i q^i_{\text{odd}}} \ . \label{eq:WW_explicit}
\end{align}
We can now verify that this agrees with the symmetric-ordering definition \eqref{eq:Weyl_ordering} of the Wigner-Weyl transform. Indeed, substituting $\hat\Psi = (\alpha_i \hat q^i)^n$ on the RHS of \eqref{eq:WW_explicit} immediately yields $\Psi(q,p) = (\alpha_i q^i)^n$ on the LHS, for any coefficients $\alpha_i$ and any power $n$. Since the polarization $\xi^I = (q^i,p_i)$ was arbitrary, the same must be true for any $\hat\Psi = (\alpha_I \hat \xi^I)^n$. The symmetric ordering prescription \eqref{eq:Weyl_ordering} then follows by linear superposition.

As a side benefit, we can read off from \eqref{eq:WW_explicit} that the operator trace is given simply by a phase space integral:
\begin{align}
 \tr\hat\Psi = \int d^\calN\!q \braket{q|\hat\Psi|q} = \int d^{2\calN}\xi\,\Psi(\xi) \ .
\end{align}

\subsection{Delta function and Gaussians} \label{sec:WW:delta_Gaussians}

In this subsection, we review two important examples of the Wigner-Weyl transform. Our first example is the operator:
\begin{align}
 \hat R = \int d^\calN\!q \ket{-q}\!\bra{q} \ , \label{eq:reversal_operator}
\end{align}
which reverses the sign of $q^i$ eigenvalues. The Wigner-Weyl transform \eqref{eq:WW_explicit} of $\hat R$ reads:
\begin{align}
 \hat R \quad \longleftrightarrow \quad \frac{\delta(\xi)}{2^\calN} \ , \label{eq:reversal_operator_WW}
\end{align}
where $\delta(\xi)$ is the phase space delta function from \eqref{eq:phase_space_integrals}. The transform \eqref{eq:reversal_operator_WW} shows that $\hat R$ is actually $Sp(2\calN)$-invariant, and can be thought of as reversing \emph{phase space} rather than just configuration space. Indeed, when acting in the adjoint, it does precisely that:
\begin{align}
 \hat R\,\hat\xi^I\hat R = -\hat\xi^I \quad \longleftrightarrow \quad \frac{\delta(\xi)}{2^\calN} \ast f(\xi^I) \ast \frac{\delta(\xi)}{2^\calN} = f(-\xi^I) \ .
\end{align}
When acting on one side, its action can be expressed as a Fourier transform in phase space:
\begin{align}
  \frac{\delta(\xi)}{2^\calN} \ast f(\xi) &= 2^\calN \int d^{2\calN}\!\xi'\,f(\xi')\,e^{-2i\Omega_{IJ}\xi'^I\xi^J} \ ; \label{eq:delta_left} \\
  f(\xi) \ast \frac{\delta(\xi)}{2^\calN} &= 2^\calN \int d^{2\calN}\!\xi'\,f(\xi')\,e^{2i\Omega_{IJ}\xi'^I\xi^J} \ . \label{eq:delta_right}
\end{align}
Thinking of $f(\xi)$ as the Wigner-Weyl transform of an operator $\hat f\in\calH\otimes\calH^*$, the operations \eqref{eq:delta_left}-\eqref{eq:delta_right} can be interpreted as ``flipping the phase space'' in just the $\calH$ or just the $\calH^*$ factor. In the terminology of section \ref{sec:WW:doubled}, this implies flipping one of the phase space copies $\xi_\pm^I$. In particular, the right-multiplication \eqref{eq:delta_right} flips the sign of $\xi_-^I$, which is equivalent to interchanging $\xi^I_{\text{even}} \leftrightarrow \xi^I_{\text{odd}}$. This leads us to a nice interpretation of the Fourier transform in \eqref{eq:delta_right}: it's just a basis change for wavefunctions in $\calH\otimes\calH^*$ between the bases \eqref{eq:Psi_even} and \eqref{eq:Psi_odd}:
\begin{align}
 \tilde\Psi(\xi) = \Psi(\xi) \ast \frac{\delta(\xi)}{2^\calN} \quad \Longleftrightarrow \quad
 \tilde\Psi(\xi_{\text{odd}}) = 2^\calN \int d^{2\calN}\!\xi_{\text{even}}\,\Psi(\xi_{\text{even}})\,e^{2i\Omega_{IJ}\xi^I_{\text{even}}\xi_{\text{odd}}^J} \ .
\end{align}

Our second example is a Gaussian operator:
\begin{align}
 \hat\Psi = e^{h_{IJ}\hat\xi^I\hat\xi^J} \ , \label{eq:Gaussian_operator}
\end{align}
where $h_{IJ}$ is some quadratic form on phase space. In terms of the Moyal star product, the corresponding phase space function is:
\begin{align}
 \Psi(\xi) = \exp_\ast\!\left(h_{IJ}\xi^I \xi^J \right) \ .
\end{align}
The result of this star-exponential is once again a Gaussian in $\xi^I$, but with modified coefficients. It is best expressed in terms of the matrix $(\Omega^{-1}h)^I{}_J = (\Omega^{-1})^{IK}h_{KJ}$, which one can plug into various functions via Taylor expansion:
\begin{align}
 \Psi(\xi) = \frac{1}{\sqrt{\det\cos(\Omega^{-1}h)}}\, \exp\left[\xi^I\Omega_{IJ}(\tan(\Omega^{-1}h))^J{}_K \xi^K\right] \ . \label{eq:WW_Gaussian}
\end{align}
To check the answer \eqref{eq:WW_Gaussian}, one can first see that it works for infinitesimal $h_{IJ}$, and then verify the correct behavior as $h_{IJ}$ is rescaled:
\begin{align}
 h_{IJ}\frac{\del\Psi(\xi)}{\del h_{IJ}} = (h_{IJ}\xi^I \xi^J) \ast \Psi(\xi) \ . 
\end{align}
Finally, it will be useful to know the Fourier transform \eqref{eq:delta_right} of the Gaussian \eqref{eq:WW_Gaussian}, i.e. the Wigner-Weyl transform of the operator $\hat\Phi = \hat\Psi\hat R$:
\begin{align}
 \hat\Phi = e^{h_{IJ}\hat\xi^I\hat\xi^J}\hat R  \quad \longleftrightarrow \quad \Phi(\xi) = \frac{1}{\sqrt{\det\sin(\Omega^{-1}h)}}\, \exp\left[-\xi^I\Omega_{IJ}(\cot(\Omega^{-1}h))^J{}_K \xi^K\right] \ . \label{eq:WW_Gaussian_Fourier}
\end{align}

\section{Individual fields in the causal patch: external spacetime as scaffolding} \label{sec:fields}

\begin{figure}%
	\centering%
	\includegraphics[scale=0.8]{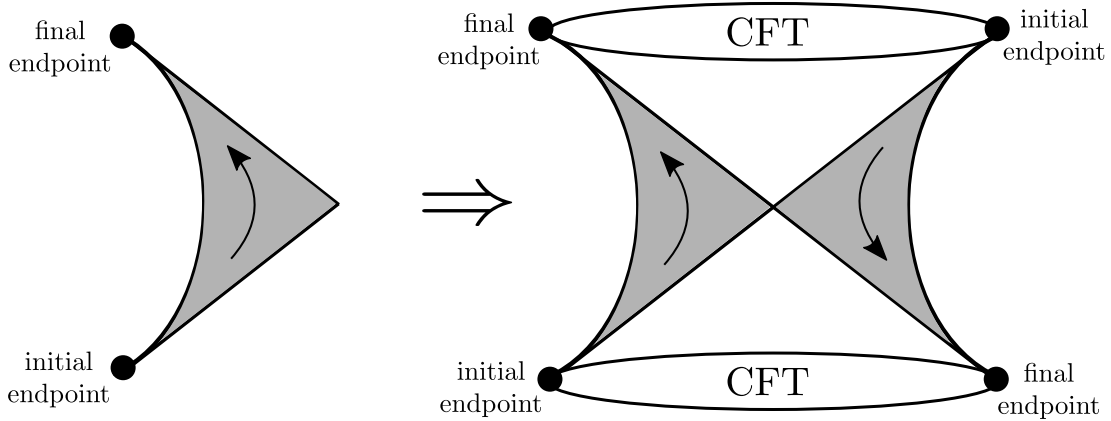} \\
	\caption{Starting from a causal patch, one can introduce the rest of de Sitter space, including its conformal boundary, via the doubled phase space formalism. First, we construct two antipodally-related copies of the causal patch, with opposite intrinsic time orientations. Next, we evolve the fields out of the doubled causal patch into the rest of de Sitter space.}
	\label{fig:Scaffolding} 
\end{figure}

In this section, we apply the general machinery of section \ref{sec:WW} to fields in de Sitter space. Following \cite{Halpern:2015zia}, we extract the Hamiltonian structure of a free massless field in the de Sitter causal patch from the 2-point function of the boundary CFT. In the process, we show how de Sitter space itself, including its unobservable conformal boundary, can be constructed out of an observer's causal patch, via the doubled phase space formalism of section \ref{sec:WW}. We illustrate this construction in figure \ref{fig:Scaffolding}.

\subsection{Spacetime in the embedding formalism: bulk, boundary, horizons, antipodes} \label{sec:fields:geometry}

In this subsection, we introduce the geometry of de Sitter space $dS_4$ and its boundary in the embedding space formalism. We define $dS_4$ as the hyperboloid of unit spacelike vectors in flat 5d Minkowski space $\bbR^{1,4}$: 
\begin{align}
dS_4 = \left\{x^\mu\in\bbR^{1,4}\, |\, x_\mu x^\mu = 1 \right\} \ . \label{eq:dS}
\end{align}
The metric $\eta_{\mu\nu}$ of $\bbR^{1,4}$ has signature $(-,+,+,+,+)$. We use indices $(\mu,\nu,\dots)$ for $\bbR^{1,4}$ vectors, which we raise and lower using $\eta_{\mu\nu}$. The isometries of $dS_4$ are just the group of rotations $O(1,4)$ in the 5d embedding space.

The $dS_4$ tangent space at a point $x^\mu$ consists of the $\bbR^{1,4}$ vectors $v^\mu$ that satisfy $x\cdot v = 0$. The $dS_4$ metric at $x$ can be identified with the projector onto this tangent space:
\begin{align}
q_{\mu\nu}(x) = \eta_{\mu\nu} - x_\mu x_\nu \ .
\end{align}
Covariant derivatives in $dS_4$ can be defined via the flat $\bbR^{1,4}$ derivative, by projecting it back onto the hyperboloid:
\begin{align}
\nabla_\mu v_\nu = q_\mu^\rho(x)\, q_\nu^\sigma(x)\, \del_\rho v_\sigma \ . \label{eq:nabla_v}
\end{align}
Spacelike and timelike geodesics in $dS_4$ are circles and hyperbolas of unit radius in $\bbR^{1,4}$, respectively. The case of null geodesics is even simpler: the lightrays of $dS_4$ are also lightrays in $\bbR^{1,4}$.

The asymptotic boundary of $dS_4$ consists of two conformal 3-spheres, representing past and future infinity. In $\bbR^{1,4}$, these are respectively the spheres of future-pointing and past-pointing null directions. Thus, boundary points are represented by null vectors $\ell^\mu\in\bbR^{1,4}$, with the equivalence $\ell^\mu\cong\lambda\ell^\mu$. The $O(1,4)$ rotations acting on these vectors form the boundary's conformal group. The limit where a bulk point $x$ approaches the boundary can be represented as an extreme boost in $\bbR^{1,4}$, where the unit vector $x^\mu$ approaches a null direction $\ell^\mu$ as:
\begin{align}
x^\mu\rightarrow \ell^\mu/z \ , \quad z\rightarrow 0 \ . \label{eq:limit}
\end{align}

One can fix a conformal frame on the boundary by choosing a section of the $\bbR^{1,4}$ lightcone. The simplest such sections have a flat $\bbR^3$ geometry. They are obtained by singling out a point on the conformal boundary 3-sphere, i.e. a null direction $n_{(\infty)}^\mu\in\bbR^{1,4}$, which becomes the ``point at infinity'' of $\bbR^3$. The flat section itself is given by the intersection of the $\bbR^{1,4}$ lightcone with the null hyperplane $\ell\cdot n_{(\infty)} = -\frac{1}{2}$. Rescalings $n_{(\infty)}^\mu\rightarrow \alpha n_{(\infty)}^\mu$ correspond to dilatations $\ell^\mu\rightarrow \ell^\mu/\alpha$ of the flat section. For example, if the ``point at infinity'' is given by:
\begin{align}
n_{(\infty)}^\mu = \left(\frac{1}{2},\mathbf{0},\frac{1}{2}\right) \ , \label{eq:n}
\end{align}
then the flat section reads:
\begin{align}
 \ell^\mu(\mathbf{r}) = \left(\frac{r^2+1}{2}, \mathbf{r}, \frac{r^2-1}{2} \right) \ ; \quad d\ell_\mu d\ell^\mu = d\mathbf{r}\cdot d\mathbf{r} \ . \label{eq:flat}
\end{align}
In this setup, a 3d displacement vector $\mathbf{dr}$ is embedded into $\bbR^{1,4}$ via:
\begin{align}
 d\ell^\mu = (\mathbf{r}\cdot\mathbf{dr}, \mathbf{dr}, \mathbf{r}\cdot\mathbf{dr}) = (0,\mathbf{dr},0) + 2(\mathbf{r}\cdot\mathbf{dr})n_{(\infty)}^\mu \ . \label{eq:flat_displacement}
\end{align}
Thus, a vector $\mathbf{v}$ on the flat $\bbR^3$ section is described by an \emph{equivalence class} of $\bbR^{1,4}$ vectors,
\begin{align}
 v\cdot n_{(\infty)} = 0 \ ; \quad v^\mu \cong v^\mu + \alpha n_{(\infty)}^\mu \ . \label{eq:tangent_vector}
\end{align}
Evaluating $\mathbf{v}$ at different points $\mathbf{r}$ produces different elements of this equivalence class.

An \emph{observer} in $dS_4$ is defined by a point $-n_{(0)}^\mu$ at past infinity and a point $n_{(\infty)}^\mu$ at future infinity, which can be thought of as the endpoints of the observer's worldline. For a detailed discussion of this assertion, see e.g. \cite{Halpern:2015zia}. Our notation refers to the fact that $n_{(0)}^\mu$ and $n_{(\infty)}^\mu$ can serve as an origin and a ``point at infinity'' for a flat conformal frame on the boundary. Under $O(1,4)$, any pair of points $n_{(0)}\neq n_{(\infty)}$ is equivalent to any other. In this paper, we consider a single observer, whose endpoints we fix without loss of generality at: 
\begin{align}
 n_{(0)}^\mu = \left(\frac{1}{2},\mathbf{0},-\frac{1}{2}\right) \ ; \quad n_{(\infty)}^\mu = \left(\frac{1}{2},\mathbf{0},\frac{1}{2}\right) \ , \label{eq:nn'}
\end{align}
which in particular implies the mutual normalization $n_{(0)}\cdot n_{(\infty)} = -1/2$. 

A cosmological horizon in de Sitter space is the lightcone of a boundary point. Its topology is a cylinder $\bbR\times S_2$, where the $S_2$'s are unit spheres, and the $\bbR$'s are lightrays orthogonal to them. In the $\bbR^{1,4}$ embedding space, these lightrays are all parallel, and point along the null vector that represents the horizon's origin point. The horizons corresponding to the boundary points \eqref{eq:nn'} can be coordinatized as:
\begin{align}
 \text{Past horizon:} \quad x^\mu &= \alpha n_{(0)}^\mu + (0,\mathbf{x},0) = \left(\frac{\alpha}{2},\mathbf{x},-\frac{\alpha}{2}\right) \ ; \label{eq:past_horizon_global} \\
 \text{Future horizon:} \quad x^\mu &= \alpha n_{(\infty)}^\mu + (0,\mathbf{x},0) = \left(\frac{\alpha}{2},\mathbf{x},\frac{\alpha}{2}\right) \ , \label{eq:future_horizon_global}
\end{align}
where $\alpha\in\bbR$ is an affine null time, and $\mathbf{x}$ is a 3d unit vector representing a point on $S_2$. The observer's \emph{causal patch} is the region enclosed by the past horizon \eqref{eq:past_horizon_global} with $\alpha < 0$ and the future horizon \eqref{eq:future_horizon_global} with $\alpha > 0$. These half-horizons can be conveniently parameterized by replacing the ``global'' null time $\alpha$ with an observer-adapted time $t$:
\begin{align}
\text{Past horizon:} \quad x^\mu &= -e^{-t} n_{(0)}^\mu + (0,\mathbf{x},0) = \left(-\frac{1}{2}e^{-t},\mathbf{x},\frac{1}{2}e^{-t} \right) \ ; \label{eq:past_horizon} \\
\text{Future horizon:} \quad x^\mu &= e^t n_{(\infty)}^\mu + (0,\mathbf{x},0) = \left(\frac{1}{2}e^t,\mathbf{x},\frac{1}{2}e^t \right) \ . \label{eq:future_horizon}
\end{align}
For the observer, the horizons \eqref{eq:past_horizon}-\eqref{eq:future_horizon} are analogous to past and future null infinity in Minkowski space. Their intersection is the 2-sphere $(0,\mathbf{x},0)$, which is the observer's analog of spacelike infinity, though it is of course an ordinary finite surface. The roles of past and future \emph{timelike} infinity are played by the boundary endpoints $-n_{(0)}^\mu$ and $n_{(\infty)}^\mu$; these two points are the only intersection between the causal patch and the true conformal boundary of $dS_4$. 

After an observer is chosen, the de Sitter symmetry group $O(1,4)$ is reduced to the group $SO(1,1)\times O(3) = \bbR\times O(3)$ of time translations and rotations. The $O(3)$ rotations leave the null vectors $n_{(0)}^\mu,n_{(\infty)}^\mu$ unchanged. The $SO(1,1)$ time translations are actually boosts in $\bbR^{1,4}$, which rescale $n_{(0)}^\mu$ and $n_{(\infty)}^\mu$ by opposite factors. On the horizons \eqref{eq:past_horizon}-\eqref{eq:future_horizon}, the $O(3)$ rotates the unit vector $\mathbf x\in S_2$, while the $SO(1,1)$ acts as translations of the time coordinate $t$.

We place particular importance on the $\bbZ_2$ ``antipodal map'' $x^\mu\rightarrow -x^\mu$, which can be used to fold de Sitter space in half, producing the ``elliptic'' de Sitter space $dS_4/\bbZ_2$ of \cite{Parikh:2002py}. This antipodal map is the unique central element of the $O(1,4)$ de Sitter group. When applied to cosmological horizons such as \eqref{eq:past_horizon_global}-\eqref{eq:future_horizon_global}, this map sends each horizon to itself via $(\alpha,\mathbf{x})\rightarrow(-\alpha,-\mathbf{x})$. When applied to the conformal boundary of dS, it interchanges past and future infinity via $\ell^\mu\rightarrow -\ell^\mu$. Thus, the conformal boundary of $dS_4/\bbZ_2$ consists of \emph{one} 3-sphere -- the sphere of null directions in $\bbR^{1,4}$, where past and future $\pm\ell^\mu$ have been identified. Note that in $dS_4$, the causal patch defined by $-n_{(0)}^\mu$ and $n_{(\infty)}^\mu$ spans ``half of space'', as defined e.g. by the $\alpha>0$ half of the horizon \eqref{eq:future_horizon_global}. The other ``half of space'' is spanned by the \emph{antipodal} causal patch, with endpoints $-n_{(\infty)}^\mu$ and $n_{(0)}^\mu$. Therefore, in $dS_4/\bbZ_2$, a causal patch spans \emph{all of space}, i.e. the fields in it can be evolved throughout the spacetime, including in particular the conformal boundary. Thus, we can have a one-to-one dictionary between field solutions in the causal patch and boundary data at conformal infinity, even though their intersection still consists of just two points.

\subsection{Focusing on a bulk field in the causal patch}

At this point, let us drop the pretense of an all-knowing deity to whom all of spacetime is visible. For an observer in de Sitter space, the empirically available spacetime is the causal patch, and the goal of physics is to describe the degrees of freedom inside it. Regions of dS that lie outside the causal patch may act as useful tools for the description of this physics. However, they remain mere fictions, and we are free to populate them with degrees of freedom at our theoretical convenience. In particular, this applies to the \emph{entire} conformal boundary of dS, with the exception of the observer's two endpoints. The challenge of dS/CFT is to relate a CFT description on this fictitious conformal boundary to the actual object of interest, i.e. causal patch physics.

In this subsection, we consider a free massless field in the de Sitter causal patch, i.e. a single member of the linearized higher-spin multiplet. For simplicity, we imagine this to be the spin-0 field, obeying the conformally-coupled massless field equation $\nabla_\mu\nabla^\mu F = 2F$; our arguments will apply equally to massless fields of any spin. In the language of section \ref{sec:WW}, we denote the phase space of our field in the causal patch as $\xi^I$, with symplectic form $\Omega_{IJ}$, where the phase space indices $(I,J,\dots)$ are infinite-dimensional. For now, we treat $\Omega_{IJ}$ as given; by the end of this section, we will \emph{derive} it from the boundary CFT's 2-point function. 

The causal patch phase space $\xi^I$ can be expressed in a variety of bases, given by the values and derivatives of our field $F$ on various hypersurfaces. In particular, the values $F$ and time derivatives $\dot F$ on a Cauchy slice provide a valid basis, though not a very useful one for holography. A more special basis is given by the values of $F$ on the observer's initial or final horizon (since the horizons are null, the time derivatives $\dot F$ need not be separately provided). In such a basis, the causal patch's symmetry group $SO(1,1)\times O(3)$ is manifest, since it maps the horizons \eqref{eq:past_horizon}-\eqref{eq:future_horizon} to themselves. What we \emph{don't} have at this point is any basis that refers to the conformal boundary of $dS_4$, or that represents the full de Sitter group $O(1,4)$. This makes it difficult to see how to make contact with the boundary CFT.

\subsection{$dS_4/\bbZ_2$: making contact with the conformal boundary} \label{sec:fields:elliptic}

To solve the above problem, we must build some scaffolding outside the causal patch. We reimagine our field in the causal patch as existing in the larger context of \emph{elliptic de Sitter space} $dS_4/\bbZ_2$. Mathematically, a field on $dS_4/\bbZ_2$ is just a field on $dS_4$ that's either even or odd under the antipodal map $x^\mu\rightarrow -x^\mu$. Starting from a solution in the causal patch, we can construct a $dS_4/\bbZ_2$ field by simply copying the same field values into the antipodal patch (with a possible sign flip, for the antipodally odd case). The advantage of this seemingly trivial step is that the causal patch in $dS_4/\bbZ_2$ (or the union of the causal patch and its antipode in $dS_4$) \emph{causally spans} the full spacetime: the field can be evolved out of the causal patch to generate a solution on all of $dS_4/\bbZ_2$, including asymptotic data at the conformal boundary. Thus, field solutions in the causal patch are in one-to-one correspondence with solutions in the whole of $dS_4/\bbZ_2$: the latter simply constitute a richer perspective on the former. 

One immediate advantage of the $dS_4/\bbZ_2$ picture is that our phase space $\xi^I$ now linearly represents the full $O(1,4)$ symmetry, rather than just the $\bbR\times O(3)$ subgroup. That being said, this larger group does \emph{not} preserve the causal patch symplectic form $\Omega_{IJ}$. In fact, there isn't \emph{any} symplectic form in $dS_4/\bbZ_2$ with the full $O(1,4)$ symmetry -- a fact closely related to the absence of a global time orientation \cite{Hackl:2014txa}.

Leaving the symplectic structure aside for now, we return our attention to the full $O(1,4)$ de Sitter group. Not only is this symmetry now represented on the phase space $\xi^I$, but we now have a particular \emph{basis} in which this larger symmetry is manifest. Such a basis is given by the field's boundary data at conformal infinity. In ordinary $dS_4$ (not folded in half), this boundary data comes in two types, which are canonically conjugate to each other. For the spin-0 field in our example, these two types of data are distinguished by their conformal weights $\Delta=1,2$. For massless gauge fields of nonzero spin, the two types of data are the electric and magnetic field strengths. In our CFT of interest, the $\Delta=1$ data for the spin-0 field (and the electric data for the gauge fields) corresponds to operator VEVs, while the $\Delta=2$ data (and the magnetic data for the gauge fields) corresponds to their sources. In $dS_4/\bbZ_2$, there is a subtlety: since the bulk degrees of freedom are halved, only \emph{one} type of boundary data can be specified independently. In the case of \emph{massless} bulk fields, as we have in higher-spin theory, the situation is particularly simple: one of the two types of boundary data in $dS_4/\bbZ_2$ simply vanishes identically \cite{Ng:2012xp,Neiman:2014npa,Vasiliev:2012vf}. The non-vanishing boundary data then constitutes our $O(1,4)$-covariant basis for the causal-patch phase space. If we choose even antipodal symmetry, then the non-vanishing boundary data is the one corresponding to CFT VEVs; if we choose odd antipodal symmetry, then it's the one corresponding to CFT sources. 

The upshot is that if we extend our causal patch fields into antipodally \emph{even} fields on $dS_4/\bbZ_2$, then we are given a precious opportunity to make contact with the boundary CFT. The phase space of causal patch fields becomes identified with the space of CFT sources, and thus \emph{the CFT partition function can be viewed as a functional on the causal patch phase space}. Since we're considering the limit of free bulk fields, the CFT partition function should be approximated as a Gaussian:
\begin{align}
 Z_{\text{CFT}}[\xi^I] \, \sim \, e^{-G_{IJ}\xi^I\xi^J} \ . \label{eq:CFT_Gaussian}
\end{align}
Here, the quadratic form $G_{IJ}$ encodes the CFT's 2-point function, and we are disregarding a $\xi^I$-independent normalization factor. A crucial question now arises: what is the meaning of the phase space function \eqref{eq:CFT_Gaussian} from the point of view of causal patch physics? In the next subsection, we will answer this question. The answer will allow us to derive the causal-patch symplectic form $\Omega_{IJ}$ out of the boundary 2-point function $G_{IJ}$.

\subsection{Full $dS_4$: the CFT partition function as a causal patch operator} \label{sec:fields:global}

We now set out to better understand the role of $Z_{\text{CFT}}$ as a functional on the causal patch phase space. To do this, we erect our next bit of scaffolding: we ``unfold'' our elliptic de Sitter space $dS_4/\bbZ_2$ into full $dS_4$. In other words, we double our phase space to include field solutions that are not antipodally symmetric. Unlike $dS_4/\bbZ_2$, which is time-orientable only inside a causal patch, $dS_4$ is time-orientable globally. Our single causal patch in $dS_4/\bbZ_2$ becomes a pair of antipodally related patches in $dS_4$. One of these patches will be ``upright'', in the sense that its intrinsic time orientation (as defined by the observer's past/future endpoints) agrees with the global time orientation of $dS_4$; the other patch will be ``time-reversed'', with opposite orientation. 

Notice that this construction of $dS_4$ maps precisely onto the doubled phase space construction of section \ref{sec:WW:doubled}. In the notation of that section, the ``upright'' copy of the causal patch in $dS_4$ is associated with the phase space $\xi_+^I$, while the ``time-reversed'' copy is associated with the phase space $\xi_-^I$. The full symplectic form on $dS_4$ takes the form \eqref{eq:Omega_future_past}, where the minus sign encodes the time-reversed nature of the $\xi_-^I$. Antipodally even and antipodally odd field solutions correspond to $\xi_{\text{even}}^I$ and $\xi_{\text{odd}}^I$ from \eqref{eq:xi_even_odd}. Under the $dS_4$ symplectic form, these are canonically conjugate to each other, as in \eqref{eq:Omega_even_odd}.

Within this setup, we recognize the CFT partition function \eqref{eq:CFT_Gaussian} as the wavefunction of some \emph{state} in the $dS_4$ Hilbert space, written in the basis $\xi_{\text{even}}^I$ of antipodally even field solutions. What is this state? According to Maldacena's insight \cite{Maldacena:2002vr,Harlow:2011ke}, we should think of $Z_{\text{CFT}}$ as the \emph{Hartle-Hawking wavefunction} \cite{Hartle:1983ai} of quantum gravity in $dS_4$, evaluated at the future conformal boundary. In the limit of free bulk fields, this becomes just the Bunch-Davies vacuum, given by the free path integral over Euclidean $AdS_4$. Thus, the partition function \eqref{eq:CFT_Gaussian} is \emph{the Bunch-Davies vacuum, expressed in the antipodally even basis}.

Now, recall, following section \ref{sec:WW:doubled}, that the Hilbert space of $dS_4$ can be thought of as a doubled version $\calH\otimes\calH^*$ of the causal patch Hilbert space. Therefore, states in $dS_4$ can be thought of as operators in the causal patch. Thus, we can ask: which causal-patch operator corresponds to the Bunch-Davies vacuum? The answer to this question is intimately related to standard derivations of the de Sitter temperature. As a first step, note that the Bunch-Davies vacuum can be evaluated by a Euclidean path integral with \emph{any} Cauchy slice as boundary, not just the one at future infinity. In particular, we can use the union of a Cauchy slice in the ``upright'' causal patch with its antipodal image in the ``time-reversed'' patch. With this boundary hypersurface, The Euclidean path integral can then be expressed as a Euclidean rotation by $\pi$, i.e. an imaginary boost by $-i\pi$, in the $n_{(0)}\wedge n_{(\infty)}$ plane. This rotation maps the ``time-reversed'' Cauchy slice onto the ``upright'' one. More precisely, since we're identifying the two patches through the antipodal map, which reverses all 5 axes in $\bbR^{1,4}$, the Euclidean rotation maps the ``time-reversed'' patch into a \emph{parity-reversed image} of the ``upright'' one. We conclude that the Bunch-Davies vacuum corresponds, in the sense of section \ref{sec:WW:doubled}, to the following operator on the causal-patch Hilbert space:
\begin{align}
 \hat\Psi_{\text{B.D.}} = e^{-\pi\hat H}\hat P \ , \label{eq:operator}
\end{align}
where $\hat H$ is the observer's Hamiltonian, which generates the $SO(1,1)$ boosts in the $n_{(0)}\wedge n_{(\infty)}$ plane, and $\hat P$ is the parity operator, which reverses the orthogonal $\bbR^3$ subspace. 

On the other hand, section \ref{sec:WW:doubled} teaches us that the map between $dS_4$ wavefunctions in the antipodally even basis and operators on the causal patch Hilbert space is nothing but the \emph{Wigner-Weyl transform}. Thus, we can calculate the Wigner-Weyl transform of \eqref{eq:operator}, and compare with \eqref{eq:CFT_Gaussian}. This will give us an equation that relates:
\begin{itemize}
	\item The CFT 2-point function.
	\item The time translation \& parity operations on the causal-patch phase space (i.e. on the space of CFT sources).
	\item The symplectic form on the causal-patch phase space.
\end{itemize}
In particular, it will allows us to express the causal-patch symplectic form in terms of boundary quantities.

\subsection{Solving for the causal patch symplectic form} \label{sec:fields:Omega}

We now set out to calculate the Wigner-Weyl transform of the operator \eqref{eq:operator}. Recall that we are dealing with free bulk fields, which are essentially collections of harmonic oscillators. For such a system, one can easily express the symmetry operators $\hat H,\hat P$ on the causal-patch Hilbert space in terms of corresponding symmetry operators \emph{on the phase space}. In particular, let $\omega^I{}_J$ be the generator of time translations on the causal-patch phase space, and let $\calP^I{}_J$ be the phase space operator that implements a parity reflection. Then the Hamiltonian can be expressed as:
\begin{align}
 \hat H = -\frac{i}{2}\,\hat\xi^I \Omega_{IJ}\omega^J{}_K \hat\xi^K \ , \label{eq:H_raw}
\end{align}
where $\Omega_{IJ}$ is the causal-patch symplectic form. The Hamiltonian \eqref{eq:H_raw} for a system of oscillators can be deduced from the example of a single oscillator with frequency $\omega_0$:
\begin{align}
 \hat\xi^I = \begin{pmatrix} \hat a \\ \hat a^\dagger \end{pmatrix} \ ; \quad \Omega_{IJ} = \begin{pmatrix} 0 & -i \\ i & 0 \end{pmatrix} \ ; \quad \omega^I{}_J = \begin{pmatrix} \omega_0 & 0 \\ 0 & -\omega_0 \end{pmatrix} \ ; \quad
 \hat H = \frac{\omega_0}{2}(\hat a\hat a^\dagger + \hat a^\dagger\hat a) \ . \label{eq:single_oscillator}
\end{align}
Note that the coefficient matrix $\Omega_{IJ}\omega^J{}_K$ in \eqref{eq:H_raw} is automatically symmetric in its indices, since time translations $\omega^I{}_J$ are a symmetry of the symplectic form $\Omega_{IJ}$. Thus, the Hamiltonian \eqref{eq:H_raw} is the symmetrically-ordered one, with zero-point energies included. 

We can now use the Wigner-Weyl transform of Gaussians \eqref{eq:Gaussian_operator}-\eqref{eq:WW_Gaussian} to write down the transform of the $e^{-\pi\hat H}$ factor in \eqref{eq:operator}:
\begin{align}
 e^{-\pi\hat H} \quad \longleftrightarrow \quad \exp\left[i\xi^I\Omega_{IJ}\left(\tanh\frac{\pi\omega}{2}\right)^J{}_K \xi^K\right] \ , \label{eq:just_H}
\end{align}
where we neglected the constant prefactor. 

What remains is to handle the parity factor $\hat P$ in \eqref{eq:operator}. First, let us note that parity commutes with time translations, and is also a symmetry of the symplectic form. Thus, we can decompose the causal-patch phase space into parity-even and parity-odd sectors, which don't get mixed by either $\Omega_{IJ}$ or $\omega^I{}_J$. One can separate these sectors using the projectors $(\delta^I_J \pm \calP^I{}_J)/2$, where $\calP^I{}_J$ is our parity operator \emph{on phase space}. Now, the \emph{Hilbert space} parity operator $\hat P$ acts as the identity on the parity-even sector of phase space, and as the reversal operator \eqref{eq:reversal_operator} on the parity-odd sector. As we recall from \eqref{eq:delta_right}, multiplication by the reversal operator corresponds to a Fourier transform on phase space functions. Thus, in the parity-even sector of phase space, the operator $e^{-\pi\hat H}\hat P$ corresponds directly to the phase space function \eqref{eq:just_H}, while in the parity-odd sector, it corresponds to its Fourier transform, as in \eqref{eq:WW_Gaussian_Fourier}. Putting the sectors back together, we obtain the Wigner-Weyl transform of \eqref{eq:operator} as:
\begin{align}
 \begin{split}
   e^{-\pi\hat H}\hat P \quad \longleftrightarrow \quad &\exp\left[i\xi^I\Omega_{IJ}\left( \frac{1 + \calP}{2}\tanh\frac{\pi\omega}{2} + \frac{1 - \calP}{2}\coth\frac{\pi\omega}{2} \right)^J{}_K \xi^K\right] \\
     = &\exp\left[i\xi^I\Omega_{IJ}\left( \coth(\pi\omega) - \frac{\calP}{\sinh(\pi\omega)} \right)^J{}_K\,\xi^K\right] \ .
 \end{split} \label{eq:H_with_P}
\end{align}
For future reference, we note that the matrix in parentheses can be inverted by the substitution $\calP\rightarrow -\calP$. 

Now, recall that $e^{-\pi\hat H}\hat P$ is the Wigner-Weyl transform of the boundary partition \eqref{eq:CFT_Gaussian}. Comparing with our expression \eqref{eq:H_with_P}, we find:
\begin{align}
 G_{IJ} = -i\Omega_{IK}\left( \coth(\pi\omega) - \frac{\calP}{\sinh(\pi\omega)} \right)^K{}_J \ ,
\end{align}
which can be inverted to give:
\begin{align}
 \Omega_{IJ} = iG_{IK}\left( \coth(\pi\omega) + \frac{\calP}{\sinh(\pi\omega)} \right)^K{}_J \ . \label{eq:Omega}
\end{align}
This simple formula expresses the causal-patch fields' symplectic form $\Omega_{IJ}$ in terms of native quantities of the Euclidean CFT: the 2-point function $G_{IJ}$, and the kinematical symmetries $\omega^I{}_J$ and $\calP^I{}_J$ (which were singled out from the full $O(1,4)$ group by our choice of observer). The significance of this formula is that $\Omega_{IJ}$ captures the Lorentzian causal structure of the bulk fields. In particular, via \eqref{eq:commutators} and \eqref{eq:H_raw}, we can read off their commutators and Hamiltonian as:
\begin{align}
 \left[\hat\xi^I,\hat\xi^J\right] &= \left( \coth(\pi\omega) - \frac{\calP}{\sinh(\pi\omega)} \right)^I{}_K\left(G^{-1}\right)^{KJ} \ ; \label{eq:commutators_G} \\
 \hat H &= \frac{1}{2}\,\hat\xi^I G_{IJ}\,\omega^J{}_K \left( \coth(\pi\omega) + \frac{\calP}{\sinh(\pi\omega)} \right)^K{}_L\,\hat\xi^L \ . \label{eq:H}
\end{align}
The results \eqref{eq:Omega}-\eqref{eq:H} of the present section are a restatement of results obtained in \cite{Hackl:2014txa,Halpern:2015zia}. The advantage of the present treatment is that we worked abstractly in phase space notation, without committing to an explicit basis. This will prove useful in our discussion of the HS multiplet below.

\section{The full higher-spin multiplet: removing the scaffolding} \label{sec:multiplet}

In this section, we repeat the construction of the causal-patch symplectic form, in a language that treats the full HS multiplet in a unified way. At the same time, we'll be able to remove the ``scaffolding'' erected in section \ref{sec:fields}: the unobservable $dS_4$ or $dS_4/\bbZ_2$ spacetime outside the causal patch. In particular, by the end of this section, we will not need the spacetime's unobservable conformal boundary: all the information of the CFT sources, as well as the partition function, can be encoded at one of the endpoints of the causal patch. In this rewriting of the boundary theory, we will effectively reduce it from a \emph{field theory}, with variables at every point of the boundary manifold, to \emph{particle mechanics}, where variables defined at one or two points will prove to be enough.

The formalism we will use is essentially the well-known on-shell version of the bilocal formalism for the free vector model \cite{Koch:2010cy,Das:2012dt,Koch:2014aqa}. Our treatment of this formalism will be somewhat novel. We will express the null momenta of the boundary fields as squares of \emph{spinors}. These are related by a Wigner-Weyl transform to the spacetime-independent \emph{twistors} of \cite{Neiman:2017mel}, which are in turn related to bulk fields via the Penrose transform. Remarkably, we'll find that the sign ambiguity of our boundary momentum spinors is directly related to the bulk antipodal map.

The discussion is structured as follows. In section \ref{sec:multiplet:twistors}, we introduce twistors as the spinors of the $\bbR^{1,4}$ embedding space, and use them to define higher-spin algebra -- the fundamental symmetry of higher-spin theory. In section \ref{sec:multiplet:bulk}, we review the Penrose transform between twistor functions and higher-spin master fields. In section \ref{sec:multiplet:mechanics}, we construct the relation between twistor space and boundary particle mechanics, and introduce momentum spinors as a basis for the boundary particle's Hilbert space. In section \ref{sec:multiplet:CFT}, we review the free vector model's partition function in the bilocal and twistor languages, and recast it in the language of particle mechanics. In section \ref{sec:multiplet:symplectic}, we plug in the machinery from section \ref{sec:fields}, and derive holographically the Hamiltonian structure of the HS multiplet in the bulk causal patch. In section \ref{sec:multiplet:symmetry}, we discuss the HS symmetry of our construction.

\subsection{Spacetime-independent twistors and higher-spin algebra} \label{sec:multiplet:twistors}

\subsubsection{Twistors}

In this section, we introduce twistors in $dS_4$ from the perspective described in \cite{Neiman:2013hca}. We refer to \cite{Neiman:2017mel} for additional identities, and to \cite{Neiman:2013hca} for a more detailed geometric picture. 

The twistors of $dS_4$ are just the complex 4-component Dirac spinors of the isometry group $SO(1,4)$. We use indices $(a,b,\dots)$ for twistors. The twistor space is equipped with a symplectic metric $I_{ab}$, which is used to raise and lower indices via:
\begin{align}
 U_a = I_{ab}U^b \ ; \quad U^a = U_b I^{ba} \ ; \quad I_{ac}I^{bc} = \delta_a^b \ .
\end{align}
Tensor and twistor indices are related through the gamma matrices $(\gamma_\mu)^a{}_b$, which satisfy the Clifford algebra $\{\gamma_\mu,\gamma_\nu\} = -2\eta_{\mu\nu}$. These 4+1d gamma matrices can be realized as the usual 3+1d ones, with the addition of $\gamma_5$ (in our notation, $\gamma_4$) for the fifth direction in $\bbR^{1,4}$. The matrices $\gamma^\mu_{ab}$ are antisymmetric and traceless in their twistor indices. We define the antisymmetric product of gamma matrices as:
\begin{align}
 \gamma^{\mu\nu}_{ab} \equiv \gamma^{[\mu}_{ac}\gamma^{\nu]c}{}_b \ .
\end{align}
The $\gamma^{\mu\nu}_{ab}$ are symmetric in their twistor indices. One can use the matrices $\gamma_\mu^{ab}$ to convert between $\bbR^{1,4}$ vectors and traceless bitwistors as:
\begin{align}
 v^{ab} = \gamma_\mu^{ab}v^\mu \ ; \quad v^\mu = -\frac{1}{4}\gamma^\mu_{ab}v^{ab} \ . \label{eq:conversion_5d}
\end{align}
Similarly, $\gamma_{\mu\nu}^{ab}$ can be used to convert between bivectors and symmetric twistor matrices:
\begin{align}
 f^{ab} = \frac{1}{2}\gamma_{\mu\nu}^{ab}f^{\mu\nu} \ ; \quad f^{\mu\nu} = \frac{1}{4}\gamma^{\mu\nu}_{ab} f^{ab} \ . \label{eq:conversion_bivectors}
\end{align}
We use the symplectic metric $I_{ab}$ to define a measure on twistor space, in the same way as in \eqref{eq:phase_space_measure}:
\begin{align}
 d^4U \equiv \frac{I_{ab}I_{cd}}{8(2\pi)^2}\,dU^a dU^b dU^c dU^d \ , \label{eq:twistor_measure}
\end{align}
where the factor of $8$ has the combinatorial origin $8 = 2^2 2!$, and we again included $2\pi$ factors, so that they won't appear explicitly in Fourier and Gaussian integrals. 

It is often convenient to use an index-free notation for products in $\bbR^{1,4}$ and in twistor space. $x\cdot x$ will represent the scalar product $x_\mu x^\mu$ in $\bbR^{1,4}$. The twistor matrices $\delta_a^b$ and $(\gamma_\mu)^a{}_b$ will be written in index-free notation as $1$ and $\gamma_\mu$. Combined with the index conversion \eqref{eq:conversion_5d}, this means that the matrix $(x^\mu\gamma_\mu)^a{}_b$ for a vector $x^\mu\in\bbR^{1,4}$ will be written simply as $x$ (this is just the Feynman slash convention, without the slash). Products in the index-free notation imply bottom-to-top index contractions. So, e.g. for two twistors $U^a,V^a$ and two vectors $\ell^\mu,x^\mu$, we have:
\begin{align}
\begin{split}
&UV \equiv U_a V^a = -I_{ab}U^a V^b \ ; \quad \ell\cdot x \equiv \ell_\mu x^\mu = -\frac{1}{4}\tr(\ell x) \ ; \\
&(xU)^a \equiv x^a{}_b U^b \ ; \quad U\ell xU \equiv U_a\ell^a{}_b x^b{}_c U^c = -\ell_\mu x_\nu\gamma^{\mu\nu}_{ab} U^a U^b \ .
\end{split}
\end{align}

The twistor space of $dS_4$ is complex. One can define a twistor complex conjugation $U^a\rightarrow \bar U^a$, under which the matrices $I_{ab}$ and $\gamma^\mu_{ab}$ are real. However, there's no notion of a \emph{real} twistor, because this conjugation is anti-idempotent: $\bar{\bar U}^a = -U^a$. Twistor functions can be complex-conjugated as:
\begin{align}
\bar f(U) \equiv \overline{f(-\bar U)} \ . \label{eq:f_bar}
\end{align}
This operation sends monomials $U^a\dots U^b$ to themselves, while complex-conjugating their coefficients. Conjugating twice yields:
\begin{align}
\bar{\bar f}(U) = f(-U) \ . \label{eq:f_bar_bar}
\end{align}
For \emph{even} twistor functions $f(U) = f(-U)$, this implies that complex conjugation is idempotent. Therefore, even functions can be real: $f(U) = \bar f(U)$. 

\subsubsection{Higher-spin algebra}

Just as the vector $\gamma^\mu$ defines the Clifford algebra, we can define \emph{higher-spin algebra} in terms of a twistor $Y^a$, which obeys the non-commutative star product:
\begin{align}
 Y^a\star Y^b = Y^a Y^b + iI^{ab} \ ; \quad \left[Y^a,Y^b\right]_\star = 2iI^{ab} \ . \label{eq:HS_algebra}
\end{align}
This product extends naturally to functions of $Y^a$ via:
\begin{align}
 f(Y)\star g(Y) = f \exp\left(iI^{ab}\overleftarrow{\frac{\del}{\del Y^a}}\overrightarrow{\frac{\del}{\del Y^b}}\right) g
 = \int d^4U d^4V f(Y+U)\, g(Y+V)\, e^{-iUV} \ . \label{eq:star_int}
\end{align}
The integral definition is the most powerful one, as it applies to very general functions and distributions. At the same time, it has the disadvantage of requiring a choice of contour in the complex twistor space. 

In its simplest form, higher-spin symmetry is the algebra of even functions $f(Y)$, which correspond to integer spins, obeying the star product \eqref{eq:star_int}. Its quadratic elements $Y^a Y^b$ generate the usual $SO(1,4)$ spacetime symmetries:
\begin{align}
  &M_{\mu\nu} = \frac{1}{8} Y\gamma_{\mu\nu}Y \ ; \label{eq:generators} \\
  &[M^{\mu\nu}, M_{\rho\sigma}]_\star = 4i\delta^{[\mu}_{[\rho}\, M^{\nu]}{}_{\sigma]} \ , \label{eq:M_algebra}
\end{align}
where we positioned the $i$ factors differently from \cite{Neiman:2017mel}. The product \eqref{eq:star_int} respects a trace operation:
\begin{align}
 \tr_\star f(Y) &= f(0) \ ; \label{eq:str} \\ 
 \tr_\star(f(Y)\star g(Y)) &= \tr_\star(g(Y)\star f(-Y)) =  \int d^4U d^4V f(U)\,g(V)\,e^{-iUV} \ . \label{eq:str_cyclic}
\end{align}
For \emph{even} functions $f(Y)$, the identity \eqref{eq:str_cyclic} tells us that $\tr_\star$ obeys the cyclic property of the trace. Another important object in HS algebra is the twistor delta function $\delta(Y)$, defined via:
\begin{align}
\delta(Y) = \int d^4U\,e^{iUY} \ ; \quad \int d^4Y f(Y)\,\delta(Y) = f(0) \ . \label{eq:delta}
\end{align}
A one-sided star product with $\delta(Y)$ implements a Fourier transform:
\begin{align}
f(Y)\star\delta(Y) = \int d^4U f(U)\,e^{iUY} \ ; \quad \delta(Y)\star f(Y) = \int d^4U f(U)\,e^{-iUY} \ , \label{eq:delta_Fourier}
\end{align}
while the two-sided product flips the sign of the twistor argument:
\begin{align}
 \delta(Y)\star f(Y)\star\delta(Y) = f(-Y) \ . \label{eq:delta_Klein}
\end{align}
In other words, while $Y^a Y^b$ act in the adjoint as the generators of $SO(1,4)$, the delta function $\delta(Y)$ acts as a $2\pi$ rotation. As a special case of \eqref{eq:delta_Klein}, we have $\delta(Y)\star\delta(Y) = 1$.

With respect to the star product, the complex conjugation \eqref{eq:f_bar} of twistor functions behaves as a Hermitian conjugation:
\begin{align}
 f(Y) = g(Y)\star h(Y) \ \Longleftrightarrow \ \bar f(Y) = \bar h(Y)\star \bar g(Y) \ . \label{eq:star_Hermit}
\end{align}
Other complex-conjugation properties include:
\begin{align}
 \tr_\star \bar f(Y) = \overline{\tr_\star f(Y)} \ ; \quad d^4\bar Y = \overline{d^4Y} \ ; \quad \bar\delta(Y) = \delta(Y) \ . \label{eq:tr_measure_delta_bar}
\end{align}

\subsection{Bulk fields and the Penrose transform} \label{sec:multiplet:bulk}

In this section, we review the Penrose transform in HS-covariant language, as described in \cite{Neiman:2017mel}. Working in the framework of the $\bbR^{1,4}$ embedding space, we will construct bulk spinors, the bulk master field $C(x;Y)$, and its mapping to a spacetime-independent twistor function $F(Y)$. The formalism introduced here won't be essential for the rest of the paper, and is provided for completeness. The results that will useful for the main discussion can be summarized as follows:
\begin{enumerate}
	\item An even twistor function $F(Y) = F(-Y)$ describes a bulk solution for the free HS multiplet, with one field for each integer spin.
	\item The restriction to \emph{even} spins corresponds to the constraint $F(iY) = -F(Y)$.
	\item The bulk antipodal map corresponds to a Fourier transform $F(Y)\rightarrow F(Y)\star\delta(Y)$ of the twistor function.
	\item For antipodally odd solutions, real bulk fields translate into real twistor functions $F(Y) = \bar F(Y)$.
\end{enumerate}

\subsubsection{Spinors and higher-spin algebra at a bulk point} \label{sec:multiplet:bulk:spinors}

When we choose a point $x\in dS_4$, the Dirac representation of $SO(1,4)$ becomes identified with the Dirac representation of the Lorentz group $SO(1,3)$ at $x$, which then decomposes into left-handed and right-handed Weyl spinors. The decomposition is accomplished by a pair of projectors:
\begin{align}
\begin{split}
P_L{}^a{}_b(x) &= \frac{1}{2}\left(\delta^a_b - ix^\mu\gamma_\mu{}^a{}_b \right) = \frac{1}{2}\left(\delta^a_b - ix^a{}_b \right) \ ; \\
P_R{}^a{}_b(x) &= \frac{1}{2}\left(\delta^a_b + ix^\mu\gamma_\mu{}^a{}_b \right) = \frac{1}{2}\left(\delta^a_b + ix^a{}_b \right) \ . \label{eq:projectors}
\end{split}
\end{align}
These serve as an $x$-dependent version of the familiar chiral projectors in $\bbR^{1,3}$. Note that $P_L$ and $P_R$ get interchanged under the antipodal map $x^\mu\rightarrow-x^\mu$. As in our treatment of tensors, we can continue using the $(a,b,\dots)$ indices for both $SO(1,4)$ and $SO(1,3)$ Dirac spinors. Covariant derivatives of left-handed and right-handed Weyl spinors in $dS_4$ can be defined in analogy to \eqref{eq:nabla_v}, by first taking the flat $\bbR^{1,4}$ derivative, and then projecting back into the appropriate subspace. Thus, for a left-handed Weyl spinor field $\chi_L^a(x)$, we define:
\begin{align}
\nabla_\mu\chi^a_L(x) = q_\mu^\nu(x) P_L{}^a{}_b(x)\,\del_\nu\chi_L^b(x) \ , \label{eq:spinor_covariant}
\end{align}
and similarly for $L \leftrightarrow R$.

Given a twistor $Y^a$, we denote its left-handed and right-handed components at $x$ as $y_{L/R}^a(x) = (P_{L/R}){}^a{}_b(x)Y^b$. The projectors $P^L_{ab}(x)$ and $P^R_{ab}(x)$ serve as the spinor metrics for the left-handed and right-handed Weyl spinor spaces. For a 2d spinor space, a symplectic metric also acts as a measure, i.e. we can define:
\begin{align}
d^2y_L \equiv \frac{P^L_{ab}(x)}{2(2\pi)}\,dY^a dY^b \ ; \quad d^2y_R \equiv \frac{P^R_{ab}(x)}{2(2\pi)}\,dY^a dY^b \ ; \quad d^4Y = d^2y_L d^2y_R \ . \label{eq:measure_chiral}
\end{align}
Analogously to $\delta(Y)$, we can define delta functions with respect to $y_L$ and $y_R$:
\begin{align}
\delta^L_x(Y) = \int_{P_L(x)} d^2u_L \,e^{iu_L Y} \ ; \quad \delta^R_x(Y) = \int_{P_R(x)} d^2u_R \,e^{iu_R Y} \ . \label{eq:delta_chiral}
\end{align}
These delta functions have star-product properties \cite{Didenko:2009td} analogous to eqs. \eqref{eq:delta_Fourier}-\eqref{eq:delta_Klein}:
\begin{align}
 \begin{split}
   f(Y)\star\delta^L_x(Y) &= \int d^2u_L\,f(u_L+y_R)\,e^{iu_L y_L} \ ; \\
   \delta^L_x(Y)\star f(Y) &= \int d^2u_L\,f(u_L+y_R)\,e^{-iu_L y_L} \ ; 
 \end{split} \label{eq:delta_Fourier_bulk_raw} \\
 \delta^L_x(Y)\star f(Y)\star\delta^L_x(Y) &= f(-y_L+y_R) = f(ixY) \ , \label{eq:delta_bulk_Klein}
\end{align}
and similarly for $\delta^R_x(y_R)$, except that the last expression becomes $f(-ixY)$. The delta functions themselves satisfy the star-product identities:
\begin{align}
\begin{split}
\delta^L_x(Y)\star\delta^L_x(Y) &= 1 \ ; \quad \delta^L_x(Y)\star\delta^R_x(Y) = \delta(Y) \ ; \\
\delta^L_x(Y)\star\delta(Y) &= \delta(Y)\star\delta^L_x(Y) = \delta^R_x(Y) \ ,
\end{split} \label{eq:delta_identities}
\end{align}
and likewise with $L\leftrightarrow R$.

Under complex conjugation, the projector $P_{L/R}(x)$ goes to $P_{R/L}(\bar x)$. For real bulk points $x\in dS_4$, this means that complex conjugation interchanges $P_L(x)\leftrightarrow P_R(x)$, just like the antipodal map. This is consistent with the fact that in the 3+1d bulk, left-handed and right-handed Weyl spinors are related by complex conjugation. For a twistor $Y^a = y_L^a + y_R^a$, the decomposition of the complex-conjugate $\bar Y^a$ into Weyl spinors reads:
\begin{align}
 \bar y_{L/R}^a \equiv (P_{L/R}){}^a{}_b(x)\bar Y^b = (\overline{P_{R/L}(x) Y})^a = (\overline{y_{R/L}})^a \ , \label{eq:y_bar}
\end{align}
again assuming real $x$. The complex conjugation of the Weyl spinor measures and delta functions then reads:
\begin{align}
\overline{d^2 y_{L/R}} = d^2\bar y_{R/L} \ ; \quad \bar\delta^{L/R}_x(Y) = \delta^{R/L}_x(Y) \ . \label{eq:delta_y_bar}
\end{align}

\subsubsection{The bulk Penrose transform} \label{sec:multiplet:bulk:Penrose}

The Penrose transform relates twistor functions $F(Y)$ to solutions of the free massless field equations (of all spins) in the $dS_4$ bulk. We restrict to integer spins, i.e. to even twistor functions $F(-Y) = F(Y)$. In HS language, the Penrose transform can be written as \cite{Neiman:2017mel}:
\begin{align}
C(x;Y) = -iF(Y)\star\delta^L_x(Y) \ , \label{eq:Penrose}
\end{align}
where $\delta^L_x(Y)$ is the bulk spinor delta function from \eqref{eq:delta_chiral}, and $C(x;Y)$ is a \emph{master field}, which encodes the higher-spin field strengths at $x$, as well as their derivatives. The factor of $-i$ follows the conventions of \cite{Neiman:2017mel}, and will serve to simplify reality conditions. The Penrose transform actually comes in both left-handed and right-handed versions, which are related by the $dS_4$ antipodal map \cite{Neiman:2013hca}. Eq. \eqref{eq:Penrose} describes the left-handed transform.

The values of the field strengths at $x$ can be extracted from the master field $C(x;Y)$ by evaluating the latter at $y_L=0$ or $y_R=0$, where $Y^a = y_L^a + y_R^a$ is the decomposition of the twistor $Y^a$ into left-handed and right-handed spinors at $x$. Explicitly, the spin-$s$ field strength can be expressed as a tensor $C^{(s)}_{\mu_1\nu_1\cdots\mu_s\nu_s}(x)$, which has the symmetries of a generalized Weyl tensor:  it is totally traceless, antisymmetric within each $\mu_k\nu_k$ index pair, symmetric under the exchange of any two such pairs, and vanishes when antisymmetrized over any three indices. For spin $s>0$, this field strength can be constructed from a pair of totally symmetric spinors $C^{(s;L)}_{a_1\dots a_{2s}}(x)$ and $C^{(s;R)}_{a_1\dots a_{2s}}(x)$, one completely left-handed and the other completely right-handed:
\begin{align}
C^{(s)}_{\mu_1\nu_1\cdots\mu_s\nu_s}(x) = \frac{1}{4^s}\,\gamma_{\mu_1\nu_1}^{a_1 b_1}\dots \gamma_{\mu_s\nu_s}^{a_s b_s}\left( C^{(s;L)}_{a_1 b_1\dots a_s b_s}(x) + C^{(s;R)}_{a_1 b_1\dots a_s b_s}(x) \right) \ .
\label{eq:Weyl_tensor}
\end{align}
Finally, the scalar field $C^{(0)}(x)$ and the chiral spin-$s$ field strengths $C^{(s;L/R)}_{a_1\dots a_{2s}}(x)$ are encoded in the master field \eqref{eq:Penrose} via:
\begin{align}
\begin{split}
C^{(0)}(x) &= C(x;0) \ ; \\
C^{(s;L)}_{a_1\dots a_{2s}}(x) &= \left.\frac{\del^{2s} C(x;y_L)}{\del y_L^{a_1}\dots\del y_L^{a_{2s}}}\right|_{y_L=0} \ ; \quad 
C^{(s;R)}_{a_1\dots a_{2s}}(x) = \left.\frac{\del^{2s} C(x;y_R)}{\del y_R^{a_1}\dots\del y_R^{a_{2s}}}\right|_{y_R=0} \ .
\end{split} \label{eq:packaging}
\end{align}
The Penrose transform \eqref{eq:Penrose} automatically ensures that these fields satisfy the free massless field equations in $dS_4$:
\begin{align}
\nabla_\mu\nabla^\mu C^{(0)}(x) = 2C^{(0)}(x) \ ; \quad \hat\nabla^{b_1 a} C^{(s;L)}_{b_1b_2\dots b_{2s}}(x) = \nabla^{ab_1} C^{(s;R)}_{b_1b_2\dots b_{2s}}(x) = 0 \ ,
\end{align}
where $\hat\nabla_{ab} \equiv P_L{}^c{}_a P_R{}^d{}_b\gamma^\mu_{cd}\nabla_\mu$ is the covariant derivative with a left-handed spinor index in the first position, and a right-handed spinor index in the second position. In addition, those Taylor coefficients of $C(x;Y)$ which don't appear in \eqref{eq:packaging} are identified by the Penrose transform with derivatives of the fields \eqref{eq:packaging}:
\begin{align}
\begin{split}
  \left.\frac{\del^{2(s+k)} C(x;Y)}{\del y_L^{a_1}\dots\del y_L^{a_{2s+k}}\del y^R_{b_1}\dots\del y^R_{b_k}}\right|_{Y=0} 
    &= \hat\nabla_{(a_1}{}^{(b_1}\dots\hat\nabla_{a_k}{}^{b_k)}C^{(s;L)}_{a_{k+1}\dots a_{2s+k})}(x) \ ; \\
  \left.\frac{\del^{2(s+k)} C(x;Y)}{\del y^L_{a_1}\dots\del y^L_{a_k}\del y_R^{b_1}\dots\del y_R^{b_{2s+k}}}\right|_{Y=0} 
    &= \hat\nabla^{(a_1}{}_{(b_1}\dots\hat\nabla^{a_k)}{}_{b_k}C^{(s;R)}_{b_{k+1}\dots b_{2s+k})}(x) \ .
\end{split} \label{eq:unfolding}
\end{align}
Thus, for given a spin $s$, the left-handed/right-handed field strength and its derivatives are encoded in the part of $C(x;Y)$ that satisfies the homogeneity condition:
\begin{align}
 y_L^a\frac{\del C(x;Y)}{\del y_L^a} - y_R^a\frac{\del C(x;Y)}{\del y_R^a} = \pm 2s \ .
\end{align}
For the twistor function $F(Y)$, we can read off from \eqref{eq:delta_Fourier_bulk_raw} that this translates into the well-known homogeneity condition:
\begin{align}
 Y^a\frac{\del F(Y)}{\del Y^a} = y_L^a\frac{\del F(Y)}{\del y_L^a} + y_R^a\frac{\del F(Y)}{\del y_R^a} = -2\mp 2s \ . \label{eq:homogeneity}
\end{align}

Now, recall that the $dS_4$ antipodal map $x^\mu\rightarrow -x^\mu$ interchanges the left-handed and right-handed spinor spaces \eqref{eq:projectors}. Therefore, it relates the left-handed Penrose transform \eqref{eq:Penrose} to its right-handed version:
\begin{align}
C(x;Y) = -iF(Y)\star\delta^L_x(Y) \ \rightarrow \ C(-x;Y) = -iF(Y)\star\delta^R_x(Y) \ .
\end{align}
We then see from \eqref{eq:delta_identities} that this antipodal map can be realized within HS algebra as:
\begin{align}
C(x;Y) \ \rightarrow \ C(-x;Y) = C(x;Y)\star\delta(Y) \ , \label{eq:antipodal_C}
\end{align}
or, equivalently:
\begin{align}
F(Y) \ \rightarrow \ F(Y)\star\delta(Y) \ . \label{eq:antipodal_F}
\end{align}
As we will see below, the bulk solutions relevant for us will be antipodally \emph{odd}. Thus, they will satisfy:
\begin{align}
C(-x;Y) = C(x;Y)\star\delta(Y) = -C(x;Y) \ ; \quad F(Y)\star\delta(Y) = -F(Y) \ . \label{eq:antip_odd_anticipation}
\end{align}

Let us now address reality conditions. For the spin-$s$ field strength \eqref{eq:Weyl_tensor} to be real, its left-handed and right-handed pieces must be complex-conjugate to each other:
\begin{align}
\overline{C^{(s;L)}_{a_1 b_1\dots a_s b_s}(x)} = C^{(s;R)}_{a_1 b_1\dots a_s b_s}(x) \ . \label{eq:bulk_reality_raw}
\end{align}
Via \eqref{eq:packaging} and \eqref{eq:unfolding}, this implies the following reality condition on the master field $C(x;Y)$:
\begin{align}
 \bar C(x;y_L + y_R) = C(x; -y_L+y_R) \ ,
\end{align}
which can be written using \eqref{eq:delta_bulk_Klein} as:
\begin{align}
 \bar C(x;Y) = \delta^L_x(Y) \star C(x;Y) \star \delta^L_x(Y) \ .  \label{eq:bulk_reality}
\end{align}
In the Penrose transform \eqref{eq:Penrose}, complex conjugation flips the sign of the $-i$ factor, reverses the order of factors in the star product, and sends $\delta^L_x(Y)\rightarrow \delta^R_x(Y)$. Thus, making use of \eqref{eq:delta_identities}, we find that the bulk fields' reality condition \eqref{eq:bulk_reality} translates into:
\begin{align}
 \bar F(Y) = -F(Y)\star\delta(Y) \ . \label{eq:Penrose_reality}
\end{align}
For antipodally odd solutions as in \eqref{eq:antip_odd_anticipation}, this becomes just the standard reality condition $\bar F(Y) = F(Y)$. 

\subsection{Boundary particle mechanics: twistors and momentum spinors} \label{sec:multiplet:mechanics}

\subsubsection{Overview}

We now turn our attention to the boundary CFT -- the free $Sp(2N)$ vector model. However, before considering it as a field theory, we will first construct the \emph{quantum mechanics} of the vector model's free massless particles. In section \ref{sec:multiplet:CFT}, we will use this particle mechanics to express the CFT partition function. Specifically, we'll find that the CFT sources can be encoded by \emph{quantum-mechanical operators} in the particle mechanics. 

We will start with a fully covariant treatment, where twistor space will play the role of phase space for the boundary particle. We then perform a partial symmetry breaking, corresponding to a choice of causal patch in the bulk. From the point of view of the boundary particle, this serves to polarize the phase space into ``configuration variables'' and ``conjugate momenta'', each parameterized by an $SO(3)$ spinor. These spinors are in fact the square roots of the particle's momentum in a pair of flat conformal frames. We will use one of these spinors as the configuration variable in our construction of the boundary particle's quantum mechanics.

Since the boundary of $dS_4$ is Euclidean, the phase space of our boundary particle will be complexified; we will not have a notion of real phase space points, or a distinction between positive and negative energies. Nevertheless, we won't have any problem defining a symplectic form and quantum-mechanical commutators: we just analytically continue the Lorentzian expressions. Here lies a key difference between mechanics and field theory: in field theory, the symplectic form and commutators are crucially linked to the Lorentzian causal structure, and the task of extracting them from a Euclidean theory is the very subject of this paper!

Given a complex phase space for the boundary particle, one should be careful with the concept of Hermitian conjugation $\hat f\rightarrow \hat f^\dagger$ in its quantum mechanics. We will find that the idempotence $(\hat f^\dagger)^\dagger = \hat f$, and thus the existence of Hermitian operators $\hat f = \hat f^\dagger$, is related to the restriction to bosons in the bulk. A consistent Hermitian conjugation for boundary particle \emph{states} will rely on the further restriction to \emph{antipodally odd} bulk fields; this will be equivalent to choosing a scalar, rather than spinor, boundary particle. The next issue of concern is the reality and positivity of the Hermitian norm. In fact, we are \emph{not} able to find a manifestly real Hermitian norm on particle states. We do find such a norm $\tr(\hat f\hat f^\dagger)$ for \emph{operators}, which will be directly related to the CFT 2-point function. This norm then turns out to have opposite signs for even and odd spins. Thus, in order to have a positive-definite norm, we must make the restriction to even bulk spins, i.e. to the $Sp(2N)$ rather than $U(N)$ vector model. To summarize, the space $\calH$ of boundary particle states is not quite a Hilbert space, since its Hermitian norm is not manifestly positive; however, with the restriction to even spins, the space of operators $\calH\otimes\calH^*$ \emph{does} have an honest Hilbert-space structure.

The connection between even spins and a positive-definite quadratic form is not surprising. Indeed, when analytically continuing 2-point functions from $AdS_4$ to $dS_4$ as in \cite{Anninos:2011ui}, one ends up with different signs for even \& odd bulk spins, which translate into opposite-sign kinetic energies. This fact is most famous in the context of $\calN=2$ supergravity, where the spin-2 graviton and its spin-1 superpartner in $dS_4$ cannot both have kinetic energy of the correct sign \cite{Pilch:1984aw}.

In our discussion of boundary mechanics, we will habitually associate twistor functions $F(Y)$ with their Penrose transform into free HS fields in the bulk. In section \ref{sec:multiplet:CFT}, we will flesh out this bulk/boundary/twistor correspondence, building on the work of \cite{Neiman:2017mel}.

\subsubsection{The phase space of a conformal boundary particle} \label{sec:multiplet:mechanics:phase_space}

To begin, consider the phase space of a spin-0, massless particle in $d$-dimensional spacetime $\bbR^{1,d-1}$. This phase space is $2(d-1)$-dimensional. Each point in the phase space represents a lightray -- the particle's worldline -- along with a magnitude of the null momentum. This phase space can be described conformally, using the embedding formalism, in which our spacetime is the projective lightcone $\{\ell^\mu\in\bbR^{2,d}\,|\,\ell_\mu\ell^\mu = 0 ; \, \ell^\mu\cong\alpha\ell^\mu\}$. In this picture, a phase space point can be encoded by a \emph{simple, totally null bivector} $M^{\mu\nu}$:
\begin{align}
 M^{\mu\nu} = M^{[\mu\nu]} \ ; \quad M^{[\mu\nu}M^{\rho]\sigma} = 0 \ ; \quad M_{\mu\nu}M^{\nu\rho} = 0 \ . \label{eq:M_symmetries}
\end{align}
Indeed, the ``direction'' of $M^{\mu\nu}$ defines a totally null 2-plane through the origin in $\bbR^{2,d}$, which, projectively, is a lightray in $\bbR^{1,d-1}$. The ``magnitude'' of $M^{\mu\nu}$ can then encode the magnitude of the null momentum. To see this more concretely, consider a conformal frame defined by the flat section $\ell\cdot n_{(\infty)} = -1/2$ of the $\bbR^{2,d}$ lightcone, where the null vector $n_{(\infty)}^\mu$ specifies a ``point at infinity'', as in \eqref{eq:n}-\eqref{eq:flat}. Then at each point $\ell^\mu$ on the particle's worldline, i.e. on the $\ell\cdot n_{(\infty)} = -1/2$ section of the 2-plane defined by $M^{\mu\nu}$, we can encode the null momentum by the vector:
\begin{align}
 p^\mu = 2M^{\mu\nu}n^{(\infty)}_\nu \ , \label{eq:p}
\end{align}
which is tangential to the worldline at $\ell^\mu$. The numerical coefficient in \eqref{eq:p} is chosen for later convenience.

Let's now construct the Poisson brackets on our particle's phase space. The Poisson bracket $\{M^{\mu\nu},M_{\rho\sigma}\}$ is actually completely fixed by its index symmetries and the totally-null property $M_{\mu\nu}M^{\nu\rho} = 0$ of $M^{\mu\nu}$:
\begin{align}
 \left\{M^{\mu\nu}, M_{\rho\sigma} \right\} = 4\delta^{[\mu}_{[\rho}\, M^{\nu]}{}_{\sigma]} \ , \label{eq:M_Poisson}
\end{align}
up to a numerical coefficient. We see that $M^{\mu\nu}$ generates the conformal group $SO(2,d)$. Let us now demonstrate that our normalizations in \eqref{eq:p}-\eqref{eq:M_Poisson} are mutually consistent. For an $SO(2,d)$ rotation along an infinitesimal bivector $\Sigma^{\mu\nu}$, the brackets \eqref{eq:M_Poisson} imply the following matrix elements in the vector representation:
\begin{align}
 \left(1 + \textstyle\frac{1}{2}\Sigma^{\mu\nu}\{M_{\mu\nu},\cdot\} \right)^\rho{}_\sigma = \delta^\rho_\sigma + \Sigma^\rho{}_\sigma \ . \label{eq:generator_elements}
\end{align}
Now, consider a particular subset of $SO(2,d)$ generators -- translations along an infinitesimal vector $\boldsymbol{\epsilon}$ in the flat conformal frame defined by $n_{(\infty)}^\mu$. In the $\bbR^{2,d}$ picture, $\boldsymbol{\epsilon}$ becomes an $\bbR^{2,d}$ vector $\epsilon^\mu$, subject to the constraint and equivalence relation \eqref{eq:tangent_vector}. A translation along $\boldsymbol{\epsilon}$ then corresponds to $\Sigma^{\mu\nu} = 4n_{(\infty)}^{[\mu}\epsilon^{\nu]}$, as we can verify by acting with the matrix \eqref{eq:generator_elements} on some point $\ell^\mu$ of the flat section $\ell\cdot n_{(\infty)} = -1/2$. Thus, the group element describing this translation can be written as:
\begin{align}
 1 + 2n_{(\infty)}^\mu\epsilon^\nu \{M_{\mu\nu},\cdot\} = 1 - \epsilon^\mu \{p_\mu,\cdot\} \ ,
\end{align}
where we used \eqref{eq:p} for the second expression. Since $1 - \epsilon^\mu \{p_\mu,\cdot\}$ is the standard group element for an infinitesimal translation, we conclude that the normalizations \eqref{eq:p}-\eqref{eq:M_Poisson} are consistent.

Now, let us specialize to our case of interest $d=3$, and analytically continue to Euclidean signature, so the embedding space becomes $\bbR^{1,4}$. In this particular dimension, two simplifications occur. First, the simplicity constraint $M^{[\mu\nu}M^{\rho]\sigma} = 0$ on a totally null bivector in embedding space is satisfied automatically. Second, the totally null bivector $M^{\mu\nu}$ can be parameterized as the square of a twistor, as in \eqref{eq:generators}:
\begin{align}
 M^{\mu\nu} = \frac{1}{8}Y\gamma^{\mu\nu}Y \ , \label{eq:M_YY}
\end{align}
Thus, the space of twistors $Y^a$ is just the double cover ($Y^a \cong -Y^a$) of the spinless, massless particle's phase space in 3d (Euclidean) spacetime! This is of course consistent with the fact that the ``direction'' of a twistor defines a lightray on the 3d conformal boundary of $dS_4$. 

The symplectic form on the boundary particle's phase space will now be some twistor matrix $\Omega_{ab}(Y)$, which must be invariant under the conformal group $SO(1,4)$. But the only such matrices are constant multiples of the twistor metric $I_{ab}$! Fixing the proportionality constant between $\Omega_{ab}$ and $I_{ab}$ is equivalent to fixing the coefficient between $M^{\mu\nu}$ and $Y^a Y^b$ in \eqref{eq:M_YY}. As we'll see momentarily, it is consistent to set:
\begin{align}
 \Omega_{ab} = -\frac{1}{2}I_{ab} \ ; \quad (\Omega^{-1})^{ab} = 2I^{ab} \ . \label{eq:Omega_I}
\end{align}
We now recognize the higher-spin algebra \eqref{eq:HS_algebra}-\eqref{eq:star_int} as the \emph{operator algebra} \eqref{eq:Moyal} of the free particle's quantum mechanics, with operators represented as phase space functions $f(Y)$ via the Wigner-Weyl transform:
\begin{align}
 \begin{split}
   Y^{a_1}\dots Y^{a_n}  \quad &\longleftrightarrow \quad  \hat Y^{(a_1}\dots \hat Y^{a_n)} \ ; \\
   f(Y)\star g(Y) \quad &\longleftrightarrow \quad \hat f\hat g \ .
 \end{split} \label{eq:HS_QM}
\end{align}
With the identification \eqref{eq:HS_QM}, we immediately see that the star-product commutators \eqref{eq:M_algebra} of $M^{\mu\nu}$ are just the quantum-mechanical commutators corresponding to the Poisson brackets \eqref{eq:M_Poisson}. This justifies our normalizations in \eqref{eq:M_YY}-\eqref{eq:Omega_I}.

To sum up, we see that twistor functions, which correspond via the Penrose transform to linearized bulk fields, and thus to boundary data at conformal infinity, can also be thought of as quantum operators in the boundary particle mechanics. This statement is still slightly imprecise, as it ignores the double-cover relationship between twistors $Y^a$ and phase space points $M^{\mu\nu}$. In the following section, we fill in this and other details.

\subsubsection{Details of the phase space/twistor dictionary} \label{sec:multiplet:mechanics:twistor_details}

Due to the different normalization \eqref{eq:Omega_I} of the symplectic form, formulas in HS algebra notation tend to be more streamlined than their phase-space counterparts from section \ref{sec:WW}. The main point requiring care is the different normalization of integration measures. The standard phase space measure \eqref{eq:phase_space_measure} for the boundary particle's mechanics, which we'll denote by $d^4Y_{\text{mech}}$, is related to the standard twistor measure \eqref{eq:twistor_measure} as $d^4Y_{\text{mech}} = \pm d^4Y/4$, where the $\frac{1}{4}$ is the square of the $-\frac{1}{2}$ in \eqref{eq:Omega_I}, and we remember the inherent sign flexibility in \eqref{eq:phase_space_measure}. In fact, we'll see that the consistent sign choice is:
\begin{align}
 d^4Y_{\text{mech}} = -\frac{1}{4}d^4Y \ . \label{eq:measures_1/4}
\end{align}
Of course, we could simply absorb this sign into our definition of $d^4Y$. However, we choose not to do this, as it would affect our sign conventions in the bulk theory, via the decomposition \eqref{eq:measure_chiral} of $d^4Y$ into bulk spinor measures.

Now, recall that twistor space is a \emph{double cover} of the 3d particle's phase space, in the sense that $\pm Y^a$ describe the same phase space point. In other words, our spin-0 boundary particle is invariant under $2\pi$ rotations. At the quantum level, we can handle this by starting from a description where twistor space \emph{is} phase space, and then restricting to states that are invariant under the phase-space reversal operator $\hat R$ from \eqref{eq:reversal_operator}. For operators $\hat f$ acting on such states, this implies invariance under both left- and right-multiplication by $\hat R$:
\begin{align}
 \hat R\hat f = \hat f\hat R = \hat f \ . \label{eq:reversal_QM}
\end{align}
As a twistor function, the reversal operator $\hat R$ is given by the Dirac delta, as in \eqref{eq:reversal_operator_WW}:
\begin{align}
 \hat R \quad \longleftrightarrow \quad \frac{1}{4}\,\delta(Y)_{\text{mech}} = -\delta(Y) \ , \label{eq:reversal_twistor}
\end{align}
where we took into account the different normalization of delta functions due to \eqref{eq:measures_1/4}. Thus, the $2\pi$ rotation symmetry \eqref{eq:reversal_QM} of states and operators is encoded by restricting to twistor functions $f(Y)$ that satisfy:
\begin{align}
 \delta(Y)\star f(Y) = f(Y)\star\delta(Y) = -f(Y) \ . \label{eq:f_antip_odd}
\end{align}
Due to \eqref{eq:delta_Klein}, this implies in particular that our twistor function is even:
\begin{align}
 f(-Y) = \delta(Y)\star f(Y)\star\delta(Y)  = f(Y) \ . \label{eq:f_even}
\end{align}
Now consider the bulk fields corresponding to $f(Y)$ via the Penrose transform. Eq. \eqref{eq:f_even} tells us that these fields are restricted to integer spins, as expected. What is interesting is that the more detailed condition \eqref{eq:f_antip_odd} fixes the bulk fields' \emph{antipodal symmetry}. Specifically, it restricts the bulk fields to be \emph{antipodally odd}, as anticipated in \eqref{eq:antip_odd_anticipation}.

For an operator $\hat f$ corresponding to a twistor function $f(Y)$, the quantum-mechanical trace takes the form:
\begin{align}
  \tr\hat f = \int d^4Y_{\text{mech}}\,f(Y) = -\frac{1}{4}\int d^4Y f(Y) = -\frac{1}{4}\tr_\star\left(f(Y)\star\delta(Y)\right) \ , \label{eq:traces_raw}
\end{align}
where $\tr_\star f(Y) = f(0)$ is the standard trace operation \eqref{eq:str} from the higher-spin literature. Under the constraint \eqref{eq:f_antip_odd}, eq. \eqref{eq:traces_raw} simplifies into:
\begin{align}
 \tr\hat f = \frac{1}{4}\tr_\star f(Y) \ . \label{eq:traces}
\end{align}

Finally, everything so far is consistent with identifying the complex conjugation \eqref{eq:f_bar} of an even twistor function $f(Y)$ with the Hermitian conjugation of the operator $\hat f$. In particular, the star-product identity \eqref{eq:star_Hermit} is consistent with $(\hat f\hat g)^\dagger = \hat g^\dagger\hat f^\dagger$, while the identities \eqref{eq:tr_measure_delta_bar} are consistent with $\tr\hat f^\dagger = \overline{\tr\hat f}$. Finally, for even (i.e. integer-spin) functions $f(Y)$, complex conjugation is idempotent \eqref{eq:f_bar_bar}, which translates into the idempotence $(\hat f^\dagger)^\dagger = \hat f$ of Hermitian conjugation. In particular, we can define Hermitian operators $\hat f^\dagger = \hat f$, which correspond to real twistor functions $\bar f(Y) = f(Y)$. Due to \eqref{eq:bulk_reality_raw}-\eqref{eq:Penrose_reality}, these in turn correspond to real bulk fields, assuming the antipodal symmetry condition \eqref{eq:f_antip_odd}.

What remains unclear at this stage is whether the above Hermitian structure is \emph{positive}, in the sense that $\tr(\hat f\hat f^\dagger)$ is positive-definite. As we'll see, this positivity requirement will force us to restrict our operators from all integer spins to just even ones.

\subsubsection{Boundary momentum spinors} \label{sec:multiplet:mechanics:spinors}

In this section, we descend from the more abstract twistor description of the boundary particle into a description in terms of plane waves with null momentum. The notion of plane waves requires a choice of flat conformal frame, which is equivalent to singling out a ``point at infinity'' $n_{(\infty)}$ on the boundary $S_3$. In addition, to fix the phases of the plane-wave basis, we will need a choice of origin in this flat frame, which amounts to choosing a second boundary point $n_{(0)}$. As we recall from section \ref{sec:fields}, such a choice of two boundary points is equivalent to a choice of observer in the $dS_4/\bbZ_2$ bulk, whose worldline begins at $n_{(0)}$ and ends at $n_{(\infty)}$. The asymmetric roles of $n_{(\infty)}$ and $n_{(0)}$ in our construction are analogous to how one might encode a bulk solution in the causal patch in terms of boundary data on the \emph{final} horizon, with the initial horizon serving ``merely'' as a placeholder for the bifurcation surface.

Let us now move on from words to equations. We fix the two boundary points $n_{(0)}^\mu$,$n_{(\infty)}^\mu$ as in \eqref{eq:nn'}, noting the relative normalization $n_{(0)}\cdot n_{(\infty)} = -1/2$ of the null vectors. The $O(1,4)$ de Sitter symmetry is broken down to $SO(1,1)\times O(3)$. For a twistor $Y^a$, this implies a decomposition into $SO(3)$ spinors:
\begin{align}
 Y^a = \begin{pmatrix} \lambda_\alpha \\ \mu^\alpha \end{pmatrix} \ ; \quad Y_a = \begin{pmatrix} -\mu^\alpha \\ \lambda_\alpha \end{pmatrix} \ . \label{eq:Y_decompose}
\end{align}
In this decomposition, 2-component upper-index spinors $u^\alpha$ form the twistor subspace spanned by $n_{(\infty)}^{ab}$, and can be thought of as the square roots of boundary vectors in the $\bbR^3$ conformal frame defined by $n_{(\infty)}$; similarly, lower-index spinors $u_\alpha$ form the subspace spanned by $n_{(0)}^{ab}$, and can be thought of as the square roots of \emph{covectors} in this $\bbR^3$. More explicitly, the decomposition \eqref{eq:Y_decompose} is consistent with the following realization of the twistor metric $I_{ab}$ and the gamma matrices $(\gamma_\mu)^a{}_b$:
\begin{align}
 \begin{split}
   &I_{ab} = \begin{pmatrix} 0 & -\delta^\alpha_\beta \\ \delta_\alpha^\beta & 0 \end{pmatrix} \ ; \quad I^{ab} = \begin{pmatrix} 0 & -\delta_\alpha^\beta \\ \delta^\alpha_\beta & 0 \end{pmatrix} \ ; \quad 
     (\gamma_k)^a{}_b = \begin{pmatrix} (\tau_k)^\beta{}_\alpha & 0 \\ 0 & (\tau_k)^\alpha{}_\beta \end{pmatrix} \ ; \\
   &(n_{(\infty)})^a{}_b = \frac{1}{2}(\gamma_0 + \gamma_4)^a{}_b = \begin{pmatrix} 0 & 0 \\ -\epsilon^{\alpha\beta} & 0 \end{pmatrix} \ ; \quad 
      (n_{(0)})^a{}_b = \frac{1}{2}(\gamma_0 - \gamma_4)^a{}_b = \begin{pmatrix} 0 & \epsilon_{\alpha\beta} \\ 0 & 0 \end{pmatrix} \ .
 \end{split} \label{eq:block_matrices}
\end{align}
Here, $\epsilon_{\alpha\beta}$ is the antisymmetric spinor metric with inverse $\epsilon^{\alpha\gamma}\epsilon_{\beta\gamma} = \delta^\alpha_\beta$, which we can use to raise and lower spinor indices as $u_\alpha = \epsilon_{\alpha\beta}u^\beta$ and $u^\alpha = u_\beta\epsilon^{\beta\alpha}$. The index $k=1,2,3$ enumerates the axes of the $\bbR^3$ subspace orthogonal to $n_{(\infty)},n_{(0)}$. Finally, the matrices $(\tau_k)^\alpha{}_\beta$ are imaginary multiples $\tau_k = -i\sigma_k$ of the Pauli matrices, which satisfy the quaternionic algebra:
\begin{align}
 \tau_i\tau_j = -\delta_{ij} + \epsilon_{ijk}\tau_k \ .
\end{align}
Under the decomposition \eqref{eq:Y_decompose}, the higher-spin algebra \eqref{eq:HS_algebra} decomposes as:
\begin{align}
 \begin{split}
   \lambda_\alpha\star\lambda_\beta = \lambda_\alpha\lambda_\beta \ &; \quad \mu^\alpha\star\mu^\beta = \mu^\alpha\mu^\beta \ ; \\
   \lambda_\alpha\star\mu^\beta = \lambda_\alpha\mu^\beta - i\delta_\alpha^\beta \ &; \quad \mu^\alpha\star\lambda_\beta = \mu^\alpha\lambda_\beta + i\delta^\alpha_\beta \ .
 \end{split} \label{eq:HS_algebra_spinors}
\end{align}
The generators $Y^a Y^b$ of $O(1,4)$ decompose as follows:
\begin{itemize}
	\item $\lambda_\alpha \lambda_\beta$ generate translations in $\bbR^3$, which are broken by the choice of $n_{(0)}$.
	\item $\mu^\alpha \mu^\beta$ generate special conformal transformations, which are broken by the choice of $n_{(\infty)}$.
	\item $\lambda_\alpha \mu^\alpha$ generates dilatations, which correspond to time translations $t\rightarrow t+\tau$ for the bulk observer, and remain unbroken. These rescale $\lambda$ and $\mu$ by opposite factors: 
	  \begin{align}
	    Y^a = \begin{pmatrix} \lambda_\alpha \\ \mu^\alpha \end{pmatrix} \, \rightarrow \, \begin{pmatrix} e^{-\tau/2}\, \lambda_\alpha \\ e^{\tau/2}\,\mu^\alpha \end{pmatrix} \ . \label{eq:spinor_time_translation}
	  \end{align} 
	\item Finally, the traceless part of $\lambda_\alpha \mu^\beta$ generates $SO(3)$ rotations in both the boundary and bulk pictures; these also remain unbroken.
\end{itemize}
As with the 4-component twistors, we adopt an index-free notation for spinor contractions, i.e. $u_\alpha v^\alpha \equiv uv$. With this convention, the inner product of two twistors reads:
\begin{align}
 YY' = \lambda\mu' - \lambda'\mu \ .
\end{align}
Now, recall that our twistors $Y^a$ form a double cover of the boundary particle's phase space, and the twistor metric $I_{ab}$ is proportional to the phase space symplectic form. We then see from \eqref{eq:block_matrices} that the spinors $\lambda_\alpha$ and $\mu^\alpha$ are canonically conjugate to each other. Moreover, their physical meaning becomes clear from the above decomposition of the $O(1,4)$ generators. $\lambda_\alpha$ is the square root of the translation generator $\lambda_\alpha \lambda_\beta$, i.e. of the boundary particle's momentum; $\lambda_\alpha$ is thus a ``momentum spinor''. Similarly, $\mu^\alpha$ is the square root of the special conformal generator $\mu^\alpha\mu^\beta$, i.e. of the particle's momentum in the \emph{inverted} $\bbR^3$ frame obtained by interchanging $n_{(\infty)}$ and $n_{(0)}$. 

The precise relation between the particle's momentum \eqref{eq:p} and $\lambda_\alpha \lambda_\beta$ can be derived as follows. The momentum \eqref{eq:p} is a vector in the embedding space $\bbR^{1,4}$. As we can see from \eqref{eq:flat_displacement}-\eqref{eq:tangent_vector}, the same $\bbR^3$ vector evaluated at different points corresponds to somewhat different $\bbR^{1,4}$ vectors. However, regardless of this ambiguity, the $\bbR^3$ vector can always be read off from the $k=1,2,3$ components of the $\bbR^{1,4}$ vector. Thus, the $\bbR^3$ momentum of the particle can be obtained by simply isolating those components of eq. \eqref{eq:p}:
\begin{align}
 p_k = 2M_{k\nu}n_{(\infty)}^\nu = \frac{1}{4}Y\gamma_k n_{(\infty)} Y = -\frac{1}{4}(\tau_k)^{\alpha\beta}\lambda_\alpha \lambda_\beta = \frac{1}{4}\lambda\tau_k \lambda \ . \label{eq:p_spinors}
\end{align}
Note that the observer's time translations \eqref{eq:spinor_time_translation} act on this momentum is dilatations, $\mathbf{p}\rightarrow e^{-\tau}\mathbf{p}$.

The twistor measure \eqref{eq:twistor_measure} decomposes into spinor measures as:
\begin{align}
 d^4Y = -d^2\lambda\, d^2\mu \ ; \quad d^2\lambda \equiv \frac{\epsilon^{\alpha\beta}d\lambda_\alpha d\lambda_\beta}{2(2\pi)} \ ; \quad d^2\mu \equiv \frac{\epsilon_{\alpha\beta}\mu^\alpha\mu^\beta}{2(2\pi)} \ . \label{eq:spinor_measures}
\end{align}
The sign in \eqref{eq:spinor_measures} is the reason for our sign choice in the proportionality \eqref{eq:measures_1/4} between the twistor and phase space measures. The phase space measure now decomposes as:
\begin{align}
 d^4Y_{\text{mech}} = d^2\lambda_{\text{mech}}\,d^2\mu_{\text{mech}} \ ; \quad d^2\lambda_{\text{mech}} = \frac{1}{2}d^2\lambda \ ; \quad d^2\mu_{\text{mech}} = \frac{1}{2}d^2\mu \ , \label{eq:spinor_measures_1/2}
\end{align}
where $d^2\lambda_{\text{mech}}$ and $d^2\mu_{\text{mech}}$ play the roles of the configuration-space \& momentum-space measures $d^\calN q$ and $d^\calN p$ from section \ref{sec:WW:explicit}. 

Using the spinor measures \eqref{eq:spinor_measures}, we define delta functions in the usual way:
\begin{align}
 \delta(\lambda) = \int d^2\mu\,e^{i\lambda\mu} \ ; \quad \int d^2\lambda\,\delta(\lambda - \lambda') f(\lambda) = f(\lambda') \ . \label{eq:spinor_deltas}
\end{align}

Finally, we come to complex conjugation. The twistor complex conjugation $Y^a\rightarrow \bar Y^a$ is related to the standard complex conjugation $u^\alpha\rightarrow \bar u^\alpha$ of $SO(3)$ spinors via:
\begin{align}
 \bar Y^a = \begin{pmatrix} \bar\lambda_\alpha \\ \bar\mu^\alpha \end{pmatrix} \ ; \quad \bar Y_a = \begin{pmatrix} -\bar\mu^\alpha \\ \bar\lambda_\alpha \end{pmatrix} \ ; \quad 
 \bar\lambda_\alpha = \left(\lambda^\alpha\right)^* \ ; \quad \bar\mu^\alpha = \left(-\mu_\alpha\right)^* \ .
\end{align}
Just like for twistors, the spinor complex conjugation is anti-idempotent: $\bar{\bar u}^\alpha = -u^\alpha$. Under this complex conjugation, the matrices $\epsilon_{\alpha\beta}$ and $\tau_k^{\alpha\beta}$ are real. The spinor measure and the spinor delta function conjugate as in \eqref{eq:tr_measure_delta_bar}:
\begin{align}
 d^2\bar u = \overline{d^2u} \ ; \quad \bar\delta(u) = \delta(u) \ .
\end{align}

\subsubsection{Boundary quantum mechanics in the spinor basis} \label{sec:multiplet:mechanics:spinor_QM}

With these ingredients in place, we define, as in \eqref{eq:q_p_basis}, a basis $\ket{\lambda}$ for the boundary particle's Hilbert space, composed of eigenstates of the momentum spinor $\lambda_\alpha$. In this basis, we can describe states $\ket{\psi}$ as wavefunctions $\psi(\lambda) = \braket{\lambda|\psi}$, which must be even under $2\pi$ rotations: $\psi(\lambda) = \psi(-\lambda)$. Similarly, an operator $\hat f$ can be expressed through its matrix elements $\braket{\lambda|\hat f|\lambda'}$, which must be even under $\lambda\rightarrow -\lambda$ and $\lambda'\rightarrow -\lambda'$ separately. Of course, one can also construct states and operators that are \emph{odd} under such $2\pi$ rotations; those would describe the quantum mechanics of a spin-$\frac{1}{2}$ particle. Below, we will initially consider states and operators with arbitrary dependence on the spinors, so that we can trace the precise roles of the various discrete symmetries. In the end, we'll restrict again to the spin-$0$ boundary particle, along with a \emph{further} restriction to CPT-invariant operators, which, from the bulk point of view, will correspond to even spins.

The matrix elements $\braket{\lambda|\hat f|\lambda'}$ of an operator in the boundary mechanics are related to a phase space function $f(Y)$ via the Wigner-Weyl transform \eqref{eq:WW_explicit}. Taking into account the factors of 2 in \eqref{eq:Omega_I} and \eqref{eq:spinor_measures_1/2}, this transform reads:
\begin{align}
 f(Y) = f(\lambda,\mu) = 2 \int d^2u \Braket{ \lambda + u | \hat f | \lambda - u } e^{-i u\mu} \ . \label{eq:WW_F}
\end{align}
This can be expressed in terms of a kernel:
\begin{align}
  f(Y) &= \int d^2u\,d^2u'\,K(u,u';Y) \braket{ u|\hat f|u' } \ ; \label{eq:F_K} \\
  K(u,u';Y) &= \frac{1}{2}\,\delta\!\left(\lambda - \frac{u + u'}{2}\right) e^{i(u - u')\mu/2} \ . \label{eq:K}
\end{align}
The kernel $K(u,u';Y)$ satisfies the identities:
\begin{align}
 K(u,u';Y)\star K(v,v';Y) &= \frac{1}{2}\delta(u'-v) K(u,v';Y) \ ; \label{eq:K_star_K} \\ 
 \delta(Y)\star K(u,u';Y) &= -K(-u,u';Y) \ ; \\ 
 K(u,u';Y)\star\delta(Y) &= -K(u,-u';Y) \ ; \\
 \tr_\star K(u,u';Y) &= 2\,\delta(u + u') \ , \label{eq:tr_K}
\end{align}
which can be verified via the integral formulas \eqref{eq:star_int},\eqref{eq:delta_Fourier} for the star product. The twistor integrals involved can be decomposed into spinor integrals via \eqref{eq:Y_decompose},\eqref{eq:spinor_measures}, and evaluated using the delta-function formulas \eqref{eq:spinor_deltas}. 

The identities \eqref{eq:K_star_K}-\eqref{eq:tr_K} ensure that the operator product, the $2\pi$ rotation symmetry and the quantum-mechanical trace are consistent between the $f(Y)$ and $\braket{u|\hat f|u'}$ representations:
\begin{align}
 f(Y) \quad \Longleftrightarrow \quad &\hat f \quad \Longleftrightarrow \quad \braket{u|\hat f|u'} \ ; \\
 f(Y)\star g(Y) \quad \Longleftrightarrow \quad &\hat f\hat g \quad \Longleftrightarrow \quad \frac{1}{2}\int d^2v \braket{u|\hat f|v}\braket{v|\hat g|u'} \ ; \\
 -\delta(Y)\star f(Y) \quad \Longleftrightarrow \quad &\hat R\hat f \quad \Longleftrightarrow \quad \Braket{-u|\hat f|u'} \ ; \label{eq:spinor_antip_1} \\
 -f(Y)\star\delta(Y) \quad \Longleftrightarrow \quad &\hat f\hat R \quad \Longleftrightarrow \quad \Braket{u|\hat f|-u'} \ ; \label{eq:spinor_antip_2} \\
 -\frac{1}{4}\tr_\star\left(f(Y)\star\delta(Y)\right) ={} &\tr\hat f = \frac{1}{2}\int d^2u \braket{u|\hat f|u} \ .
\end{align}
Notice the particular elegance of eqs. \eqref{eq:spinor_antip_1}-\eqref{eq:spinor_antip_2}. Flipping the sign of one of the spinors $u,u'$ describes a $2\pi$ rotation on one of the ``legs'' of the operator $\hat f$. The boundary theory is usually blind to this; in particular, the null momenta $\mathbf{p}\sim u\boldsymbol{\tau}u$ and $\mathbf{p'}\sim u'\boldsymbol{\tau}u'$ remain unaffected. At the same time, the boundary theory is usually blind to the fact that ``de Sitter space is twice too big'': it's unaware of the existence of two boundaries, or the difference between $dS_4$ and $dS_4/\bbZ_2$. Eqs. \eqref{eq:spinor_antip_1}-\eqref{eq:spinor_antip_2} are telling us that these two ``double cover issues'' are the same: flipping the sign of $u$ or $u'$ is just (minus) the bulk antipodal map. In other words, replacing the momentum vector with its spinorial square root gains us access to the ``doubled'' nature of the $dS_4$ bulk!

Let us now address Hermitian conjugation. The transformation kernel \eqref{eq:K} satisfies:
\begin{align}
 \overline{K(u,u';Y)} = K(\bar u', \bar u; \bar Y) \ ,
\end{align}
which establishes that complex conjugation of the twistor function is realized on operator matrix elements as follows:
\begin{align}
f(Y) \ \rightarrow \ \bar f(Y) \quad \Longleftrightarrow \quad \Braket{u|\hat f|u'} \ \rightarrow \ \overline{\Braket{-\bar u'|\hat f|-\bar u}} \ . \label{eq:conjugation}
\end{align}
In section \ref{sec:multiplet:mechanics:twistor_details}, we saw that $f(Y)\rightarrow \bar f(Y)$ has the appropriate algebraic properties for a Hermitian conjugation $\hat f\rightarrow \hat f^\dagger$, and is idempotent for integer bulk spins. In the language of operator matrix elements, integer bulk spins are characterized by $\braket{u|\hat f|u'} = \braket{-u|\hat f|{-}u'}$, which indeed ensures that \eqref{eq:conjugation} is idempotent. To complete the correspondence between \eqref{eq:conjugation} and Hermitian conjugation, we should identify the last expression in \eqref{eq:conjugation} with the matrix element $\braket{u|\hat f^\dagger|u'}$ of $\hat f^\dagger$. For this to make sense, we must define Hermitian conjugation not just on operators, but also on the bra and ket states. The most sensible definition is:
\begin{align}
\ket{u}^\dagger = \bra{\eta\bar u} \ ; \quad \bra{u}^\dagger = \ket{-\bar\eta\bar u} \ , \label{eq:state_conjugation} 
\end{align}
where $\eta$ is a possible phase factor. The minus sign in \eqref{eq:state_conjugation} makes sure that this Hermitian conjugation is idempotent. With this definition, the matrix elements of $\hat f^\dagger$ read:
\begin{align}
 \Braket{u|\hat f^\dagger|u'} = \overline{\Braket{\eta\bar u'|\hat f|-\bar\eta\bar u}} \ .
\end{align}
This coincides with \eqref{eq:conjugation} only if we choose $\eta=\pm 1$, \emph{and} restrict $\braket{u|\hat f|u'}$ to be even not only under $(u,u')\rightarrow (-u,-u')$, but under $u\rightarrow -u$ and $u'\rightarrow -u'$ separately. That is of course just our original restriction to states and operators on the \emph{spin-0} boundary particle, which corresponds to antipodally odd bosonic fields in the bulk. As we recall from \eqref{eq:Penrose_reality}, under the same restriction, real twistor functions $\bar f(Y) = f(Y)$, i.e. Hermitian operators $\hat f^\dagger = \hat f$, correspond to real bulk fields. Once everything is even under flipping spinor signs, there is no difference between choosing $\eta=+1$ or $\eta=-1$ in \eqref{eq:state_conjugation}. For concreteness, we choose $\eta=+1$, i.e.:
\begin{align}
 \ket{u}^\dagger = \bra{\bar u} \ ; \quad \bra{u}^\dagger = \ket{-\bar u} \ .
\end{align}

Next, we must address the reality and positivity of the Hermitian norm. For a state $\ket{\psi}$, the norm formally reads:
\begin{align}
 \braket{\psi|\psi} = \frac{1}{2}\int d^2u \braket{\psi|u}\braket{u|\psi} \ . \label{eq:state_norm}
\end{align}
There are two problems with this expression. First $u_\alpha$ is a complex variable, and the integration contour in \eqref{eq:state_norm} isn't specified. Second, $\braket{\psi|u}$ and $\braket{u|\psi}$ aren't actually related by complex conjugation, since the Hermitian conjugate of $\ket{u}$ is $\bra{\bar u}$, not $\bra{u}$. The situation is better for the Hermitian norm of \emph{operators}, which reads:
\begin{align}
 \tr(\hat f\hat f^\dagger) = \frac{1}{4}\int d^2u\,d^2u' \braket{u|\hat f|u'}\braket{u'|\hat f^\dagger|u} = \frac{1}{4}\int d^2u\,d^2u' \Braket{u|\hat f|u'}\overline{\Braket{\bar u|\hat f|-\bar u'}} \ .
\end{align}
This time, since we are integrating over \emph{two} spinors, we can choose a real contour of the form $u' = \zeta\bar u$, where $\zeta$ is another phase factor. We then have:
\begin{align}
 \tr(\hat f\hat f^\dagger) = -\zeta^2\int d^4u \Braket{u|\hat f|\zeta\bar u}\overline{\Braket{\bar u|\hat f|\bar\zeta u}} \ , \label{eq:operator_norm_raw}
\end{align}
where we defined a 4-real-dimensional measure over spinor space:
\begin{align}
 d^4u \equiv -\frac{1}{4}\,d^2u\,d^2\bar u \ , \label{eq:d4u}
\end{align}
and chose the contour's orientation such that the integral over this measure is positive. Now, the two factors in \eqref{eq:operator_norm_raw} are still not complex conjugates of each other, \emph{unless} $\hat f$ satisfies an additional discrete symmetry:
\begin{align}
 \Braket{u|\hat f|u'} = \Braket{\zeta u'|\hat f|\zeta u} \ . \label{eq:even_spins_raw}
\end{align}
For most choices of $\zeta$, such a symmetry is not conformally invariant, i.e. does not have a covariant expression in terms of the twistor function $f(Y)$. The important exception is $\zeta = \pm i$, in which case \eqref{eq:even_spins_raw} takes the form:
\begin{align}
\Braket{u|\hat f|u'} = \Braket{iu'|\hat f|iu} \quad \Longleftrightarrow \quad f(Y) = -f(iY) \ , \label{eq:even_spins}
\end{align}
thanks to the following property of the transformation kernel \eqref{eq:K}:
\begin{align}
 K(u,u';Y) = -K(iu',iu,iY) \ , \label{eq:K_even_spins}
\end{align}
As we can see from \eqref{eq:homogeneity}, the new discrete symmetry \eqref{eq:even_spins} is just the restriction to \emph{even bulk spins}. With this restriction, the norm \eqref{eq:operator_norm_raw} becomes manifestly positive:
\begin{align}
  \tr(\hat f\hat f^\dagger) = \int d^4u \Braket{u|\hat f|i\bar u}\overline{\Braket{u|\hat f|i\bar u}} \ . \label{eq:operator_norm}
\end{align}
Flipping the overall sign in \eqref{eq:even_spins} would correspond to \emph{odd} bulk spins, and yield a norm of the \emph{opposite sign}.

Our somewhat strange contour choice $u' = \pm i\bar u$ is actually quite natural, once we recall that $u$ is the square root of the boundary particle's momentum $\mathbf{p}$. The ``reality condition'' $u' = \pm i\bar u$ means simply $\mathbf{p'} = -\bar{\mathbf{p}}$, which implies in particular that the \emph{total momentum} $\mathbf{p} - \mathbf{p'}$ in the matrix element $\braket{u|\hat f|u'}$ is real. As for the symmetry \eqref{eq:even_spins}, it interchanges the initial and final momenta, while also changing their signs: $(\mathbf{p},\mathbf{p'})\rightarrow (-\mathbf{p'},-\mathbf{p})$. If our 3d boundary were Lorentzian, this would be called a CPT reflection! Thus, the restriction to even bulk spins is simply a restriction to \emph{CPT-invariant operators} in the boundary particle mechanics.

Finally, let us address more fully the decomposition of $\braket{u|\hat f|u'}$ into bulk spins. The symmetry \eqref{eq:K_even_spins} has a continuous generalization:
\begin{align}
 K(u,u';Y) = e^{2i\theta}K(u\cos\theta + iu'\sin\theta, u'\cos\theta + iu\sin\theta; e^{i\theta}Y) \ .
\end{align}
This allows us to write the condition \eqref{eq:homogeneity} for left-handed/right-handed bulk spin $s$ as:
\begin{align}
 \left(u'_\alpha\frac{\del}{\del u_\alpha} + u_\alpha\frac{\del}{\del u'_\alpha} \right)\braket{u|\hat F|u'} = \mp 2s \ .
\end{align}
In particular, for the real contour $u' = i\bar u$, we get:
\begin{align}
 \left(\bar u_\alpha\frac{\del}{\del u_\alpha} - u_\alpha\frac{\del}{\del \bar u_\alpha} \right)\braket{u|\hat F|i\bar u} = \pm 2is \ .
\end{align}
In other words, \emph{helicity is generated by $SO(2)$ rotations between $u$ and $\bar u$}:
\begin{align}
 u_\alpha \rightarrow u_\alpha\cos\theta - \bar u_\alpha\sin\theta \ ; \quad \bar u_\alpha \rightarrow \bar u_\alpha\cos\theta + u_\alpha\sin\theta \ . \label{eq:helicity_rotation}
\end{align}

\subsection{The CFT partition function} \label{sec:multiplet:CFT}

In the above, we constructed a language for the (complexified) quantum mechanics of a boundary particle. In the standard formulation of holography, one of course talks instead about the \emph{field theory} of these particles. However, since we're dealing with a \emph{free} CFT, one should expect that it can be completely captured by the particle mechanics. In this section, we will realize that expectation explicitly. 

\subsubsection{Bilocal and twistor formulations}

Consider the free vector model at the boundary of $dS_4$. We'll begin by discussing the $U(N)$ model, corresponding to all integer spins in the bulk. The truncation to the $Sp(2N)$ model, i.e. to even bulk spins, can be made at the end. The $U(N)$ vector model is given by the action:
\begin{align}
S = -\int d^3\ell\,\bar\phi_I\Box\phi^I \ . \label{eq:S_free}
\end{align}
Here, $\phi^I(\ell)$ and $\bar\phi_I(\ell)$ are spin-0 fields on the 3d boundary with conformal weight $1/2$, and $\Box$ is the conformal Laplacian (which, in a flat conformal frame, is just the ordinary Laplacian). The internal index $I$ runs from $1\dots N$. For dS/CFT (as opposed to AdS/CFT), to have a 2-point function of the correct sign, we must take $\phi^I$ and $\bar\phi_I$ to be anticommuting, in violation of ordinary spin-statistics \cite{Anninos:2011ui,Anninos:2012ft}. We do not worry about this violation, since the Euclidean CFT \eqref{eq:S_free} is never meant to be analytically continued into a sensible Lorentzian theory: Lorentzian physics should instead emerge holographically in the bulk, via a mechanism that we're exploring in this very paper.

The single-trace primaries of the theory \eqref{eq:S_free} are quadratic in the fundamental fields. They form a tower of conserved currents, one for each integer spin, which are dual to the massless higher-spin fields of the bulk theory. On the boundary, it is simpler to treat these currents, together with their descendants, as the Taylor expansion of a \emph{bilocal} operator $\bar\phi_I(\ell')\phi^I(\ell)$ \cite{Das:2003vw}. Coupling bilocal sources $\Pi(\ell',\ell)$ to these operators, we get the action:
\begin{align}
S[\Pi(\ell',\ell)] = -\int d^3\ell\,\bar\phi_I\Box\phi^I - \int d^3\ell' d^3\ell\,\bar\phi_I(\ell')\Pi(\ell',\ell)\phi^I(\ell) \ , \label{eq:S_sources}
\end{align}
from which one immediately reads off the partition function:
\begin{align}
Z_{\text{CFT}}[\Pi(\ell',\ell)] \sim \left(\det{(\Box + \Pi)}\right)^N \ . \label{eq:Z_bilocal}
\end{align}
Here, we are treating $\Box$ and $\Pi$ as matrices with continuous ``indices'' $\ell',\ell$, and the exponent is $N$ rather than $-N$ due to the fields' anticommuting nature. For the action \eqref{eq:S_sources} to be real, the bilocal source should satisfy:
\begin{align}
 \overline{\Pi(\ell',\ell)} = \Pi(\ell,\ell') \ . \label{eq:reality_Pi}
\end{align}

Now, in \cite{Neiman:2017mel}, we introduced a transform from the bilocal sources $\Pi(\ell',\ell)$ into twistor functions $F(Y)$:
\begin{align}
 F(Y) = \frac{1}{\pi i}\int \frac{d^3\ell\,d^3\ell'}{\sqrt{-2\ell\cdot\ell'}}\,e^{\frac{iY\ell\ell' Y}{2\ell\cdot\ell'}}\,\Pi(\ell',\ell) \ , \label{eq:holographic_Penrose}
\end{align}
where we're inserting an extra $i$ in the denominator, to make the corresponding bulk fields real in $dS_4$ (rather than $EAdS_4$, as in \cite{Neiman:2017mel}). The partition function \eqref{eq:Z_bilocal} is now rewritten in terms of higher-spin algebra, as:
\begin{align}
 Z_{\text{CFT}}[F(Y)] \sim \exp\left(-\frac{N}{4}\tr_\star\ln_\star[1 + iF(Y)] \right) \ , \label{eq:Z_twistor}
\end{align}
where $\ln_\star[1+f(Y)]$ is defined by substituting star products into the Taylor expansion of $\ln(1+x)$.

\subsubsection{The connection to boundary particle mechanics} \label{sec:multiplet:CFT:connection_to_mechanics}

Let us now connect this picture of the CFT to our previous discussion of boundary particle mechanics. First, let us just notice that the CFT sources are packaged in \eqref{eq:holographic_Penrose} into a twistor function $F(Y)$, which corresponds, via the boundary Wigner-Weyl transform \eqref{eq:WW_F}, to a quantum operator $\hat F$ in the boundary particle mechanics. Thus, the CFT partition function \eqref{eq:Z_twistor} becomes a functional of this particle-mechanics operator:
\begin{align}
 Z_{\text{CFT}}[\hat F] = \left(\det[1 + i\hat F]\right)^{-N} = \exp\left(-N\tr\ln[1 + i\hat F] \right) \ , \label{eq:Z_twistor_QM}
\end{align}
where the numerical factor in the exponent arises from \eqref{eq:traces}. More explicitly, we can decompose into $n$-point functions:
\begin{align}
  \ln Z_{\text{CFT}}[\hat F] = -N\tr\ln[1 + i\hat F] = N\sum_{n = 1}^{\infty} \frac{(-i)^n}{n}\tr\hat F^n \ , \label{eq:ln_Z}
\end{align}
where each $n$-point function takes a trivial form in the momentum-spinor basis:
\begin{align}
 \tr\hat F^n = \frac{1}{2^n}\int d^2u^{(1)}\dots d^2u^{(n)} \braket{u^{(1)}|\hat F|u^{(2)}}\dots\braket{u^{(n-1)}|\hat F|u^{(n)}}\braket{u^{(n)}|\hat F|u^{(1)}} \ . \label{eq:n_point}
\end{align}

Abstractly, the twistor function $F(Y)$ and the mechanics operator $\hat F$ are one and the same, so that \eqref{eq:Z_twistor_QM} is just a trivial rewriting. It becomes more significant once we represent the operator $\hat F$ concretely, via matrix elements $\braket{u|\hat F|u'}$, as in \eqref{eq:n_point}. Let us now understand the meaning of these matrix elements from the perspective of CFT sources $\Pi(\ell',\ell)$. Our claim is that the momentum-spinor picture of the particle mechanics corresponds to a particular \emph{gauge choice} for $\Pi(\ell',\ell)$ -- a choice in which the sources are all pushed into the single point $n_{(\infty)}$. Thus, it's a gauge in which the sources \emph{vanish almost everywhere}. For the individual local spin-$s$ sources, such a gauge is not possible; it only becomes accessible in the bilocal language, which rearranges the entire HS multiplet in a non-local way.

To understand this claim, let us recall the origin of gauge redundancy in the bilocal setup. The ``current conservation laws'' behind the gauge redundancy of the bilocal source $\Pi(\ell',\ell)$ are just the field equations acting on each factor in $\bar\phi_I(\ell')\phi^I(\ell)$. One can thus get rid of the redundancy by simply switching from the off-shell fields $\bar\phi_I(\ell'),\phi^I(\ell)$ to \emph{solutions of the field equations}, i.e. by taking the bilocal operator on-shell. The problem generally is that the field equations themselves depend on the sources $\Pi(\ell',\ell)$, and thus cannot be solved in a source-independent way. One approach then is to ``cheat'' by simply setting the source to zero. At first sight, this is foolish: on the compact Euclidean $S_3$ boundary of de Sitter space, the source-free massless field equation just \emph{doesn't have} non-trivial solutions! In other words, zero sources lead to zero VEVs. There is, however, a loophole. We can remove one point $n_{(\infty)}$ from the $S_3$ boundary, which is equivalent to choosing a flat $\bbR^3$ conformal frame. On this $\bbR^3$, the free massless field equation \emph{does} have solutions -- plane waves with null momenta. These solutions can be thought of as \emph{wavefunctions} for the free boundary particle -- this is our connection to the particle mechanics of section \ref{sec:multiplet:mechanics}. Note that the null momenta here are complex, which means the corresponding ``plane waves'' have a complex wavevector; however, we already took this carefully into account in section \ref{sec:multiplet:mechanics:spinor_QM}. In the resulting picture, bilocal products $\bar\phi_I(\ell')\phi^I(\ell)$ are replaced by ket/bra products $\ket{u'}\!\bra{u}$, which, instead of coupling to a bilocal source $\Pi(\ell',\ell)$, couple to a particle-mechanics operator $\hat F$.

What happened here to our original bilocal source $\Pi(\ell',\ell)$? On one hand, we assumed that it vanishes everywhere. On the other hand, we still have sources, parameterized by the mechanics operator $\hat F$. And, as we've seen through the transforms \eqref{eq:holographic_Penrose},\eqref{eq:WW_F}, this $\hat F$ is sufficient to describe the \emph{most general} boundary source! The answer to the paradox is that the sources $\Pi(\ell',\ell)$ vanish everywhere \emph{except} the ``point at infinity'' $n_{(\infty)}$. Indeed, at that point, a plane wave does \emph{not} satisfy the conformal free field equation. Thus, our description of the sources via the matrix elements $\braket{u|\hat F|u'}$ is really a \emph{gauge choice}: it preserves all the physical information, while conveniently setting $\Pi(\ell',\ell)$ to zero almost everywhere.

Let's now identify the precise subclass of twistor functions $F(Y)$ (or boundary-particle operators, or bulk solutions) that are generated by this procedure. From the point of view of boundary particle mechanics, we've seen that operators $\hat F$ on the spin-0 boundary particle are even under sign flips of the momentum spinors \eqref{eq:reversal_QM}. For the twistor function $F(Y)$, this corresponds to the condition \eqref{eq:f_antip_odd}, which, in the bulk, implies integer spins with odd antipodal symmetry. Crucially, this is consistent with almost-everywhere-vanishing boundary sources: as we've seen in section \ref{sec:fields}, antipodally odd bulk solutions are indeed associated with the vanishing of source-type boundary data, i.e. their boundary data is purely VEV-type. Thus, the restriction to antipodally odd integer-spin fields is consistent across the boundary and bulk pictures. On top of this, we can now restrict from the $U(N)$ vector model to the $Sp(2N)$ model, i.e. to even bulk spins, so as to ensure a positive quadratic norm \eqref{eq:operator_norm}; recall that this implies restricting to $F(iY) = -F(Y)$, which is equivalent to boundary-CPT invariance \eqref{eq:even_spins} for $\hat F$. Finally, we can impose the reality condition $\bar F(Y) = F(Y)$, which, in the antipodally odd integer-spin sector, is equivalent to real bulk fields and Hermitian $\hat F$. In summary, we end up with real, antipodally odd bulk fields of even spin, which are described by twistor functions subject to the symmetries:
\begin{align}
 F(Y) = -F(Y)\star\delta(Y) = -F(iY) = \bar F(Y) \ , \label{eq:discrete_symm_F}
\end{align}
or by Hermitian, CPT-invariant operators on the spin-0 boundary particle:
\begin{align}
 \Braket{u|\hat F|u'} = \Braket{u|\hat F|-u'} = \Braket{iu'|\hat F|iu} = \overline{\Braket{\bar u'|\hat F|\bar u}} \ . \label{eq:discrete_symm}
\end{align}

\subsubsection{Revised relation to bulk fields}

Let us now examine more closely the asymptotic boundary data of the bulk fields described by $F(Y)$. As we've seen above, with the exception of the singular point $n_{(\infty)}$, these are purely VEV-type boundary data. On the other hand, recall that our construction is just an exotic gauge fixing for some CFT \emph{sources}, which were initially described by the bilocal source $\Pi(\ell',\ell)$. What is the relationship between these sources on one hand and VEVs on the other? The answer is straightforward, and was worked out in \cite{Neiman:2017mel}: in regions of the boundary where the CFT sources vanish (which, in our case, is the entire boundary except $n_{(\infty)}$), the boundary data of the bulk fields described by $F(Y)$ are just the linearized VEVs induced by the sources via the CFT 2-point function (or, equivalently, via regularity on Euclidean $AdS_4$). 

Thus, the bulk interpretation of the CFT sources in our twistor and spinor languages is somewhat subtle. The Penrose transform of our twistor function $F(Y)$ is not the (antipodally even) bulk solution with the corresponding source-type boundary data, but the antipodally \emph{odd} solution with the corresponding \emph{VEV-type} boundary data. However, $Z_{\text{CFT}}$ is still the Hartle-Hawking wavefunction with source-type boundary conditions! This mismatch requires a revision of the construction in section \ref{sec:fields}. Fortunately, the required modification is minimal. First, recall that we are using just the 2-point piece of $Z_{\text{CFT}}$, i.e. the Bunch-Davies wavefunction of free fields in $dS_4$. Now, free fields in $dS_4$ respect a Kahler structure, in which the role of ``multiplication by $i$'' is played precisely by the switch between source-type and VEV-type boundary data via the 2-point function. Under this operation, the quadratic form of the Kahler structure is invariant. But the Bunch-Davies wavefunction in any configuration basis is just the exponent of this quadratic form, restricted to the chosen configuration space! All this is to say that the Bunch-Davies wavefunction is invariant under our switch between source-type and VEV-type boundary data. Thus, the 2-point piece of the partition function \eqref{eq:Z_twistor_QM} can be interpreted directly as the Bunch-Davies wavefunction in the \emph{antipodally odd} basis described by $\hat F$. 

As a result, our derivation of the causal-patch symplectic structure in section \ref{sec:fields} can remain intact, apart from a simple flip of the antipodal symmetry. Recall that in section \ref{sec:fields:Omega}, we decomposed the antipodal map between the ``upright'' and ``time-reversed'' causal patches into a $-\pi i$ time translation and a parity reflection. Thus, to obtain the results for the opposite antipodal symmetry, we can just flip the sign of the phase-space parity operator $\calP^I{}_J$. The symplectic form, commutators and Hamiltonian \eqref{eq:Omega}-\eqref{eq:H} become:
\begin{align}
 \Omega_{IJ} &= iG_{IK}\left( \coth(\pi\omega) - \frac{\calP}{\sinh(\pi\omega)} \right)^K{}_J \ ; \label{eq:Omega_flipped} \\
 \left[\hat{\hat\xi}^I,\hat{\hat\xi}^J\right] &= \left( \coth(\pi\omega) + \frac{\calP}{\sinh(\pi\omega)} \right)^I{}_K\left(G^{-1}\right)^{KJ} \ ; \label{eq:commutators_flipped} \\
 \hat{\hat H} &= \frac{1}{2}\,\hat{\hat\xi}^I G_{IJ}\,\omega^J{}_K \left( \coth(\pi\omega) - \frac{\calP}{\sinh(\pi\omega)} \right)^K{}_L \,\hat{\hat\xi}^L \ , \label{eq:H_flipped}
\end{align}
where $\xi^I$ is now the causal-patch phase space as encoded by the \emph{antipodally odd} twistor function $F(Y)$ (or by the boundary mechanics operator $\hat F$), and $G_{IJ}$ is the quadratic piece of $-\ln Z_{\text{CFT}}$. Note that we placed double hats on the bulk operators in \eqref{eq:commutators_flipped}-\eqref{eq:H_flipped}, to express the fact that they are ``second-quantized'': the \emph{classical} bulk phase space already corresponds to quantum operators $\hat F$ on the boundary particle.

\subsubsection{Sign ambiguities}

Our brave discussion of the precise sign of antipodal symmetry should be contrasted with the more cautious attitude of \cite{Neiman:2017mel}. There, the star products relevant for the partition function \eqref{eq:Z_twistor}, or for examining the corresponding bulk fields' antipodal symmetry, always reduced to complex Gaussian integrals, which could only be defined up to sign. Though we fixed some of these signs by various arguments, the situation remained rather murky. In particular, while the kernel of the transform \eqref{eq:holographic_Penrose} formally looked antipodally odd, this was in clear contrast with the presence of boundary sources. We were therefore careful in \cite{Neiman:2017mel} to restrict some of our central statements to regions where the boundary sources vanish. 

In contrast, in the present paper, we haven't run into any such difficulty. There are two reasons for this, one technical and one conceptual. Conceptually, our spinor-momentum construction is cleaner, in that it pushes all the boundary sources into a singular point. As a result, we should \emph{expect} unambiguously odd antipodal symmetry. At the technical level, the reason that the sign ambiguities disappear is that the relevant star products \emph{no longer look like Gaussians}. This comes about due to our partial breaking of $O(1,4)$ symmetry, via the choice of the observer's boundary endpoints $n_{(\infty)},n_{(0)}$. The latter leads to the decomposition \eqref{eq:Y_decompose} of twistors $Y^a$ into a pair $(\lambda_\alpha,\mu^\alpha)$ of $SO(3)$ spinors. With this decomposition, a quadratic form $Y^a A_{ab} Y^b$ can sometimes be rewritten as $\lambda_\alpha A^\alpha{}_\beta \mu^\beta$, which is now bilinear in the \emph{two different} variables $\lambda_\alpha$ and $\mu^\alpha$. In such a case, a would-be Gaussian integral $\int e^{-YAY}  d^4Y$ becomes a delta-function-type integral of the form \eqref{eq:spinor_deltas}, which can be evaluated without any sign ambiguity. This is what happened in our derivation of eqs. \eqref{eq:K_star_K}-\eqref{eq:tr_K}. 

[NOTE: this manuscript has been revised after understanding that one of the sign ambiguities in \cite{Neiman:2017mel} was actually resolved incorrectly, and propagated into the present paper. In the previous version, there was no minus sign and no $i$ factor in \eqref{eq:Z_twistor}-\eqref{eq:Z_twistor_QM}. Furthermore, the reality properties of \eqref{eq:holographic_Penrose} were misinterpreted, and are indeed confusing: the twistor function in \eqref{eq:holographic_Penrose} does result in real $dS_4$ fields, but is \emph{not} itself real, in contradiction to the discussion in section \ref{sec:multiplet:bulk:Penrose}. In the end, it is the reality of the $dS_4$ fields that matters; the reality of $F(Y)$ follows, as long as one uses a more well-behaved basis such as \eqref{eq:F_K}-\eqref{eq:K}, instead of the sign-ambiguity-ridden \eqref{eq:holographic_Penrose}. Finally, note the potentially confusing fact that the coefficients of all the odd orders $n$ in the effective action \eqref{eq:ln_Z} are imaginary. In fact, it turns out that the traces \eqref{eq:n_point} \emph{vanish} for odd $n$, for any choice of $F(Y)$ that describes a sensible bulk solution (specifically, for any superposition of $SO(4)$ harmonics on the boundary 3-sphere). See \cite{Local} for the detailed clarification of these issues.]

\subsection{Hamiltonian structure of the causal-patch fields in spinor language} \label{sec:multiplet:symplectic}
 
In this section, we apply the prescription \eqref{eq:Omega_flipped}-\eqref{eq:H_flipped} to explicitly extract the Hamiltonian structure in the causal patch from the partition function \eqref{eq:Z_twistor_QM}-\eqref{eq:n_point}.
 
\subsubsection{Results in the momentum-spinor basis} \label{sec:multiplet:symplectic:spinors}
 
We begin with the expression \eqref{eq:n_point} for the CFT $n$-point function. In general, this expression is not completely well-defined: since the ``configuration space'' of momentum spinors $u_\alpha$ is complex, the integral in \eqref{eq:n_point} requires a contour choice. However, in section \ref{sec:multiplet:mechanics:spinor_QM}, we already resolved this problem for the special case $n=2$: choosing the ``real'' contour $u'_\alpha = i\bar u_\alpha$, we obtained the manifestly real and positive expression \eqref{eq:operator_norm}. With this choice of contour, the quadratic piece of $\ln Z_{\text{CFT}}$ takes the form:
\begin{align}
G[\hat F,\hat F] = \left.-\ln Z_{\text{CFT}}[\hat F]\right|_{\text{quadratic}} = \frac{N}{2}\tr\hat F^2 = \frac{N}{2}\int d^4u \Braket{u|\hat F|i\bar u}\Braket{iu|\hat F|\bar u} \ , \label{eq:G_explicit}
\end{align}
where we used the notation $G[\hat F,\hat F]$ to invoke the quadratic form $G_{IJ}$ from \eqref{eq:CFT_Gaussian}, and $d^4u$ is the measure \eqref{eq:d4u} over the 4 real components of the 2-complex-component spinor $u_\alpha$. 

From the 2-point function \eqref{eq:G_explicit}, we can now extract the Hamiltonian structure of bulk fields in the causal patch. Before we begin, recall that the \emph{phase space} of the bulk causal patch is equated with the CFT's \emph{space of source configurations}, which we are presently equating with the \emph{space of operators} of the boundary particle mechanics. These operators are in turn parameterized by matrix elements $\braket{u|\hat F|u'}$, where only the values on some real contour of $(u,u')$ are independent. Following eq. \eqref{eq:G_explicit} and section \ref{sec:multiplet:mechanics:spinor_QM}, we choose the real contour $u' = i\bar u$. Thus, the bulk phase space is coordinatized by the matrix elements $\braket{u|\hat F|i\bar u}$, viewed as a function of the 4 real components of $u_\alpha$. The discrete symmetries \eqref{eq:discrete_symm} imply that this function is invariant under $u_\alpha\rightarrow -u_\alpha$ and $u_\alpha\rightarrow \bar u_\alpha$:
\begin{align}
 \Braket{u|\hat F|i\bar u} = \Braket{-u|\hat F|-i\bar u} = \Braket{\bar u|\hat F|-iu} = \Braket{-\bar u|\hat F|iu} \ . \label{eq:contour_symm}
\end{align}

To apply eqs. \eqref{eq:Omega_flipped}-\eqref{eq:H_flipped}, we will need the action of the time translation generator $\omega$ on the bulk phase space, as well as that of parity $\calP$. Instead of $\omega$ itself, it will be helpful to first work with finite time translations $\calD(t) = e^{-i\omega t}$; we will return to an $\omega$ basis in section \ref{sec:multiplet:symplectic:normal}. The action of $\calD(t)$ takes the form:
\begin{align}
\calD(t) \ : \quad \hat F \ \rightarrow \ \hat D(t)\hat F\hat D(-t) \ , \label{eq:time_translation_adjoint}
\end{align}
where $\hat D(t)$ is the time translation operator on the boundary particle's Hilbert space:
\begin{align}
 \hat D(t) = \frac{1}{2}\,e^{t/2}\int d^2u \Ket{e^{t/2}u}\!\Bra{\vphantom{e^{t/2}}u} \ . \label{eq:time_translation_operator}
\end{align}
Here, the rescaling $u_\alpha\rightarrow e^{t/2}u_{\alpha}$ is read off from \eqref{eq:spinor_time_translation}, and the normalization is to ensure $\hat D^\dagger(t)\hat D(t) = 1$. Altogether, the effect of the time translation \eqref{eq:time_translation_adjoint} on the matrix elements of $\hat F$ is:
\begin{align}
 \Braket{u|\calD(t)\hat F|u'} = e^{-t}\Braket{e^{-t/2}u|\hat F|e^{-t/2}u'} \ . \label{eq:time_translation_matrix_elems}
\end{align}
We can now use these finite time translations to express the functions of $\omega$ appearing in \eqref{eq:Omega_flipped}, which have very simple Fourier transforms:
\begin{align}
\coth(\pi\omega) = \frac{i}{2\pi}\int_{-\infty}^\infty dt\,\coth\frac{t}{2}\,\calD(t) \ ; \quad \frac{1}{\sinh(\pi\omega)} = \frac{i}{2\pi}\int_{-\infty}^\infty dt\,\tanh\frac{t}{2}\,\calD(t) \ , \label{eq:Fourier}
\end{align}
It remains to express the parity operation $\calP$ on the bulk phase space. In the $\bbR^{1,4}$ picture, this operation amounts to flipping the 3 axes orthogonal to the $n_{(0)}\wedge n_{(\infty)}$ plane. This can be accomplished in two steps. First, we use the antipodal map to flip all the 5 axes; since $\hat F$ is antipodally odd, this simply sends $\hat F\rightarrow -\hat F$. Second, we flip back the 2 axes of the $n_{(0)}\wedge n_{(\infty)}$ plane, which can be accomplished by a boost through an imaginary angle $\pi i$. Recalling that boosts in the $n_{(0)}\wedge n_{(\infty)}$ plane are just the observer's time translations, we conclude that this second step is just an imaginary time translation $\calD(\pi i)$. Combining the two steps together, we obtain the parity operation as:
\begin{align}
 \begin{split}
   \calP &= -\calD(\pi i) \ ; \\ 
   \Braket{u|\calP\hat F|u'} &= -\Braket{u|\calD(\pi i)\hat F|u'} = \Braket{-iu|\hat F|-iu'} = \Braket{-iu|\hat F|iu'} \ .
 \end{split} \label{eq:parity_matrix_elems}
\end{align}
In the last step, we used the $2\pi$ rotation symmetry $\braket{u|\hat F|u'}=\braket{u|\hat F|-u'}$ to restore the reality condition $u' = i\bar u$. As expected, parity reverses the boundary particle's momentum \eqref{eq:p_spinors}. 

Plugging everything into eq. \eqref{eq:Omega_flipped}, we obtain the bulk symplectic form as:
\begin{align}
 \begin{split}
  &\Omega_{\text{bulk}}[\hat F_1,\hat F_2] = iG\!\left[\hat F_1, \left( \coth(\pi\omega) - \frac{\calP}{\sinh(\pi\omega)} \right)\!\hat F_2 \right] \\
  &\quad = -\frac{N}{4\pi}\int_{-\infty}^\infty dt \left( \coth\frac{t}{2}\tr\!\left[\hat F_1\calD(t)\hat F_2\right] + \tanh\frac{t}{2}\tr\!\left[\hat F_1\calD(t + \pi i)\hat F_2\right] \right) \\
  &\quad = \frac{N}{4\pi}\int d^4u \Braket{u|\hat F_1|i\bar u}
     \int_{-\infty}^\infty e^t dt \left( \coth\frac{t}{2}\Braket{ie^{t/2}u|\hat F_2|e^{t/2}\bar u} - \tanh\frac{t}{2}\Braket{e^{t/2}u|\hat F_2|ie^{t/2}\bar u} \right) \\
  &\quad = -\frac{N}{4\pi}\int d^4u \Braket{u|\hat F_1|i\bar u}
     \int_{-\infty}^\infty d\alpha\sign(\alpha)\,\frac{\alpha - 1}{\alpha + 1}\Braket{\sqrt{\alpha}\, u|\hat F_2|\sqrt{\alpha}\,i\bar u} \ .
 \end{split} \label{eq:Omega_spinors}
\end{align}
Here, in the third line, we switched integration variables from $t$ to $-t$. In the fourth line, we switched from $t$ to $\alpha = -e^t$ and $\alpha = e^t$ respectively in the first and second terms, which allowed us to combine these terms into a single integral; since the matrix elements $\braket{u|\hat F|u'}$ are insensitive to the spinors' signs, we do not worry about the sign of $\sqrt{\alpha}$. 

Roughly speaking, the symplectic form \eqref{eq:Omega_spinors} couples \emph{parallel or anti-parallel} null momenta $(\mathbf{p},\alpha\mathbf{p})$, with weights $\pm\frac{\alpha - 1}{\alpha + 1}$. Note that, while the momentum $\mathbf{p}$ is complex, the proportionality factor $\alpha$ runs only over \emph{real} values. One can verify, by changing integration variables, that the symplectic form \eqref{eq:Omega_spinors} is antisymmetric under $\hat F_1\leftrightarrow\hat F_2$.

With the symplectic form in hand, we can similarly derive the causal-patch commutators and Hamiltonian. To obtain the commutators \eqref{eq:commutators_flipped}, we write the inverse of the quadratic form \eqref{eq:G_explicit} as:
\begin{align}
G^{-1}\!\left(\braket{u|\hat F|i\bar u}, \braket{v|\hat F|i\bar v}\right) = \frac{1}{2N}\left[ \delta^4(v - iu) + \delta^4(v + iu) + \delta^4(v - i\bar u) + \delta^4(v + i\bar u) \right] \ , \label{eq:G_inverse}
\end{align}
where $\delta^4(u)$ is a delta function with respect to the 4-real-dimensional spinor measure \eqref{eq:d4u}:
\begin{align}
\delta^4(u) \equiv -4\,\delta(u)\delta(\bar u) \ ,
\end{align}
and we took into account the discrete symmetries \eqref{eq:contour_symm}. One can verify that \eqref{eq:G_inverse} is indeed the inverse of \eqref{eq:G_explicit}, by checking the ``matrix inverse'' relation:
\begin{align}
G^{-1}\!\left(\braket{u|\hat F|i\bar u}, G[\hat F,\hat f] \right) = \braket{u|\hat f|i\bar u} \ .
\end{align}
The causal-patch commutators \eqref{eq:commutators_flipped} can now be evaluated as:
\begin{align}
 \begin{split}
   &\left[ \widehat{\braket{u|\hat F|i\bar u}}, \widehat{\braket{v|\hat F|i\bar v}} \right] = G^{-1}\!\left(\Braket{u|\left( \coth(\pi\omega) + \frac{\calP}{\sinh(\pi\omega)} \right)\!\hat F|i\bar u}, \Braket{v|\hat F|i\bar v}\right) \\
   &\quad = \frac{1}{2\pi i}\int_{-\infty}^\infty d\alpha\,\frac{\alpha - 1}{\alpha + 1}\,G^{-1}\!\left(\Braket{\sqrt{\alpha}\,iu|\hat F|\sqrt{\alpha}\,\bar u}, \Braket{v|\hat F|i\bar v} \right) \\
   &\quad = \frac{1}{4\pi iN}\int_{-\infty}^\infty d\alpha\,\frac{\alpha - 1}{\alpha + 1}
     \left[\delta^4\!\left(v - \sqrt{\alpha}\,u\right) + \delta^4\!\left(v + \sqrt{\alpha}\,u\right) + \delta^4\!\left(v - \sqrt{\alpha}\,\bar u\right) + \delta^4\!\left(v + \sqrt{\alpha}\,\bar u\right) \right] \ . \\
 \end{split}
\end{align}
Finally, we turn to the causal-patch Hamiltonian \eqref{eq:H_flipped}. To evaluate it, we need the Fourier transforms of $\omega\coth(\pi\omega)$ and $\omega/\sinh(\pi\omega)$, which can be obtained from \eqref{eq:Fourier} by taking $t$ derivatives:
\begin{align}
 \omega\coth(\pi\omega) = -\frac{1}{4\pi}\int_{-\infty}^\infty \frac{dt}{\sinh^2(t/2)}\,\calD(t) \ ; \quad \frac{\omega}{\sinh(\pi\omega)} = \frac{1}{4\pi}\int_{-\infty}^\infty  \frac{dt}{\cosh^2(t/2)}\,\calD(t) \ .
\end{align}
The Hamiltonian now follows as:
\begin{align}
  \hat{\hat H} &= \frac{N}{16\pi}\int d^4u\,\widehat{\Braket{u|\hat F|i\bar u}}\int_{-\infty}^\infty e^t dt 
    \left( \frac{\widehat{\Braket{ie^{t/2}u|\hat F|e^{t/2}\bar u}}}{\sinh^2(t/2)} + \frac{\widehat{\Braket{e^{t/2}u|\hat F|ie^{t/2}\bar u}}}{\cosh^2(t/2)} \right) \nonumber \\
     &= \frac{N}{4\pi}\int d^4u\,\widehat{\Braket{u|\hat F|i\bar u}} \int_{-\infty}^\infty \frac{|\alpha|d\alpha}{(\alpha + 1)^2}\,\widehat{\Braket{\sqrt{\alpha}\,u|\hat F|\sqrt{\alpha}\,i\bar u}} \ . \label{eq:H_spinors}
\end{align}
Recall that this is the \emph{symmetrically ordered} Hamiltonian, i.e. with zero-point energies included. We will construct the normal-ordered Hamiltonian in the next section, using a more convenient basis.

\subsubsection{Frequency basis and normal ordering} \label{sec:multiplet:symplectic:normal}

We can streamline the above derivation of the bulk Hamiltonian structure, by Fourier-transforming the bulk phase space with respect to the observer's time $t$. To do this, we recall that time translations \eqref{eq:time_translation_matrix_elems} act by rescaling the spinor $u$ in $\braket{u|\hat F|i\bar u}$. Let us then write $u_\alpha = e^{-t/2}\nu_\alpha$, where $\nu_\alpha$ is a unit spinor $\bar\nu\nu = 1$, and $t \equiv -\ln(\bar u u)$. The Fourier transform between time $t$ and energy $E$ reads:
\begin{align}
 \tilde F(E, \nu) &= \int_{-\infty}^\infty dt\,e^{iEt} \Braket{\nu|\calD(t)\hat F|i\bar\nu} = \int_{-\infty}^\infty dt\,e^{(1 - iE)t} \Braket{e^{t/2}\nu|\hat F|ie^{t/2}\bar\nu} \ ; \label{eq:annihilation} \\
 \tilde F^*(E, \nu) &= \int_{-\infty}^\infty dt\,e^{-iEt}\,\overline{\Braket{\nu|\calD(t)\hat F|i\bar\nu}} = \int_{-\infty}^\infty dt\,e^{(1 + iE)t} \Braket{ie^{t/2}\nu|\hat F|e^{t/2}\bar\nu} \ . \label{eq:creation}
\end{align}
Here, $E > 0$ is a positive energy, and we used the discrete symmetries \eqref{eq:discrete_symm}. Upon second quantization, the quantities \eqref{eq:annihilation} will become bulk annihilation operators, with \eqref{eq:creation} the corresponding creation operators. $\tilde F$ is invariant under the action of the discrete symmetries \eqref{eq:discrete_symm} on $\nu_\alpha$:
\begin{align}
 \tilde F(E,\nu) = \tilde F(E,-\nu) = \tilde F(E,\bar\nu) = \tilde F(E,-\bar\nu) \ , \label{eq:discrete_symm_tilde}
\end{align}
and likewise for $\tilde F^*$. The time translations \eqref{eq:time_translation_matrix_elems} act on these quantities as:
\begin{align}
 \calD(t) \ : \quad \tilde F(E, \nu) \ \rightarrow \ e^{-iEt}\tilde F(E, \nu) \ ; \quad \tilde F^*(E, \nu) \ \rightarrow \ e^{iEt}\tilde F^*(E, \nu) \ ,
\end{align}
implying the desired action of their generator $\omega$:
\begin{align}
 \omega \ : \quad \tilde F(E, \nu) \ \rightarrow \ E\tilde F(E, \nu) \ ; \quad \tilde F^*(E, \nu) \ \rightarrow \ -E\tilde F^*(E, \nu) \ .
\end{align}
Finally, let us consider parity \eqref{eq:parity_matrix_elems}. In the present context, the expression $\calP = -\calD(\pi i)$ is not sensible: once we commit to real frequencies $E$, complex-analyticity is violated, so that a time translation by $\pi i$ no longer acts as a reflection. However, we can still use the final expressions in \eqref{eq:parity_matrix_elems}, which say that parity simply sends $u\rightarrow \pm iu$. In our present basis, this becomes:
\begin{align}
 \calP \ : \quad \tilde F(E, \nu) \ \rightarrow \ \tilde F(E, i\nu) \ ; \quad \tilde F^*(E, \nu) \ \rightarrow \ \tilde F^*(E, i\nu) \ ,
\end{align}
which allows us to decompose $\tilde F$ into parity-even and parity-odd parts:
\begin{align}
 \tilde F_\pm(E, \nu) = \frac{1}{2}\left(\tilde F(E, \nu) \pm \tilde F(E, i\nu) \right) \ ; \quad 
 \tilde F^*_\pm(E, \nu) = \frac{1}{2}\left(\tilde F^*(E, \nu) \pm \tilde F^*(E, i\nu) \right) \ .
\end{align}

The inverse of the Fourier transform \eqref{eq:annihilation}-\eqref{eq:creation} reads:
\begin{align}
 \braket{u|\hat F|i\bar u} = \frac{1}{\bar u u}\int_0^\infty \frac{dE}{2\pi}
   \left((\bar u u)^{iE}\tilde F\!\left(E, \frac{u}{\sqrt{\bar u u}}\right) + (\bar u u)^{-iE}\tilde F^*\!\left(E, \frac{iu}{\sqrt{\bar u u}}\right) \right) \ .
\end{align}
Plugging this into the boundary 2-point function \eqref{eq:G_explicit}, we get simply:
\begin{align}
 \begin{split}
   G[\tilde F,\tilde F] &= \frac{N}{2}\int_0^\infty \frac{dE}{2\pi} \int d^3\nu\,\tilde F^*(E,\nu)\,\tilde F(E,\nu) \\
     &= \frac{N}{2}\int_0^\infty \frac{dE}{2\pi} \int d^3\nu \left(\tilde F^*_+(E,\nu)\,\tilde F_+(E,\nu) + \tilde F^*_-(E,\nu)\,\tilde F_-(E,\nu)  \right) \ . 
 \end{split} \label{eq:G_tilde}
\end{align}
Here, $d^3\nu$ is the measure on the 3-sphere of unit spinors, defined via:
\begin{align}
 d^4u = \rho^3 d\rho\,d^3\nu \ , \quad \text{where} \quad u_\alpha = \rho\nu_\alpha \ .
\end{align}
The inverse of the quadratic form \eqref{eq:G_tilde} reads:
\begin{align}
 \begin{split}
   G^{-1}\!\left(\tilde F(E,\nu), \tilde F^*(E',\nu') \right) &= \frac{8\pi}{N}\,\delta(E' - E)\,\delta^3(\nu',\nu)_{\text{symm}} \ ; \\
   G^{-1}\!\left(\tilde F(E,\nu), \tilde F(E',\nu') \right) &= G^{-1}\!\left(\tilde F^*(E,\nu), \tilde F^*(E',\nu') \right) = 0 \ ,
 \end{split} \label{eq:G_inverse_tilde}
\end{align}
where $\delta^3(\nu',\nu)_{\text{symm}}$ is the delta function on the 3-sphere of unit spinors, symmetrized under \eqref{eq:discrete_symm_tilde}:
\begin{align}
 \delta^3(\nu',\nu)_{\text{symm}} = \frac{1}{4}\left[ \delta^3(\nu',\nu) + \delta^3(\nu',-\nu) + \delta^3(\nu',\bar\nu) + \delta^3(\nu',-\bar\nu) \right] \ .
\end{align}

The bulk symplectic form \eqref{eq:Omega_flipped} and the commutators \eqref{eq:commutators_flipped} can now be evaluated immediately:
\begin{align}
  &\Omega_{\text{bulk}}[\tilde F_1,\tilde F_2] = \frac{iN}{4}\int_0^\infty \frac{dE}{2\pi} \int d^3\nu\,\tilde F_1^*(E, \nu)
    \left(\coth(\pi E) \tilde F_2(E,\nu) - \frac{\tilde F_2(E,i\nu)}{\sinh(\pi E)} \right) - (1\leftrightarrow 2) \ ; \nonumber \\
  &\left[\hat{\tilde F}(E,\nu), \hat{\tilde F}^\dagger(E',\nu') \right] = \frac{8\pi}{N}\,\delta(E' - E) 
    \left( \coth(\pi E)\,\delta^3(\nu',\nu)_{\text{symm}} + \frac{\delta^3(\nu',i\nu)_{\text{symm}}}{\sinh(\pi E)} \right) \ ,
\end{align}
with $[\hat{\tilde F},\hat{\tilde F}]$ and $[\hat{\tilde F}^\dagger,\hat{\tilde F}^\dagger]$ vanishing. Note that we're using single hats on the bulk operators to reduce clutter, but the context is still that of second quantization. In parity-even and parity-odd components, the commutators \eqref{eq:commutators_flipped} read:
\begin{align}
 \begin{split}
   \left[\hat{\tilde F}_+(E,\nu), \hat{\tilde F}_+^\dagger(E',\nu') \right] &= \frac{4\pi}{N}\,\delta(E' - E)\coth\frac{\pi E}{2} \left( \delta^3(\nu',\nu)_{\text{symm}} + \delta^3(\nu',i\nu)_{\text{symm}} \right) \ ; \\
   \left[\hat{\tilde F}_-(E,\nu), \hat{\tilde F}_-^\dagger(E',\nu') \right] &= \frac{4\pi}{N}\,\delta(E' - E)\tanh\frac{\pi E}{2} \left( \delta^3(\nu',\nu)_{\text{symm}} - \delta^3(\nu',i\nu)_{\text{symm}} \right) \ ,
 \end{split}
\end{align}
with all other commutators vanishing. Finally, the causal-patch Hamiltonian \eqref{eq:H_flipped} is now easy to write in normal ordering:
\begin{align}
 {:\!\hat H\!:} = \frac{N}{4}\int_0^\infty \frac{E dE}{2\pi} \int d^3\nu \left( \tanh\frac{\pi E}{2}\,\hat{\tilde F}_+^\dagger(E, \nu) \hat{\tilde F}_+(E, \nu) + \coth\frac{\pi E}{2}\,\hat{\tilde F}_-^\dagger(E, \nu) \hat{\tilde F}_-(E, \nu) \right) \ .
\end{align}
These results are remarkably similar to the results of \cite{Halpern:2015zia} for a single bulk scalar. Essentially, the only difference is that our unit spinor $\nu_\alpha$ is integrated over $S_3$, whereas in \cite{Halpern:2015zia} we integrated a unit \emph{vector} over $S_2$ (using spherical harmonics). The two pictures are related by the Hopf fibration of $S_3$ over $S_2$, in which the $S_1$ fibers encode the helicity degree of freedom of the bulk fields. For the bulk scalar, the helicity doesn't come into play, so we only see harmonics on the spatial $S_2$. For a single bulk field with nonzero spin, we should expect \emph{spin-weighted} spherical harmonics on $S_2$. The HS multiplet just rearranges all these spin-weighted harmonics into simple \emph{scalar} harmonics on $S_3$. 

Note that in standard  presentations of the Hopf fibration, the $S_1$ fibers correspond to phase rotations $\nu_\alpha\rightarrow e^{i\theta}\nu_\alpha$ of the spinor. In our case, that role is played instead by the rotations \eqref{eq:helicity_rotation}, which mix $\nu_\alpha$ with its complex conjugate.
 
\subsection{Higher spin symmetry -- full and residual} \label{sec:multiplet:symmetry}

In the above, we hardly discussed the HS symmetry of our construction. This symmetry will of course be important for any future treatment of the bulk interactions. In fact, HS symmetry, appropriately reduced by the choice of observer, was present through much of our discussion. However, it is only present at the complex-analytic level, and is broken by our choice of the real contour $u' = i\bar u$. 

In the language of twistor functions, before reducing to even spins, the infinitesimal generators of HS symmetry consist of even functions $g(Y)$, acting in the adjoint:
\begin{align}
 \delta F(Y) = g(Y)\star F(Y) - F(Y)\star g(Y) \ . \label{eq:HS_symm_star}
\end{align}
On the bulk master field \eqref{eq:Penrose}, this results in the so-called twisted adjoint action:
\begin{align}
 \delta C(x;Y) = g(Y)\star C(x;Y) - C(x;Y)\star g(ixY) \ .
\end{align}
Let's now apply our interpretation of $F(Y)$ as the Wigner-Weyl transform of an operator $\hat F$ on the boundary particle's Hilbert space. The HS transformation \eqref{eq:HS_symm_star} then takes the form:
\begin{align}
 \delta\hat F = \hat g\hat F - \hat F\hat g \ , \label{eq:HS_symm_QM}
\end{align} 
where the operator $\hat g$ is the Wigner-Weyl transform of $g(Y)$. The $2\pi$ rotation symmetry $\hat R\hat F = \hat F\hat R = \hat F$ of the boundary particle, or its twistor version $F(Y)\star\delta(Y) = \delta(Y)\star F(Y) = -F(Y)$, is preserved by the transformation \eqref{eq:HS_symm_star}-\eqref{eq:HS_symm_QM}. This is a result of $g(Y)$ being even $g(-Y) = g(Y)$, which means that $\hat g$ satisfies $\hat R\hat g\hat R = \hat g$.

Thus, HS symmetry is just the group of basis transformations on the boundary particle's Hilbert space. Both the full partition function \eqref{eq:Z_twistor_QM} and the individual $n$-point functions \eqref{eq:n_point} are manifestly invariant under this symmetry. The reduction \eqref{eq:even_spins} to even spins restricts us further to generators satisfying $g(iY) = -g(Y)$, which correspond to CPT-invariant basis transformations $\hat g$ in the boundary mechanics.

The causal-patch symplectic form \eqref{eq:Omega_spinors} is \emph{not} invariant under the full symmetry algebra \eqref{eq:HS_symm_star}-\eqref{eq:HS_symm_QM}. This is to be expected: already among the geometric $O(1,4)$ symmetries, we only expect $\Omega_{\text{bulk}}$ to be invariant under the observer's subgroup $SO(1,1)\times O(3) = \bbR\times O(3)$ of time translations and rotations. The relevant question, then, is whether $\Omega_{\text{bulk}}$ is invariant under the higher-spin extension $\mathfrak{hs}[SO(1,1)\times O(3)]$ of the \emph{observer's} geometric symmetries? As we will see immediately, the answer is yes. 

First, let's recall from section \ref{sec:multiplet:mechanics:spinors} that the observer's $SO(1,1)\times O(3)$ symmetry is generated by the $\lambda_\alpha\mu^\beta$ components of the $O(1,4)$ generators $Y^a Y^b$. Therefore, its natural HS extension consists of arbitrary products of these generators. From the algebra \eqref{eq:HS_algebra_spinors}, we see that these are just the functions $g(Y^a) = g(\lambda_\alpha,\mu^\alpha)$ that consist of equal powers of $\lambda_\alpha$ and $\mu^\alpha$. This can be encoded as a homogeneity condition:
\begin{align}
 g(\rho\lambda_\alpha,\rho^{-1}\mu^\alpha) = g(\lambda_\alpha,\mu^\alpha) \ . \label{eq:g_homogeneity}
\end{align}
Now, recall that a rescaling of the form \eqref{eq:g_homogeneity} is precisely the effect of time translations \eqref{eq:spinor_time_translation}. Thus, the residual HS symmetry after choosing an observer can be characterized very simply: it consists of those algebra elements $g(Y)$, or operators $\hat g$, which commute with the observer's time translations! Under this symmetry, the symplectic form as written on the second line of \eqref{eq:Omega_spinors} is manifestly invariant.

There is one more useful way to organize the generators of the residual HS symmetry. Instead of \emph{products} of infinitesimal $SO(1,1)\times O(3)$ generators, we can use their \emph{exponentials}, i.e. finite \emph{group elements} of $SO(1,1)\times O(3)$, as an (overcomplete) basis for the residual HS generators. In \eqref{eq:time_translation_operator}, we expressed a finite time translation as an operator in the boundary particle mechanics. The same can be done with a rotation by a vector of angles $\boldsymbol{\theta}$, which we can write as an exponential $e^{i\boldsymbol{\theta}\cdot\boldsymbol{\sigma}/2} = e^{-\boldsymbol{\theta}\cdot\boldsymbol{\tau}/2}$ of Pauli matrices. Overall, the boundary particle mechanics operator describing a general $SO(1,1)\times SO(3)$ transformation reads:
\begin{align}
 \hat D(t,\boldsymbol{\theta}) &= \frac{1}{2}\,e^{t/2}\int d^2u \Ket{e^{(t - \boldsymbol{\theta}\cdot\boldsymbol{\tau})/2}u}\!\Bra{\vphantom{e^{t/2}}u} \ .
\end{align}
This operator can now play two different roles. First, it can act in the usual way as a \emph{finite group element} of $SO(1,1)\times SO(3)$, in a simple generalization of \eqref{eq:time_translation_matrix_elems}:
\begin{align}
 \begin{split}
   \hat F \ &\rightarrow \ \hat D(t,\boldsymbol{\theta})\hat F\hat D(-t,-\boldsymbol{\theta}) \ ; \\ 
   \Braket{u|\hat F|u'} \ &\rightarrow \ e^{-t}\Braket{e^{(\boldsymbol{\theta}\cdot\boldsymbol{\tau} - t)/2}u|\hat F|e^{(\boldsymbol{\theta}\cdot\boldsymbol{\tau}-t)/2}u'} \ . 
 \end{split} \label{eq:geometric_symmetry}
\end{align}
Second, as mentioned above, it can act as an \emph{infinitesimal generator} of the residual HS symmetry $\mathfrak{hs}[O(1,1)\times O(3)]$:
\begin{align}
 \begin{split}
   \delta\hat F &= \varepsilon\left[\hat D(t,\boldsymbol{\theta}), \hat F\right] \ ; \\
   \Braket{u|\delta\hat F|u'} &= \varepsilon\left( e^{-t/2}\Braket{e^{(\boldsymbol{\theta}\cdot\boldsymbol{\tau} - t)/2} u|\hat F|u'} - e^{t/2}\Braket{u|\hat F|e^{(t-\boldsymbol{\theta}\cdot\boldsymbol{\tau})/2} u'} \right) \ ,
 \end{split} \label{eq:HS_generator}
\end{align}
where $\varepsilon$ is an infinitesimal number. In this form, the residual HS symmetry \eqref{eq:HS_generator} is clearly broken by our reality condition $u' = i\bar u$, in contrast with the geometric symmetry \eqref{eq:geometric_symmetry}, which is preserved. Thus, our expressions for e.g. the 2-point function $G$ and the bulk symplectic form $\Omega_{\text{bulk}}$ are invariant under the residual HS symmetry, but only when written in terms of unspecified contour integrals. The expressions with the fixed $u' = i\bar u$ contour are \emph{not} manifestly HS-invariant, and therefore neither are the frequency-basis expressions from section \ref{sec:multiplet:symplectic:normal}. Thus, we are forced to choose between manifest HS symmetry and manifest reality/positivity.

\section{Discussion} \label{sec:discuss}

In this paper, we constructed the symplectic structure of linearized higher-spin fields in a de Sitter causal patch out of the boundary CFT partition function. From the point of view of dS/CFT, this is the first derivation of bulk causal structure (as encoded in the quantum-mechanical commutators) from the timeless, Euclidean boundary theory. From the point of view of HS gravity, this is the first time that a manifestly HS-invariant symplectic structure has been written down for a bulk region (though, as we've seen, manifest HS symmetry is in conflict with manifest reality/positivity). In addition, our construction is a proof-of-concept for a new kind of holography, in which the conformal boundary manifold is not essential: instead of boundary \emph{field theory}, we can make do with on-shell \emph{particle mechanics}, in which the degrees of freedom live just at the endpoints of the causal patch.

One element that's missing is a more explicit dictionary between our boundary momentum spinor variables and the bulk higher-spin fields. At the moment, this dictionary consists of two steps: the boundary Wigner-Weyl transform \eqref{eq:F_K} between momentum spinors and twistors, followed by the Penrose transform \eqref{eq:Penrose},\eqref{eq:packaging} between twistors and bulk fields. An important exercise will be to calculate explicitly the transform between the boundary momentum spinors and the \emph{lightlike initial data} of bulk HS fields on one of the observer's horizons. Since the symplectic structure on such lightlike initial data is well-known, one can then perform a full consistency check on the construction in this paper. Our plan is to perform this check in a separate publication, which will be concerned more broadly with bulk applications of the momentum-spinor variables.

As discussed in section \ref{sec:intro:HS}, our construction of an HS-symmetric symplectic structure in the causal patch does not have a global analog: there is no symplectic form with the full $\mathfrak{hs}[O(1,4)]$ symmetry. There is, however, another case of partially broken HS symmetry, which, to our knowledge, has not received much attention: one can try and construct a symplectic structure for the $Y$-dependence of the master field $C(x;Y)$ at a given bulk point $x$. Via the unfolding scheme \eqref{eq:unfolding}, this $Y$-dependence essentially encodes the lightlike initial data on the lightcone of $x$, where, once again, the symplectic structure for individual spins is well-known. One might even expect that this case should be \emph{easier} than the dS causal patch, since the residual spacetime symmetry $O(1,3)$ is larger than the residual symmetry $SO(1,1)\times O(3)$ in the causal-patch case. 

Of course, the main open question is how to extend our treatment of \emph{free} HS theory in the de Sitter causal patch to the interacting level. Our main reason for hope on this front is that the CFT partition function \eqref{eq:Z_twistor_QM}-\eqref{eq:n_point} in the momentum-spinor variables is fully known, including all the higher $n$-point functions which correspond to bulk interactions. On the other hand, just as with the 2-point function in the present paper, a new layer of dictionary will be required to translate these Euclidean $n$-point functions into information about Lorentzian physics in the causal patch. Here, new insight will be needed. 

A plausible intermediate goal would be to try and find the \emph{classical S-matrix} of interacting HS theory in the de Sitter causal patch, i.e. the mapping of bulk field values between the initial and final horizons. In \cite{Halpern:2015zia}, it was argued that, even for free fields, this S-matrix cannot be read off from the CFT. However, the free-field S-matrix can be found more directly. In \cite{Hackl:2014txa,Halpern:2015zia}, this was achieved for the spin-0 field using boundary-to-bulk propagators. For the full HS multiplet, such an approach would be cumbersome, but one could use instead the various transforms developed here and in \cite{Neiman:2017mel} between bulk fields and the non-local twistor \& momentum-spinor languages. With the free S-matrix thus obtained, one could try and extract corrections from the CFT's higher $n$-point functions, by extending the technique of section \ref{sec:fields}. 

Eventually, we'll need to simultaneously take into account the theory's interactions and its quantum nature. The main difficulty here is that, unlike for free bulk fields, we do not have an independently existing understanding of the causal, Hamiltonian or commutator structure for interacting HS gravity. Somehow, we'll need to simultaneously construct the holographic derivation of these structures \emph{and} the correct way to think about them in the bulk theory. 
One strategy is to begin by gathering experience with interacting Yang-Mills and GR, within a similar setup \& variables to the ones we have constructed here for HS gravity.

\section*{Acknowledgements}

I am grateful to Eugene Skvortsov, Juan Maldacena, Massimo Taronna, Dionysios Anninos, Adrian David, LinQing Chen and Tomonori Ugajin for discussions. This work was supported by the Quantum Gravity Unit of the Okinawa Institute of Science and Technology Graduate University (OIST). Much of the writing took place under the kind hospitality of the ``New Frontiers in String Theory 2018'' workshop at YITP and the ``Recent Developments in Gauge Theory and String Theory'' workshop at Keio University.


\begin{thebibliography}{99}

\bibitem{Vasiliev:1995dn} 
M.~A.~Vasiliev,
``Higher spin gauge theories in four-dimensions, three-dimensions, and two-dimensions,''
Int.\ J.\ Mod.\ Phys.\ D {\bf 5}, 763 (1996)
[hep-th/9611024].

\bibitem{Vasiliev:1999ba} 
M.~A.~Vasiliev,
``Higher spin gauge theories: Star product and AdS space,''
In *Shifman, M.A. (ed.): The many faces of the superworld* 533-610
[hep-th/9910096].

\bibitem{Klebanov:2002ja} 
I.~R.~Klebanov and A.~M.~Polyakov,
``AdS dual of the critical O(N) vector model,''
Phys.\ Lett.\ B {\bf 550}, 213 (2002)
[hep-th/0210114].

\bibitem{Maldacena:1997re} 
J.~M.~Maldacena,
``The Large N limit of superconformal field theories and supergravity,''
Int.\ J.\ Theor.\ Phys.\  {\bf 38}, 1113 (1999)
[Adv.\ Theor.\ Math.\ Phys.\  {\bf 2}, 231 (1998)]
doi:10.1023/A:1026654312961
[hep-th/9711200].

\bibitem{Witten:1998qj} 
E.~Witten,
``Anti-de Sitter space and holography,''
Adv.\ Theor.\ Math.\ Phys.\  {\bf 2}, 253 (1998)
[hep-th/9802150].

\bibitem{Aharony:1999ti} 
O.~Aharony, S.~S.~Gubser, J.~M.~Maldacena, H.~Ooguri and Y.~Oz,
``Large N field theories, string theory and gravity,''
Phys.\ Rept.\  {\bf 323}, 183 (2000)
doi:10.1016/S0370-1573(99)00083-6
[hep-th/9905111].

\bibitem{Anninos:2011ui} 
D.~Anninos, T.~Hartman and A.~Strominger,
``Higher Spin Realization of the dS/CFT Correspondence,''
Class.\ Quant.\ Grav.\  {\bf 34}, no. 1, 015009 (2017)
doi:10.1088/1361-6382/34/1/015009
[arXiv:1108.5735 [hep-th]].

\bibitem{Maldacena:2002vr} 
J.~M.~Maldacena,
``Non-Gaussian features of primordial fluctuations in single field inflationary models,''
JHEP {\bf 0305}, 013 (2003)
[astro-ph/0210603].

\bibitem{Anninos:2017eib} 
D.~Anninos, F.~Denef, R.~Monten and Z.~Sun,
``Higher Spin de Sitter Hilbert Space,''
arXiv:1711.10037 [hep-th].

\bibitem{Halpern:2015zia} 
I.~F.~Halpern and Y.~Neiman,
``Holography and quantum states in elliptic de Sitter space,''
JHEP {\bf 1512}, 057 (2015)
doi:10.1007/JHEP12(2015)057
[arXiv:1509.05890 [hep-th]].

\bibitem{Parikh:2002py} 
M.~K.~Parikh, I.~Savonije and E.~P.~Verlinde,
``Elliptic de Sitter space: dS/Z(2),''
Phys.\ Rev.\ D {\bf 67}, 064005 (2003)
[hep-th/0209120].

\bibitem{Folacci:1986gr} 
A.~Folacci and N.~G.~Sanchez,
``Quantum Field Theory and the 'Elliptic Interpretation' of De Sitter Space-time,''
Nucl.\ Phys.\ B {\bf 294}, 1111 (1987).

\bibitem{Koch:2010cy} 
R.~de Mello Koch, A.~Jevicki, K.~Jin and J.~P.~Rodrigues,
``$AdS_4/CFT_3$ Construction from Collective Fields,''
Phys.\ Rev.\ D {\bf 83}, 025006 (2011)
doi:10.1103/PhysRevD.83.025006
[arXiv:1008.0633 [hep-th]].

\bibitem{Das:2012dt} 
D.~Das, S.~R.~Das, A.~Jevicki and Q.~Ye,
``Bi-local Construction of Sp(2N)/dS Higher Spin Correspondence,''
JHEP {\bf 1301}, 107 (2013)
doi:10.1007/JHEP01(2013)107
[arXiv:1205.5776 [hep-th]].

\bibitem{Koch:2014aqa} 
R.~de Mello Koch, A.~Jevicki, J.~P.~Rodrigues and J.~Yoon,
``Canonical Formulation of $O(N)$ Vector/Higher Spin Correspondence,''
J.\ Phys.\ A {\bf 48}, no. 10, 105403 (2015)
doi:10.1088/1751-8113/48/10/105403
[arXiv:1408.4800 [hep-th]].

\bibitem{Das:2003vw} 
S.~R.~Das and A.~Jevicki,
``Large N collective fields and holography,''
Phys.\ Rev.\ D {\bf 68}, 044011 (2003)
doi:10.1103/PhysRevD.68.044011
[hep-th/0304093].

\bibitem{Vasiliev:2012vf} 
M.~A.~Vasiliev,
``Holography, Unfolding and Higher-Spin Theory,''
J.\ Phys.\ A {\bf 46}, 214013 (2013)
[arXiv:1203.5554 [hep-th]].

\bibitem{Gelfond:2008ur} 
O.~A.~Gelfond and M.~A.~Vasiliev,
``Higher Spin Fields in Siegel Space, Currents and Theta Functions,''
JHEP {\bf 0903}, 125 (2009)
doi:10.1088/1126-6708/2009/03/125
[arXiv:0801.2191 [hep-th]].

\bibitem{Anninos:2011af} 
D.~Anninos, S.~A.~Hartnoll and D.~M.~Hofman,
``Static Patch Solipsism: Conformal Symmetry of the de Sitter Worldline,''
Class.\ Quant.\ Grav.\  {\bf 29}, 075002 (2012)
doi:10.1088/0264-9381/29/7/075002
[arXiv:1109.4942 [hep-th]].

\bibitem{Neiman:2017mel} 
Y.~Neiman,
``The holographic dual of the Penrose transform,''
JHEP {\bf 1801}, 100 (2018)
doi:10.1007/JHEP01(2018)100
[arXiv:1709.08050 [hep-th]].

\bibitem{Hackl:2014txa} 
L.~Hackl and Y.~Neiman,
``Horizon complementarity in elliptic de Sitter space,''
Phys.\ Rev.\ D {\bf 91}, no. 4, 044016 (2015)
[arXiv:1409.6753 [hep-th]].

\bibitem{Gelfond:2013xt} 
O.~A.~Gelfond and M.~A.~Vasiliev,
``Operator algebra of free conformal currents via twistors,''
Nucl.\ Phys.\ B {\bf 876}, 871 (2013)
doi:10.1016/j.nuclphysb.2013.09.001
[arXiv:1301.3123 [hep-th]].

\bibitem{Penrose:1986ca} 
R.~Penrose and W.~Rindler,
``Spinors And Space-time. Vol. 2: Spinor And Twistor Methods In Space-time Geometry,''
Cambridge, Uk: Univ. Pr. (1986) 501p

\bibitem{Ward:1990vs} 
R.~S.~Ward and R.~O.~Wells,
``Twistor geometry and field theory,''
Cambridge, UK: Univ. Pr. (1990) 520p

\bibitem{Wigner} 
W.~B.~Case,
``Wigner functions and Weyl transforms for pedestrians,''
Am.\ J.\ Phys.\  {\bf 76}, 937 (2008).

\bibitem{Neiman:2014npa} 
Y.~Neiman,
``Antipodally symmetric gauge fields and higher-spin gravity in de Sitter space,''
JHEP {\bf 1410}, 153 (2014)
[arXiv:1406.3291 [hep-th]].

\bibitem{Ng:2012xp} 
G.~S.~Ng and A.~Strominger,
``State/Operator Correspondence in Higher-Spin dS/CFT,''
Class.\ Quant.\ Grav.\  {\bf 30}, 104002 (2013)
[arXiv:1204.1057 [hep-th]].

\bibitem{Harlow:2011ke} 
D.~Harlow and D.~Stanford,
``Operator Dictionaries and Wave Functions in AdS/CFT and dS/CFT,''
arXiv:1104.2621 [hep-th].

\bibitem{Hartle:1983ai} 
J.~B.~Hartle and S.~W.~Hawking,
``Wave Function of the Universe,''
Phys.\ Rev.\ D {\bf 28}, 2960 (1983).

\bibitem{Neiman:2013hca} 
Y.~Neiman,
``Twistors and antipodes in de Sitter space,''
Phys.\ Rev.\ D {\bf 89}, no. 6, 063521 (2014)
[arXiv:1312.7842 [hep-th]].

\bibitem{Didenko:2009td} 
V.~E.~Didenko and M.~A.~Vasiliev,
``Static BPS black hole in 4d higher-spin gauge theory,''
Phys.\ Lett.\ B {\bf 682}, 305 (2009)
Erratum: [Phys.\ Lett.\ B {\bf 722}, 389 (2013)]
doi:10.1016/j.physletb.2013.04.021, 10.1016/j.physletb.2009.11.023
[arXiv:0906.3898 [hep-th]].

\bibitem{Pilch:1984aw} 
K.~Pilch, P.~van Nieuwenhuizen and M.~F.~Sohnius,
``De Sitter Superalgebras and Supergravity,''
Commun.\ Math.\ Phys.\  {\bf 98}, 105 (1985).
doi:10.1007/BF01211046

\bibitem{Anninos:2012ft} 
D.~Anninos, F.~Denef and D.~Harlow,
``Wave function of Vasiliev’s universe: A few slices thereof,''
Phys.\ Rev.\ D {\bf 88}, no. 8, 084049 (2013)
doi:10.1103/PhysRevD.88.084049
[arXiv:1207.5517 [hep-th]].

\bibitem{Local}
A.~David and Y.~Neiman,
``Higher-spin symmetry vs. boundary locality, and a rehabilitation of dS/CFT,''
to appear.

\end{thebibliography}
\end{document}